\documentclass[elsart12,epsf,eqsecnum,graphics,cite,nofootinbib]{revtex4}

\usepackage{epsfig,epsf}
\usepackage{graphicx}
\usepackage{grffile}
\usepackage{amsmath,amsfonts,amssymb,amsthm,nccmath,latexsym,mathtools} \usepackage[mathscr]{euscript}

\newcommand{\beq}{\begin{eqnarray}}
\newcommand{\eeq}{\end{eqnarray}}
\newcommand{\be}{\begin{eqnarray*}}
\newcommand{\ee}{\end{eqnarray*}}
\DeclareMathOperator{\tr}{tr}

\def\pp{p_{\perp}}
\def\by{\bar{y}}
\def\bz{\bar{z}}
\def\bx{\bar{x}}

\def\bq{\bar{q}}
\def\bb{\bar{b}}

\def\bc{\bar{c}}
\def\Ai{A^i_{\xi,x_B}}
\def\Am{A^m_{\xi,x_B}}
\usepackage{hyperref}

\begin{document}

\voffset1.5cm

\title{Single-inclusive particle production in proton-nucleus collisions at next-to-leading order in the hybrid formalism}
\author{Tolga Altinoluk$^1$, N\'estor Armesto$^1$, Guillaume Beuf$^{1,2}$, Alex Kovner$^{3}$ and Michael Lublinsky$^2$}

\affiliation{
$^1$ Departamento de F\'{i}õsica de Part\'{i}culas and IGFAE, Universidade de Santiago de Compostela, 15706 Santiago de Compostela, Galicia-Spain\\
$^2$ Department of Physics, Ben-Gurion University of the Negev,
Beer Sheva 84105, Israel\\
$^3$ Physics Department, University of Connecticut, 2152 Hillside
Road, Storrs, CT 06269-3046, USA}

\begin{abstract}
We reconsider the perturbative next-to-leading calculation of the single inclusive hadron production in the framework of the hybrid formalism, applied  to hadron production in proton-nucleus collisions.
Our analysis, performed in the wave function approach, differs from the previous works in three points. First, we are careful to specify unambiguously the rapidity interval that has to be included in the evolution of the leading-order eikonal scattering amplitude. This is important, since varying this interval by a number of order unity changes the next-to-leading order correction that the calculation is meant to determine. Second, we introduce the explicit requirement that fast fluctuations in the projectile wave function which only exist for a short time are not resolved by the target. This Ioffe time cutoff also strongly affects the next-to-leading order terms. Third, our result does not employ the approximation of a large number of colors. Our final expressions are unambiguous and do not coincide at next-to-leading order with the results available in the literature.

\end{abstract}
\maketitle
\section{Introduction and conclusions}
It has been suggested thirty years ago \cite{glr}, that at high energies hadronic structure is considerably different from that at lower energies as hadrons exhibit perturbative saturation. Observation of saturation is of course a very interesting possibility, as it would open a door for exploring a qualitative new regime of QCD - the regime of dense saturated states, which is nevertheless perturbative in the sense that the relevant coupling constant remains small.

There has been a lot of activity in the last 20 years to try to better understand this regime theoretically. With the advent of the Relativistic Heavy Ion Collider and, later, the Large Hadron Collider, many attempts to describe available data in the framework of saturation have also been made.

It is fair to say that at the moment we do not have a clear understanding, whether effects of saturation (or Color Glass Condensate (CGC), as its weak coupling implementation
 \cite{Mueller,balitsky,balitsky1,Kovchegov,JIMWLK,cgc})
  have already been seen in the current experiments, although some saturation-based calculations provide good description of data
  (e.g. \cite{KLN,GLLM,cronin,cronin1,LR,DV}). One of the major reasons for this, is that the calculational precision of the saturation based approaches is still far from satisfactory. For example, only a small (although important) part of next-to-leading order corrections (the running coupling effects) is presently included in numerical implementations of high-energy evolution \cite{bknumerics} even if the full result is already available \cite{BKNLO,Grabovsky:2013mba,Balitsky:2013fea,Kovner:2013ona}. Calculation of various observables, like inclusive hadroproduction \cite{AM},
 photoproduction \cite{photo}, etc. has also been mostly confined to leading order in the strong coupling constant $\alpha_s$.

There is therefore an urgent need to improve the accuracy of the CGC-based calculations. Efforts in this direction have been made in recent years, with the calculation of NLO corrections to several observables, like deep inelastic scattering \cite{Balitsky:2010ze,Beuf:2011xd}, or single hadroproduction cross section at forward rapidities \cite{production,bowen} in the so called "hybrid'' formalism \cite{hybrid}. However, numerical studies have yet been performed only in the latter case, and indicate very strong effects of the NLO corrections, with cross sections even becoming negative at moderate transverse momenta \cite{amir,ana1}. The recent followup \cite{ana2} to \cite{ana1} underscores the problem even more, since a change which is supposed to affect the result only at next-to-next-to-leading (NNLO) order, modifies the NLO results significantly. There is also an ongoing discussion on the correct choice of the factorization scale for the high-energy evolution \cite{vitev,Xiao:2014uba}, and on the eventual relevance of additional collinear resummations at small $x$  \cite{Beuf:2014uia}.

The purpose of this paper is to reanalyze the NLO calculation of inclusive hadron production using the wave function approach employed in \cite{production}. Such an analysis is necessary, since the calculations of \cite{production} and \cite{bowen} are not quite complete. The most important element missing in \cite{production,bowen} is a  treatment of the limitation on the phase space of emissions due to finite life time of the low-$x$ fluctuations, the so-called Ioffe time \cite{ioffetime}. We
rectify this deficiency of the previous calculations and provide formulae which explicitly take into account this constraint.

We also address the question what is the rapidity to which the eikonal scattering amplitudes have to be evolved. This point has not been addressed explicitly in \cite{production}, while \cite{bowen} uses an heuristic argument based on the kinematics of $2\rightarrow 1$ processes and \cite{vitev} proposes a different solution.

The result of this work is a complete set of formulae for hadron production in the hybrid formalism, at NLO accuracy, including all channels. We implement in our formulae the Ioffe time restriction which ensures that only fluctuations that live long enough are included in the NLO calculation. This Ioffe time provides a scale  that  allows a clean separation between the collinear and soft divergencies. As in previous calculations, collinear divergencies are absorbed in the parton densities of the projectile and the fragmentation functions of the final state partons into hadrons, evolved according to the Dokshitzer-Gribov-Lipatov-Altarelli-Parisi (DGLAP) evolution equations \cite{dglap}.

The soft divergencies are regulated by the Ioffe time. We show that the correct scaling of the Ioffe time under Lorentz boost leads to the leading-order Balitsky-Kovchegov (BK) evolution \cite{balitsky,Kovchegov} of the eikonal scattering amplitude on the target, with a well defined prescription for the scale up to which this scattering amplitude has to be evolved. This scale differs from those suggested in \cite{bowen} and \cite{vitev}. Our conclusion is that the prescriptions adopted in \cite{bowen,vitev} are not strictly consistent with the NLO accuracy of the rest of the calculation. We provide the corrected expressions and discuss to what extent they differ from those in \cite{bowen}.
We also present the results beyond the limit of large number of colors. 

 At the end of the day, our formulae are somewhat different from those given in \cite{production} and \cite{bowen} which were used as the basis of numerical calculations of \cite{amir} and \cite{ana1}. The differences may be significant in some kinematical regions. Whether these differences can lead to stabilization of numerical results will have to be determined by numerical calculations.

We note that the collinear factorization scheme that we use, inside which the parton densities and fragmentation functions are defined, does not coincide with the standard $\overline{MS}$ one. But the relation between these functions in both schemes amounts to a mild rescaling of factorization scales.

The outline of our paper is as follows: In Section II we discuss the basic setup of our calculation, explaining what is the relevant rapidity interval for the evolution, and  paying special attention to the question of the kinematic restriction due to finite Ioffe time of the fluctuations. In Section III we present the results of the calculation and discuss the difference between our results and those of \cite{bowen}. In these two sections, for simplicity of presentation, we limit ourselves to hadron production from a projectile quark. In Section IV we present the results of the full calculation, including the gluon channel. The details of the calculation are given in the Appendices.

\section{The basics}

We consider inclusive hadron production at forward rapidities in pA scattering. We use the "hybrid" formalism \cite{hybrid} as our calculational framework. This means that we treat the wave function of the projectile proton in the spirit of collinear factorization, as an assembly of partons with zero intrinsic transverse momentum. Perturbative corrections to this wave function are provided by the usual QCD perturbative splitting processes. On the other hand the target is treated as distribution of strong color fields which during the scattering event transfer transverse momentum to the propagating partonic configuration of the projectile.

For simplicity of presentation, in this and the next section we consider in detail only one channel for hadron production: an incoming quark from the projectile wave function, which propagates through the target and fragments into the observed hadron in the final state. The general discussion pertaining to all important aspects of the calculation carries over almost verbatim to other production channels as well. Complete results for all production channels are presented in Section IV.

\subsection{The kinematics and the choice of frame}

First, let us specify the kinematics of the process. We require the production of a quasi-on-shell parton with momentum $p$, at a forward rapidity $\eta$ (in the projectile-going diection) and with a sizable transverse momentum $p_\perp$.
By definition, one has
\begin{equation}\label{eta}
\eta=\frac{1}{2}\ln \frac{p^+}{p^-}=\ln\frac{\sqrt{2}p^+}{|p_\perp|}\ .
\end{equation}
This parton then fragments into a hadron of momentum $p_h$. The fragmentation is treated as collinear, so that all components of the momentum are reduced by the same factor $\zeta$: $p_h^+=\zeta\, p^+$ and $p_{h\perp}=\zeta\, p_\perp$. Hence, the rapidity $\eta$ is both the rapidity of the produced parton and of the hadron it fragments into.
Let us define the fractions $x_p$ and $x_F$ of the light-cone momentum $P^+_P$ of the projectile carried by the produced parton and hadron respectively, as
\begin{equation}
x_p=\frac{p^+}{P^+_P}   \qquad \textrm{and} \qquad x_F=\frac{p^+_h}{P^+_P}\, .
\end{equation}
Notice that  the standard Feynman-$x$ variable $x_F=x_p \zeta$. Since $p^+$, $p^+_h$ and $P^+_P$ are scaled by the same factor under a longitudinal boost, $x_p$ and $x_F$ are boost invariant.

Ignoring the masses of the target and projectile, in a frame where the projectile has large momentum $P^+_P$, and the target a large momentum $P^-_T$, one has
\begin{equation}
2P^+_PP^-_T=s.
\end{equation}
Thus in the center of mass frame
\begin{equation}
P^+_{P,\ CM}=P^-_{T,\ CM}=\sqrt{\frac{s}{2}}, \ \ \ \ \ p^+_{CM}=x_p\sqrt{\frac{s}{2}}, \ \ \ \ \   p^+_{h,CM}=x_F\sqrt{\frac{s}{2}} \,.
\end{equation}
The rapidity of the produced parton and hadron in the center of mass frame is related to $x_p$ or $x_F$ as
\begin{equation}\label{eta_CM}
\eta_{CM}=\ln\frac{x_p \sqrt {s}}{|p_\perp|}=\ln\frac{x_F  \sqrt {s}}{\zeta\, |p_\perp|}\ .
\end{equation}

As will be explained momentarily, we will find it convenient to work in the frame where most of the energy of the process is carried by the target. We refer to it as PROJ (projectile frame). In this frame we have
\begin{equation}
P^+_{P,\ PROJ}=\frac {s}{2P^-_{T,\ PROJ}}\,.
\end{equation}
The momenta $P^+_{P,\ CM}$ and $P^+_{P,\ PROJ}$ scale differently with total energy of the process.
\begin{equation}
P^+_{P,\ CM}\propto s^{1/2}; \ \ \ \ \ P^+_{P,\ PROJ}={\rm const}.
\end{equation}

We will be interested in deriving the evolution of production cross section with the total energy of the process. While deriving the evolution we will find it convenient to think of the change in energy $s\rightarrow se^{\Delta Y}$ as due to the slight boost of the projectile. In this case $P^+_{P,\ PROJ}\propto e^{\Delta Y}$.
Also, we should keep in mind that if we increase the energy of the process (by boosting the projectile) but still measure particle production at fixed center of mass rapidity, the value of $x_p$ has to be changed according to
\begin{equation}
x_p\propto s^{-1/2}.
\end{equation}

Now, although the cross section can be calculated in any Lorentz  frame, and the result should not depend on the choice of frame, it is advantageous to perform the calculation in a frame where it is simplest.
Clearly, if we choose a frame in which the projectile moves very fast, this is far from optimal. In such a frame the wave function of the projectile itself has many gluons, and one needs to calculate it with high precision (high order in perturbation theory) in order to calculate the production probability correctly.
On the other hand we would like to treat the parton that produces the outgoing hadron as part of the projectile wave function. Thus it is most convenient to choose such a frame in which the target moves fast and carries almost all the energy of the process, while the projectile moves fast enough to be able to accommodate partons with momentum fraction $x_p$ but not so fast that it develops a large low $x$ tail. Since the observed hadron has large rapidity, the relevant values of $x_p$ are not small, and thus such a choice is possible.

For the scattering of the leading parton on the target to be eikonal in our chosen frame we need
\begin{equation}
x_pP^+_{PROJ}\gg M,
\end{equation}
where the scale $M$ is of the order of the typical longitudinal recoil that the target can impart. This scale may slowly depend on energy, and at high energy can be of order $M\sim \frac{Q_s^2}{\Lambda_{QCD}}$. At relevant energies this is of the order of the typical hadronic scale, and we will treat it as such.

Another auxiliary quantity we introduce is the initial energy $s_0$. Our final results do not depend on it explicitly but it turns out to be a useful concept.
This energy is arbitrary, except it is required to be high enough, so that the eikonal approach is valid at $s>s_0$.
Starting from this energy we can evolve the target according to the high-energy evolution. The energy $s_0$ is achieved by boosting the projectile from its rest frame to rapidity $Y_P$, and the target from its rest frame by rapidity $Y^0_T$, so that
\begin{equation}
s_0=2P^+_{P,PROJ}P^{-0}_T;\ \ \ \ \ \ P^+_{P,PROJ}=\frac{M_P}{\sqrt{2}}\, e^{Y_P}; \ \ \ \ \ \ \ P^{-0}_T=\frac{M_T}{\sqrt{2}}\, e^{Y^0_T}\,.
\end{equation}
Starting from this initial energy $s_0$, the energy of the process is increased further by boosting the target.

Thus, in our setup the projectile wave function at any energy is evolved only to rapidity $Y_p=\ln \frac{1}{x_p}+Y_0$,
with $Y_0$ being a fixed number of order one.

The target on the other hand is evolved by
\begin{equation}\label{YT}
Y_T=\ln \frac{s}{s_0}\,,
\end{equation}
where $s$ is the total energy of the process. The initial condition for the evolution of the target wave function has to be specified at $Y^0_T$.

\begin{figure}[h]
\vskip -1cm
\begin{center}
 \includegraphics[width=0.75\textwidth]{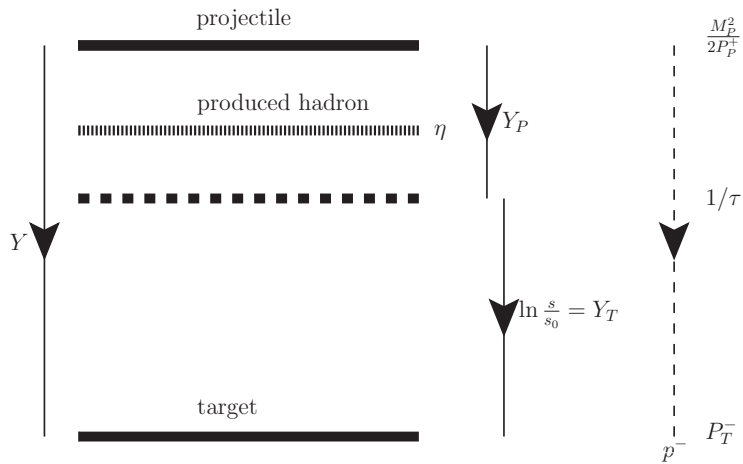}
 \end{center}
 \vskip -9.5cm
 \caption{Illustration of the different rapidity and momentum scales in our setup.}
 \label{fig1}
\end{figure}

At first sight this setup looks similar to that in \cite{vitev} and very different from the one in \cite{bowen}, where the interval of the target evolution is taken to depend on the transverse momentum of the final state hadron.  In fact our setup differs from that adopted in both works. Nevertheless, as we will show in
subsection III.C, in a certain kinematic regime our evolution interval turns out to be effectively similar to the one in \cite{bowen, Mueller:2012uf}. The different scales are illustrated in fig. \ref{fig1}.

With this partition of degrees of freedom between the projectile and the target, our setup is fixed. Any projectile parton scatters on a member of the same target field ensemble. Averaging over this ensemble leads to the dipole scattering matrix $s_{Y_T}(x,y)$, which at fixed energy of the process does not depend on the transverse momentum or rapidity of the final state hadron.

Note that at this point we do not have to specify what is exactly the evolution equation that governs the evolution of the target. This equation is self-consistently determined from the calculation itself. Unsurprisingly, we will find that at the accuracy of our calculation the relevant evolution is the leading-order BK equation.

\subsection{$Y_T$ {\it vs.} $Y_g$}

Importantly, the above discussion  does not uphold the prescription used in \cite{bowen} and in current numerical implementations \cite{amir,ana1,ana2}. The procedure set out in \cite{bowen} is to evolve the target to rapidity $Y_g=\ln\frac{1}{x_g}$ with $x_g=\frac{p_\perp}{\sqrt s}e^{-\eta}$. The reason for choosing this particular value of $Y_g$ in \cite{bowen} is based on the following kinematic argument. At leading order the incoming projectile parton carries momentum $(p^+,0,0)$. The parton measured in the final state has the same $+$ component of momentum, transverse momentum $p_\perp$ and is on shell. This means that during the scattering it picks up $-$-component of momentum  $p^-=\frac{p_\perp^2}{2p^+}=e^{-\eta}\frac{p_\perp}{\sqrt 2}$ from the target.
If one assumes that this momentum has been transferred to the projectile parton by a single gluon of the target, the  gluon in question must have carried at least this amount of $p^-$, and therefore had to have the longitudinal momentum fraction of the target
\begin{equation}\label{xg}
x_g=\frac{p^-}{P^-}=e^{-\eta}\frac{p_\perp}{\sqrt {s}}\,.
\end{equation}
On the other hand, the high-energy evolution (in the dilute regime) has the property that any hadronic wave function is dominated by softest gluons. One thus may conclude that $x_g$ is the longitudinal momentum fraction of the softest gluons in the target wave function, and thus the target has to be evolved to $Y_g$.

On closer examination, however, it transpires that this argument does not hold water. It overlooks the fact that the target is in fact dense. For the dense target, the projectile parton undergoes multiple scatterings, and therefore picks up momentum $p^-$ not from a single target gluon, but from several. This means that $x_g$ is actually an upper bound on the momentum fraction of the target gluons, and therefore $Y_g$ only gives a lower bound on the rapidity up to which the target wave function has to be evolved. In fact, it is very natural that the total rapidity $Y_T$ should not depend on the transverse momentum of the produced particle rather than depend on it as in (\ref{xg}). Recall that in the dense scattering regime, the transverse momentum of the scattered parton "`random walks"' as the parton propagates through the target. Thus the total transverse momentum is proportional to the square root of the number of collisions with the target gluons, $p^2_\perp\propto N_g$. On the other hand the  transferred $p^-$ does not random walk, since all the gluons in the target have  $p^-$ of the same sign. Thus  $p^-\propto N_g$, which is perfectly consistent with the relation between $p^-$ and $p_\perp$ that follows from the onshellness condition of the outgoing parton. Therefore, increasing $p_\perp$ of the observed parton (at fixed $p^+$), while increasing the total $p^-$ acquired by the projectile parton, does not change the fraction of longitudinal momentum of individual gluons in the target wave function that participate in the scattering, and therefore does not affect the value of $Y_T$.

In the leading-logarithmic approximation it is not important what exactly is the value of the evolution parameter for the target as long as it is of the order of total rapidity. However, since we are interested in the next-to-leading perturbative corrections, this question becomes important, as changing the value of the evolution parameter affects the result at NLO. It is thus important to use $Y_T$ rather than $Y_g$.

\subsection{What scatters? The Ioffe time restriction}

While in the hybrid approach one assumes that the projectile partons scatter eikonally on the target fields, clearly this assumption can only be valid for partonic configurations of the projectile wave function which exist long enough to traverse the longitudinal extent of the target \cite{ioffetime}.

Consider for example scattering of a projectile quark.
The parton level production cross section at leading order is \cite{production,bowen}
\begin{equation}
\frac{d\sigma^q}{d^2p_\perp d\eta}=\frac{1}{(2\pi)^2}\int d^2xd^2y e^{ip_\perp(x-y)} s_{Y_T}(x,y),
\end{equation}
where $s_{Y_T}(x,y)$ is the fundamental dipole scattering amplitude at rapidity $Y_T$, defined in terms of the eikonal scattering factors as $s(x,y)=\frac{1}{N_c}tr \ [S_F(x)S_F^{\dagger}(y)]$, with $S_F(x)$ the  Wilson line for propagation of a high-energy parton in the fundamental representation of $SU(N_c)$ at transverse position $x$.

At next-to-leading order the quark splits in the projectile wave function with probability of order $\alpha_s$ into a quark-gluon configuration.
The wave function of the ``dressed'' quark state with transverse momentum $n_\perp$ and $+$ momentum $x_BP^+$ ($x_B$ refers to Bjorken $x$) to order $g$ is
\beq
\label{pair}
&&\hspace{-0.9cm}\vert ({\rm q}) \, x_BP^+, n_\perp,\alpha,s\rangle_D=\int d^2x e^{in_\perp x}\Big\{A^q\vert ({\rm q})\, x_BP^+,x,\alpha, s\rangle\nonumber\\
&&\hspace{-0.9cm}+g\int \frac{dLPS}{2\pi} \int_{y \,z} \,F_{(\rm qg)}(x_BP^+,\xi,y-x,z-x)_{s\bar s; j}\,t^a_{\alpha\beta}\vert ({\rm q})\ p^+=(1-\xi)x_BP^+,y ,\beta,\bar s; ({\rm g})\  q^+=\xi x_BP^+,z,a,j\rangle\Big\}\, .
\eeq
Here $s$ and $\bar s$ are the quark spin indices; $j$ - the gluon polarization index, $\alpha$, $\beta$ are fundamental and $a$ are adjoint color indices. We use the notation $dLPS$ to denote the longitudinal phase space in $+$ components for the splitting. This corresponds to the $+$ component of the parent parton for the real terms, $dLPS=d\left[\frac{x_pP^+}{1-\xi}\right]$, and the $+$ momentum running in the loop for the virtual ones, $dLPS=d\left[\xi x_pP^+\right]$, with $\xi$ being the $+$-momentum fraction taken by the emitted gluon.
The constant $A^q$ differs from unity by an amount of order $g^2$ and is needed to preserve the normalization of the state at order $\alpha_s$.
At this point we do not need to know explicitly the form of the function $F_{(qg)}$ nor the normalization constant $A^q$.

The dressed quark now scatters on the target and produces final state particles. In the spirit of the hybrid approximation, we treat the scattering of the $qg$ configuration as a completely coherent process where each parton picks an eikonal phase during the interaction with the target. However this can only apply to configurations that have a coherence time (Ioffe time) greater than the propagation time through the target.  The usual argument for splitting of a parton with vanishing transverse momentum into two partons, with transverse momenta $\pm k_\perp$ and longitudinal fractions
$\xi$ and $1-\xi$ gives
\begin{equation}
t_c\sim \frac{2\xi(1-\xi) p^+}{k_\perp^2}=\frac{2\xi(1-\xi) x_BP^+}{k_\perp^2}\,.
\end{equation}
Note that $P^+\equiv P^+_{P,\ PROJ}$ in this formula is the momentum of the projectile in the frame defined by eq. (\ref{YT}) rather than in the center-of-mass frame. We will use this simplified notation throughout the rest of the paper.

Only the $qg$ pairs that satisfy the relation
\begin{equation}
\frac{2(1-\xi)\xi x_BP^+}{k_\perp^2}>\tau\ ,
\end{equation}
where $\tau$ is a fixed time scale determined by the longitudinal size of the target, scatter coherently.
As we will see later, the time $\tau$ actually stays constant through the evolution, it thus has to be identified with the size of the target at the initial energy $s_0$.
This is approximately given by the inverse of $P^{-0}_T$. Thus the parameter that enters our calculation is in fact the initial energy
\begin{equation}
P^+/\tau=s_0/2
\end{equation}
and the Ioffe time restriction can be written as
\begin{equation}\label{Ioffe}
\frac{(1-\xi)\xi x_B}{k_\perp^2}>s_0^{-1}\ .
\end{equation}
This point is discussed in detail in Appendix A.

The pairs that do not exist long enough are not resolved. Those pairs have large $k_\perp$, and have a small transverse size. The scattering and particle production from those two parton configurations must be indistinguishable from that of a single parent quark. Thus only those quark-gluon components of the dressed quark wave function eq. (\ref{pair}) that satisfy the condition eq. (\ref{Ioffe}) scatter eikonally. For the rest of the components  their scattering matrix should be taken  identical to that of a single bare quark.

Thus, for the purposes of the calculation of scattering cross section the dressed quark wave function should be taken to be
\begin{eqnarray}
\label{pair1}
&&\vert ({\rm q})\ x_BP^+,0,\alpha,s\rangle_{D \ \Omega}=\int d^2x \, \Big\{A^q\vert ({\rm q})\, x_BP^+,x,\alpha, s\rangle\\
&&+g\int_\Omega \frac{dLPS}{2\pi} d^2 y\, d^2z\, F_{({\rm qg})}(x_BP^+,\xi,y-x,z-x)_{s\bar s; j}\,t^a_{\alpha\beta}\,\vert ({\rm q})\ p^+=(1-\xi)x_BP^+,y,\beta,\bar s; ({\rm g})\  q^+=\xi x_BP^+,z,a,j\rangle\Big\},\nonumber
\end{eqnarray}
where a quark with $+$-momentum $p^++q^+=x_BP^+$ and zero transverse momentum splits into a quark with $p^+=(1-\xi)x_BP^+$ and  $p_\perp$ and a gluon with $q^+=\xi x_BP^+$ and $q_\perp$.
The normalization constant reads
\begin{eqnarray}\label{A1}
A^q&=&1-g^2\frac{N_c^2-1}{4N_c S}x_BP^+\int_\Omega d\xi\,\int_{\, x\, \bar x\, y \,z} \,e^{in_\perp(x-\bar x)} F_{({\rm qg})}(x_BP^+,\xi,y-x,z-x)\,F_{({\rm qg})}^*(x_BP^+,\xi,y-\bar x,z-\bar x)\\
&=&
1-g^2\frac{N_c^2-1}{4N_c S}x_BP^+\int_\Omega d\xi \int_{ x \, \bar x \,  y \, z} \, F_{({\rm qg})}(x_BP^+,\xi,y-x,z-x)\,F_{({\rm qg})}^*(x_BP^+,\xi,y-\bar x,z-\bar x)\,,\nonumber
\end{eqnarray}

where the summation over $\bar s$ and $j$ is understood in $F^2_{({\rm qg})}$.
Here $\Omega$ is the part of the  phase space for the splitting defined by
\begin{equation}
\Omega: \ \ \frac{2(1-\xi)\xi p^+}{k_\perp^2}=\frac{2(1-\xi)\xi x_BP^+}{k_\perp^2}>\tau.
\label{eq:Omega}
\end{equation}

The second equality in eq. (\ref{A1}) holds due to the explicit coordinate dependence of $F_{({\rm qg})}$ which sets $x$ equal to $\bar x$, eq.(\ref{f}). Note that the normalization constant $A^q$ is UV finite. The constant $S$ is the total transverse area. As we will see later it cancels against the integral over the coordinate $z$.
The longitudinal momentum factor $x_BP^+$ in eq.(\ref{A1}) appears due to the normalization of the longitudinal momentum eigenstates.

An analogous formula holds for a quark with arbitrary transverse momentum.

The function $F_{({\rm qg})}$ can be read off the known formulae, for example in \cite{production}, where to order $g$ the single ``dressed'' quark state was written as
\begin{eqnarray}\label{delta}
&&\delta\vert ({\rm q}) p^++q^+,p_\perp+q_\perp,\alpha,s\rangle\rightarrow \vert ({\rm q})\,p^+,p_\perp,\beta,\bar s;({\rm g})\,q^+,q_\perp,a,j\rangle \,t^a_{\alpha\beta}\\
&&\times\frac{1}{2\sqrt {2q^+}}\frac{1}{\frac{(p_\perp+q_\perp)^2}{2(p^++q^+)}-\frac{p_\perp^2}{2p^+}-\frac{q_\perp^2}{2q^+}}
\left\{\delta_{s\bar s}\delta_{ij}\frac{2p^++q^+}{p^++q^+}\left[\frac{p^i_\perp}{p^+}-\frac{q^i_\perp}{q^+}\right]-i\epsilon_{ij}\sigma^3_{s\bar s}\frac{q^+}{p^++q^+}\left[\frac{p^i_\perp}{p^+}-\frac{q^i_\perp}{q^+}\right]\right\}\,.\nonumber
\end{eqnarray}
It is convenient to define
\begin{equation}
l_\perp=p_\perp-(1-\xi)(q_\perp+p_\perp)=\xi p_\perp-(1-\xi)q_\perp\, ,
\end{equation}
which is the transverse momentum of the daughter quark relative to the one of the parent quark.

The Ioffe time constraint for a quark state with an arbitrary initial transverse momentum then reads
\begin{equation}
l_\perp^2<2\xi(1-\xi)\frac{p^++q^+}{\tau}\,.
\end{equation}
eq. (\ref{delta}) can be rewritten as
\beq
\label{delta1}
&&\hspace{-4cm}\delta\vert ({\rm q})\, p^++q^+,p_\perp+q_\perp,\alpha,s\rangle\rightarrow \vert ({\rm q})\,p^+,p_\perp,\beta,\bar s;({\rm g})\,q^+,q_\perp,a,j\rangle \, t^a_{\alpha\beta}\nonumber
\\
&&\hspace{2cm}\times
\frac{-1}{\sqrt {2\xi(p^++q^+)}}\frac{1}{l^2_\perp}
\Big\{\delta_{s\bar s}\delta_{ij}(2-\xi)l^i_\perp-i\epsilon_{ij}\sigma^3_{s\bar s}\xi l^i_\perp\Big\}\,.
\eeq
To Fourier transform $F$ into coordinate space, it is convenient to perform the change of variables with unit Jacobian
\begin{equation}
l_\perp=\xi p_\perp-(1-\xi)q_\perp; \ \ \ \ \ m_\perp=\xi q_\perp+(1+\xi)p_\perp
\end{equation}
or
\begin{equation}
p_\perp=\xi l_\perp+(1-\xi)m_\perp; \ \ \ \ \ q_\perp=\xi m_\perp-(1+\xi)l_\perp\,.
\end{equation}
We then have
\begin{eqnarray}\label{F}
\hspace{-1.7cm}F_{({\rm qg})}(x_BP^+,\xi,y-x,z-x)_{s\bar s; j} \ & =&\int \frac{d^2p_\perp}{(2\pi)^2} \frac{d^2 q_\perp}{(2\pi)^2} F_{({\rm qg})}(x_BP^+,\xi,p_\perp, q_\perp)e^{-ip_\perp(y-x)-iq_\perp(z-x)}\nonumber\\
&=&-\frac{1}{\sqrt {2\xi(p^++q^+)}}\Big\{\delta_{s\bar s}\delta_{ij}(2-\xi)-i\epsilon_{ij}\sigma^3_{s\bar s}\xi\Big\}\,
\delta^2\Big[(1-\xi)(y-x)+\xi(z-x)\Big]
\nonumber\\
&\times&
\int_{
l_\perp^2<2\xi(1-\xi)\frac{p^++q^+}{\tau}}\frac{d^2l_\perp}{(2\pi)^2}\frac{l_\perp^i}{l_\perp^2} e^{-il_\perp\left[\xi(y-x)-(1+\xi)(z-x)\right]}\,.
\end{eqnarray}
We can get rid of $x$ by realising the $\delta$-function with the result
\begin{eqnarray}\label{f}
F_{({\rm qg})}(x_BP^+,\xi,y-x,z-x)_{s\bar s; j} \ &=&-\frac{1}{\sqrt {2\xi x_BP^+}}\Big\{\delta_{s\bar s}\delta_{ij}(2-\xi)-i\epsilon_{ij}\sigma^3_{s\bar s}\xi\Big\}\nonumber\\
&\times&\delta^2\Big(x-[(1-\xi)y+\xi z]\Big)\int_{
l_\perp^2<2\xi(1-\xi)\frac{p^++q^+}{\tau}}\frac{d^2l_\perp}{(2\pi)^2} \frac{l_\perp^i}{l_\perp^2}e^{-il_\perp(y-z)}\\
&=&\frac{-i}{\sqrt {2\xi x_BP^+}}\Big\{\delta_{s\bar s}\delta_{ij}(2-\xi)-i\epsilon_{ij}\sigma^3_{s\bar s}\xi\Big\}\delta^2\Big(x-[(1-\xi)y+\xi z]\Big)A_{\xi,x_B}^i(y-z),\nonumber
\end{eqnarray}
where we have used $p^++q^+=x_BP^+$, and the modified Weizs\"acker-Williams field is defined as
\begin{eqnarray}
A_{\xi,x_B}^i(y-z)&\equiv &-i\int_{l_\perp^2<2\xi(1-\xi) \frac{x_BP^+}{\tau}}
\frac{d^2l_\perp}{(2\pi)^2} \frac{l_\perp^i}{l_\perp^2} e^{-il_\perp(y-z)}\nonumber\\
&=&-\frac{1}{2\pi} \; \frac{(y-z)^i}{(y-z)^2}\; \left[1- \textrm{J}_0\left(|y-z| \sqrt{2\xi(1-\xi)\frac{x_BP^+}{\tau}}\right)\right]\, .
\end{eqnarray}
For future use we also define
\begin{eqnarray}\label{axi}
A_{\xi}^i(y-z)&\equiv & A_{\xi,\frac{x_p}{1-\xi}}^i(y-z)\nonumber\\
&=& -\frac{1}{2\pi} \; \frac{(y-z)^i}{(y-z)^2}\; \left[1- \textrm{J}_0\left(|y-z| \sqrt{2\xi\frac{x_p P^+}{\tau}}\right)\right]\, .
\end{eqnarray}
When we discard the Ioffe-time cut-off, or equivalently in the $P^+/\tau\rightarrow+\infty$ limit, both $A_{\xi,x_B}^i(y-z)$ and $A_{\xi}^i(y-z)$ reduce to the standard Weizs\"acker-Williams field
\beq
A^i(y-z)=-\frac{1}{2\pi}\frac{(y-z)^i}{(y-z)^2}\, .
\eeq

Using the result \eqref{f} we can write the normalization factor eq.(\ref{A1}) as
\begin{equation}\label{A}
A^q=1-g^2\frac{N_c^2-1}{4N_c}\int_{0}^{1} \frac{d\xi}{2\pi}\frac{1+(1-\xi)^2}{\xi}\int_z A^i_{\xi,x_p}(z)A^i_{\xi,x_p}(z).
\end{equation}
Two comments are in order here. First, the function $F_{({\rm qg})}$, in addition to transverse coordinate variables and the momentum fraction $\xi$, also depends on the longitudinal momentum of the parent quark, and therefore on $x_B$ at fixed $P^+$.
This dependence has to be kept in mind especially whenever $x_B$ has to be integrated over.

Second, note that we have now implemented the Ioffe time constraint on the phase space $\{k_\perp, \xi\}$ in the definition of $F_{({\rm qg})}(x_BP^+,\xi,y-x,z-x)$ rather than in the integral over $\xi$ as in eqs. (\ref{pair1},\ref{A1}). This is of course equivalent, but is in fact mathematically more appropriate, since the energy fraction $\xi$ in eq.(\ref{pair1}) is integrated last. Thus, in the following we will not denote the restricted phase space by $\Omega$ but will instead directly use the analog of the Weiszacker-Williams field $A_{\xi,x_B}$, which explicitly depends on the energy fraction $\xi$ due to implementation of the Ioffe time constraint in the transverse phase space.

As noted above, neglecting the Ioffe time constraint on $l_\perp$,  one gets for  $A^i(y-z)$ the standard Weizsacker-Williams field at point $z$. The only difference with the soft approximation then is in the ``recoil'' position of the emitter after the emission, which does not remain at the same transverse coordinate. With the constraint, on the other hand, the relative contribution of short distances in $A^i(y-z)$ which is the reflection of the short Ioffe time of small dipoles, is suppressed.

\section{The results: production from quarks.}
The calculation of the production cross section from this starting point proceeds along the standard lines. We use the approach developed in \cite{production} and consider scattering of dressed parton states. Compared to the calculation of \cite{production} we now include virtual corrections, which were not important for the purposes of \cite{production}. Although our technique is slightly different than the straightforward summation of diagrams as in \cite{bowen}, we have checked that  modulo the Ioffe time constraint and the value of $Y_T$ in the leading-order term, our results agree with those of \cite{bowen}.

We present the detailed calculation in the appendix for completeness. In the body of the paper we only present the final result and discuss some of its features.

Our result for the inclusive hadroproduction of hadron $H$ from an initial state quark via hadronization of the final state quark in pA scattering, is
\begin{eqnarray}
\frac{d\sigma^{q\rightarrow q\rightarrow H}}{d^2p_{h\perp} d\eta}&=&\frac{1}{(2\pi)^2} \int_{x_F}^1\frac{d\zeta}{\zeta^2}\; D^q_{H,\mu^2}(\zeta)\; \frac{x_F}{\zeta}\; f^q_{\mu^2}\left(\frac{x_F}{\zeta}\right) \int d^2x d^2y\; e^{i\frac{p_{h\perp}}{\zeta}(x-y)} s_{Y_T}(x,y)\nonumber\\
& &+\int_{x_F}^{1}\frac{d\zeta}{\zeta^2}\; D^q_{H,\mu^2}(\zeta)\;
\frac{d\bar\sigma_1^{q\rightarrow q}}{d^2p_\perp d\eta}\left(\frac{p_{h\perp}}{\zeta},\frac{x_F}{\zeta}\right)\label{sigmah2}
\end{eqnarray}
with
\begin{equation}\label{qh}
\frac{d\bar\sigma_1^{q\rightarrow q}}{d^2p_\perp d\eta}(p_\perp,x_p)=
\frac{d\bar{\sigma}_1^{q\rightarrow q, r}}{d^2p_\perp d\eta}(p_\perp,x_p)
+
\frac{d\bar{\sigma}_1^{q\rightarrow q, v}}{d^2p_\perp d\eta}(p_\perp,x_p)
\end{equation}
where the real and the virtual contributions to the quark production cross section are given by

\beq\label{dsigmaqcol1}
\frac{d\bar{\sigma}_1^{q\rightarrow q, r}}{d^2\pp d\eta}(p_\perp, x_p)&=&\frac{g^2}{(2\pi)^3} C_F \;\int_0^{1-x_p} d\xi \; \frac{x_p}{1-\xi} f^q_{\mu^2}\left(\frac{x_p}{1-\xi}\right)\;  \left[ \frac{1+(1-\xi)^2}{\xi} \right]\int _{y \by z} e^{i\pp(y-\by)} \nonumber\\
&&\times \Big[A_{\xi}^i(y-z)A_{\xi}^i(\bar y-z)-
C_{\mu^2}\left(\xi,\frac{x_p}{1-\xi},z\right)\Big] \, \bigg\{  s[y,\by] +(1-\xi)^2 \,  s\big[(1-\xi)y,(1-\xi)\by\big] \bigg\} \nonumber\\
&&+\frac{g^2}{(2\pi)^3}  \; \int_0^{1-x_p}d\xi \; \frac{x_p}{1-\xi}f^q_{\pp}\left(\frac{x_p}{1-\xi}\right)\;  \left[ \frac{1+(1-\xi)^2}{\xi} \right]\int _{y \by z} e^{i\pp(y-\by)} \nonumber\\
&&\times A_{\xi}^i(y-z)A_{\xi}^i(\bar y-z) \, \bigg\{ -\frac{N_c}{2}\bigg( s\big[ y-\xi(y-z), z \big] \, s[z,\by] + s\big[ z, \by-\xi(\by-z)\big] \, s[y,z] \bigg) \nonumber\\
&&\hspace{5cm}+\frac{1}{2N_c} \bigg( s \big[y-\xi(y-z),\by \big] + s\big[ y, \by-\xi(\by-z)\big] \bigg) \bigg\}
\eeq

and

\beq\label{sigmavirtual2}
\frac{d\bar{\sigma}_1^{q\to q,v}}{d^2\pp d\eta}(p_\perp,x_p)&=&-\frac{g^2}{(2\pi)^3}C_F\; x_p\; f^{q}_{\mu^2}(x_p) \;  \int_0^1 d\xi  \left[ \frac{1+(1-\xi)^2}{\xi} \right]
\int_{y\by z} e^{i\pp(y-\by)} s[y,\by]\nonumber\\
&&\times \Big[A_{\xi,x_p}^i(y-z)A_{\xi,x_p}^i(y-z)+A_{\xi,x_p}^i(\bar y-z)A_{\xi,x_p}^i(\bar y-z)-2C
_{\mu^2}(\xi,x_p,z)\Big]
\nonumber\\
&&+\frac{g^2}{(2\pi)^3}\; x_p f^{q}_{\pp}(x_p) \;  \int_0^1 d\xi  \left[ \frac{1+(1-\xi)^2}{\xi} \right]
\int_{y\by z} e^{i\pp(y-\by)}\nonumber\\
&&\times\Bigg\{
A_{\xi,x_p}^i(y-z)A_{\xi,x_p}^i(y-z) \bigg[ \frac{N_c}{2} \, s\Big[ y+\xi(y-z), z+\xi(y-z) \Big] \, s\Big[ z+\xi(y-z),\by\Big]\nonumber\\
&&-\frac{1}{2N_c}\, s\Big[ y+\xi(y-z), \by\Big] \Bigg]\nonumber\\
&&+ A_{\xi,x_p}^i(\bar y-z)A_{\xi,x_p}^i(\bar y-z) \bigg[ \frac{N_c}{2} \, s\Big[ z+\xi(\by-z), \by+\xi(\by-z) \Big] \, s\Big[ y,z+\xi(\by-z)\Big]\nonumber\\
&&-\frac{1}{2N_c}\, s\Big[ y, \by+\xi(\by-z)\Big] \Bigg] \Bigg\}\; .
\eeq
In the above, $D_{H,\mu^2}^q$ and $f^q_{\mu^2}$ are the quark fragmentation (FF) and proton quark distribution (PDF) functions respectively defined below
in  (\ref{fq}). Both are calculated with a generic factorizations scale (resolution) $\mu$. The natural value of the factorization scale to use in the present problem is
$\mu=|p_{h\perp}|$. The function $s_{Y_T}(x,y)\equiv \frac{1}{N_c}{\rm Tr[S_F(x)S^\dagger_F(y)]}$ is the eikonal dipole scattering amplitude on the heavy nuclear target.
The eikonal dipole amplitude in eq.(\ref{sigmah2}) is evolved according to BK equation up to rapidity $Y_T$ as explained above. The collinear subtraction term is defined as
\begin{equation}\label{fact1}
\tilde{C}_{\mu^2}(\xi,x_B)\equiv\int d^2z\; C_{\mu^2}(\xi,x_B,z)\equiv\int d^2z \; A_{\xi,x_B}^i(z)\,A_{\xi,x_B}^i(z)\,\theta(z^2\mu^2-1) \; .
\end{equation}

Note that the PDFs and FFs are defined in a given factorization scheme that, in our framework, does not coincide with the standard $\overline{MS}$ one. But their relation, as discussed in detail in Appendix C, amounts to a mild rescaling of factorization scales.

These expressions can be somewhat simplified. Before getting to that we discuss the region of applicability of these formulae.

\subsection{How much should we believe it?}

Here is the run down on the approximations employed in our derivation.

\noindent 1. The eikonal approximation for the partonic scattering matrix at initial energy. This approximation is tantamount to neglecting power corrections of the type $Q_s^2/s_0$ and $p_\perp^2/s_0$ in the production cross section.  We do not expect this to be a major concern, since with a reasonable choice of $s_0$ and at not astronomical transverse momenta, both ratios in question should be very small. Consistently, the rapidity evolution is also required to be accurate only up to such power suppressed terms. The BK evolution for amplitude indeed arises when we omit such power corrections in the derivation (see Appendix A). Equations (\ref{dsigmaqcol1},\ref{sigmavirtual2}) contain explicitly some such power suppressed terms. While formally this is not a problem, within our accuracy we are able to discard these terms and thus simplify the evaluation of the cross section. This will be done in the next subsection.

\noindent 2. Collinear approximation for the projectile wave function. As usual, this approximation is expected to be good up to power corrections of the type $\Lambda_{QCD}^2/p_\perp^2$. Note that the $O(\alpha_s)$ correction we have included seems to go beyond the collinear approximation, as discussed in \cite{production}. It includes the contribution from the partons in the projectile wave function which have high transverse momentum, of the order of $p_\perp$, and is not power suppressed. This contribution is not part of the collinear DGLAP evolution, but is rather a genuine NLO fixed order correction to the cross section. As discussed above, the Ioffe time constraint eliminates contribution of very high tranverse momentum partons, and it remains to be understood how important is the remaining contribution from this kinematical region.

\noindent 3. Use of the light cone PDF while not working in the infinite momentum frame. As we have discussed in detail, we work in the frame where the projectile has a large but finite longitudinal momentum. We have nevertheless used the quark distribution function $f_{\mu^2}(x_p)$ in our formulae. Strictly speaking, our function $f_{\mu^2}(x_p)$ also depends on the momentum $P^+$
. If we had to take the dependence on $P^+$ into account, this would complicate things considerably. In particular it would affect the derivation of the evolution equation in Appendix A, since changing $P^+$ through the boost of the projectile would also change the distribution
$f_{\mu^2}(x_p)$. We have neglected such terms while deriving the evolution equations eqs.(\ref{e2},\ref{e1}) for the following reason. We do not expect $f_{\mu^2}(x_p)$ to depend significantly on $P^+$ as long as the value of $x_p$ is not too small, namely $x_p>\frac{M_P}{P^+}$. For small values of $x$ one indeed has to take into account the $P^+$ dependence, and in fact one expects $f_{\mu^2}(x_p)$ to drop to zero quickly in this range. However for $x_p\sim 1$, all the partons involved are moving very fast, and the correction due to finite $P^+$ should again be a power correction in the ratio $\frac{M_P}{P^+}$. It is thus the same type of power correction as discussed above.

\noindent 4. Leading order in $\alpha_s$ BK evolution. This is probably the most important approximation of all and we expect it to be the most important limiting factor of our results. The point here is the following. While writing our expressions we have assumed that the dipole scattering amplitude is $O(1)$, and under this assumption have calculated the order $O(\alpha_s)$ terms in the cross section. However, if we really want to know the production cross section to $O(\alpha_s)$ we also need to know the amplitude $s_{Y_T}$ with the accuracy $O(\alpha_s)$. The question is how far in energy can we evolve the amplitude from the initial $s_0$ using the {\it leading-order} evolution and still correctly account for all $O(\alpha_s)$ terms.

To answer this question imagine calculating higher order corrections without resumming them into the evolution of $s_{Y_T}$. The result will have the form
\begin{equation}
\frac{d\sigma}{d^2p_\perp d\eta}=B_{00}+\alpha_s B_{10}+\Sigma_{n=1}^\infty\alpha_s^nY^n_TB_{nn}+\alpha_s\Sigma_{n=1}^\infty\alpha_s^nY^n_TB_{n+1,n}+\alpha_s^2B_{20}+...
\end{equation}
For small $Y_T$ this is a genuinely perturbative expansion. However when $Y_T$ is large enough, so that $Y_T\sim 1/\alpha_s$, one needs to resum all terms where the power of $Y_T$ is the same as the power of $\alpha_s$, namely all terms $B_{nn}$. This resummation is equivalent to solving the leading-order evolution equation for $s_{Y_T}$. Our calculation with {\it leading-order evolved} $s_{Y_T}$ is equivalent to this resummation and the inclusion of the term $B_{10}$. The problem is however, that for these values of $Y_T$ any one of the terms $B_{n+1,n}$ is as large as $B_{10}$ due to the enhancement by the right number of factors of $Y_T$.
Thus if we want to keep the accuracy $O(\alpha_s)$ all the way up to rapidities $Y_T\sim 1/\alpha_s$, we need also to resum the terms $B_{n+1,n}$, which is equivalent to solving {\it next-to-leading-order evolution} for $s_{Y_T}$. This sets the limit on the applicability of the current calculation with the LO evolution. Within this range it still makes sense to resum $B_{nn}$ terms, as some of them may be parametrically larger than $B_{10}$. For instance if we push up to rapidities $Y_T\sim 1/\alpha_s^{1/2}$, the terms $B_{11}$ and $B_{22}$ are at least as large as $B_{10}$ and both are resummed in the LO evolution, while for $Y_T\sim 1/\alpha_s^{3/4}$ the terms $B_{11}$, $B_{22}$, $B_{33}$ and $B_{44}$ are all important and are all resummed. On the other hand all the $B_{n+1,n}$ terms are parametrically suppressed with respect to $B_{10}$ and do not have to be resummed.

Also on the positive side, even for $Y_T\sim 1/\alpha_s$ we can modify the formalism by solving the evolution of $s_{Y_T}$ at NLO, but still keeping only $B_{10}$ as the only unenhanced contribution. This will indeed make the overall accuracy of the result $O(\alpha_s)$ in all of this larger rapidity range without ever needing to calculate $B_{20}$ etc.

To summarize, even though we have calculated NLO terms in the production cross section, our result can be considered as a complete NLO result only for rapidities $Y_T\ll 1/\alpha_s$. For larger rapidities, the accuracy of our calculation is the same as of the leading order. To have NLO accuracy for all rapidities one needs to include NLO terms also in the evolution of $s_{Y_T}$ \cite{BKNLO}.

\subsection{Discarding power corrections: The quark channel}
As discussed in the previous section, our results only hold up to power corrections in powers of $p^2_\perp/s_0$ and $Q^2_s/s_0$. We can thus simplify our final expression somewhat by  discarding  these terms. Our final expression for the production cross section is given in terms of the real and virtual part of production cross section eqs.(\ref{dsigmaqcol1},\ref{sigmavirtual2}) where the dipole amplitude $s_{Y_T}(x,y)$ is evolved to rapidity $Y_T$. We now add and subtract from  eq.(\ref{qh}) the following expression, which is obtained from the quark production amplitude by setting $\xi=0$ everywhere except in  $A_\xi$ and the explicit factor $1/\xi$:
\beq\label{regs}
&&\frac{g^2}{(2\pi)^3} N_c x_p\: f^q_{\mu^2}(x_p)\int_0^1\frac{d\xi}{\xi}\int_{y,\bar y,z}e^{ip_\perp(y-\bar y)}\Bigg[A_{\xi}^i(y-z)A_{\xi}^i(y-z)+A_{\xi}^i(\bar y-z)A_{\xi}^i(\bar y-z)-2A_{\xi}^i(y-z)A_{\xi}^i(\bar y-z)\Bigg]\nonumber\\
&&\hspace{5.8cm}\times\;\Big[s(y,z)s(z,\bar y)-s(y,\bar y)\Big] \; .
\eeq
This simple form follows if we take the same modified WW field in the real and virtual terms to be $A_\xi$, and set $\xi=0$ in all other entries. Now the term that contains the difference between the quark cross section and the expression  eq.(\ref{regs}) is finite at $\xi\rightarrow 0$ even if we remove the Ioffe time cutoff in the WW field and take $A^i_\xi(x)\rightarrow A^i(x)$, where $A^i(x)$ is the standard WW field with unrestricted integral over the transverse momentum $l_{\perp}$. The difference between keeping $A^i_\xi(x)$ and modifying it to $A^i(x)$ is clearly a power correction in inverse powers of $s_0$, and the same holds for $C_{\mu^2}$. We are therefore allowed to simplify this term by dropping these power corrections. On the other hand we cannot simplify the stand alone term of the form eq.(\ref{regs}), and we leave it as is. The final result of this operation can be written as

\beq\label{sigmafinal}
&&\frac{d\bar{\sigma}_1^{q\to q}}{d^2\pp d\eta}(p_\perp, x_p)=\frac{g^2}{(2\pi)^3}
\int_0^{1-x_p} d\xi  \int_{y,\bar y,z} e^{ip_\perp(y-\bar y)}\frac{1+(1-\xi)^2}{[\xi]_+}\Bigg[ \\
 &&\;\frac{x_p}{1-\xi} f^q_{\mu^2}\left(\frac{x_p}{1-\xi}\right)\;\Bigg\{  C_F\Big[A^i(y-z)A^i(\bar y-z) -\Big(A^i(z)A^i(z)\Big)_{\mu^2}\Big] \, \bigg\{  s[y,\by] +(1-\xi)^2 \,  s\big[(1-\xi)y,(1-\xi)\by\big] \bigg\} \nonumber\\
&&+A^i(y-z)A^i(\bar y-z) \, \bigg\{ -\frac{N_c}{2}\Big[s\big[ y-\xi(y-z), z \big] \, s[z,\by] + s\big[ z, \by-\xi(\by-z)\big] \, s[y,z] \Big] \nonumber\\
&&\hspace{5cm}+\frac{1}{2N_c} \Big[ s \big[y-\xi(y-z),\by \big] + s\big[ y, \by-\xi(\by-z)\big] \Big]\bigg\}\Bigg\}\Bigg]\nonumber\\
&&-\frac{g^2}{(2\pi)^3}
\int_0^1 d\xi  \int_{y,\bar y,z} e^{ip_\perp(y-\bar y)}\frac{1+(1-\xi)^2}{[\xi]_+}\Bigg[ \nonumber\\
&&x_p\: f^{q}_{\mu^2}(x_p) \; \Bigg\{  C_F \Big[A^i(y-z)A^i(y-z) + A^i(\bar y-z)A^i(\bar y-z) - 2 \Big(A^i(z)A^i(z)\Big)_{\mu^2}\Big]s[y,\by]
\nonumber\\
&&-\;\bigg\{
A^i(y-z)A^i(y-z) \left[ \frac{N_c}{2} \, s\Big[ y+\xi(y-z), z+\xi(y-z) \Big] \, s\Big[ z+\xi(y-z),\by\Big]-\frac{1}{2N_c}\, s\Big[ y+\xi(y-z), \by\Big] \right]\nonumber\\
&&+ A^i(\bar y-z)A^i(\bar y-z) \left[ \frac{N_c}{2} \, s\Big[ z+\xi(\by-z), \by+\xi(\by-z) \Big] \, s\Big[ y,z+\xi(\by-z)\Big]-\frac{1}{2N_c}\, s\Big[ y, \by+\xi(\by-z)\Big] \right] \bigg\}\Bigg\}\Bigg]\nonumber\\
&+&\frac{g^2}{(2\pi)^3}N_c x_p\: f^q_{\mu^2}(x_p)\int_0^1\frac{d\xi}{\xi}\int_{y,\bar y,z}e^{ip_\perp(y-\bar y)}[A_{\xi}^i(y-z)-A_{\xi}^i(\bar y-z)][A_{\xi}^i(y-z)-A_{\xi}^i(\bar y-z)]\Big[s(y,z)s(z,\bar y)-s(y,\bar y)\Big]\nonumber
\eeq
where, as usual, the $+$ prescription is defined so that for any smooth function $F(\xi)$,
\begin{equation}
\int d\xi F(\xi)\frac{1}{[\xi]_+}\equiv \int d\xi \frac{F(\xi)-F(0)}{\xi} \;
\end{equation}
and
\begin{equation}
\Big(A^i(z)A^i(z)\Big)_{\mu^2}\equiv A^i(z)A^i(z) \theta(z^2\mu^2-1).
\end{equation}

\subsection{Discussion.}
The last term in eq.(\ref{sigmafinal}) looks somewhat similar to an extra contribution to the evolution of the leading-order term. It would be equivalent to extra evolution if the Ioffe time modified WW field $A^i_\xi$ could be substituted by $A^i$. This substitution is of course not possible, since the Ioffe time regulates the pole at $\xi=0$ in this term. Nevertheless let us try to interpret this term in the language of additional evolution. To do that we interchange the order of integration over $\xi$ and the momentum in the Fourier transform of the WW field, similarly to eqs.(\ref{exch},\ref{xmin}).
Performing, in the last term in eq.(\ref{sigmafinal}), the integral over rapidity first  we can write it in terms of the Fourier transform of the derivative of the dipole cross section
\begin{equation}\label{extra}
\int d^2l_{\perp}\int d^2m_{\perp} \ln\left(\frac{1}{\xi_{min}}\right)\frac{d}{dY}s(l_{\perp}+p_{\perp},m_{\perp}-p_{\perp})
\end{equation}
with
\begin{equation}
\xi_{min}={\rm max}\left\{\frac{l_{\perp}^2}{x_ps_0},\frac{m_{\perp}^2}{x_ps_0}\right\} \; .
\end{equation}
If $\xi_{min}$ did not depend on the momentum $l_\perp$ and $m_\perp$, this would be equivalent to extra evolution of the leading term by an additional rapidity interval $\Delta Y=\ln\frac{1}{\xi_{min}}$. In this case we could forget about this term, and instead evolve $s$ in the leading-order term to $Y_T+\ln  \frac{1}{\xi_{min}}$.
However $\xi_{min}$ itself depends on the momentum $l_{\perp}$ (or $m_{\perp}$). Note that momentum $l_{\perp}$ is the transverse momentum of the gluon in the projectile wave function.
Since this extra emitted gluon is not present in the leading-order term at all, it is impossible to ascribe the contribution eq.(\ref{extra}) to extra evolution.
As discussed in \cite{production}, there are two distinct regions of $l_{\perp}$ that contribute to the production, $l_{\perp}\sim Q_s$ and $l_{\perp}\sim p_{\perp}$.
Thus it is important to keep the full dependence of $l_{\perp}$ in the calculation.

Nevertheless, it may be instructive to think of this term as extending the evolution interval of the leading-order term as this makes it easier to compare our final expression with those of \cite{bowen} and \cite{vitev}. The expression in \cite{bowen} does not contain the last term in eq.(\ref{sigmafinal}), but instead evolves the leading-order amplitude to rapidity $Y_g=\ln \frac{1}{x_g}$. On the other hand \cite{vitev} argues that the leading term instead has to be evolved to $Y_M=\ln\frac{s}{M^2}$ with $M^2$ being typical hadronic scale. Our expression, if interpreted in terms of the evolution of the leading term, means that the correct interval of the evolution depends on the transverse momentum of the gluon, and is given by
\begin{equation}
Y_{l_{\perp}}=Y_T+\ln \frac{x_ps_0}{l_{\perp}^2}=\ln \frac{1}{x_g}+\ln\frac{p_{\perp}^2}{l_{\perp}^2} \; .
\end{equation}
Thus the correct evolution interval is neither the one suggested in \cite{bowen} nor in \cite{vitev}.

Note that for the contribution of high transverse momentum gluons $l_{\perp}\sim p_{\perp}$, and thus $\ln \frac{1}{x_g}$ is indeed a reasonable approximation to $Y_{l_{\perp}}$. These high transverse momentum gluons contribute to the ``inelastic partonic scattering'' in the language of \cite{production}. However the ``elastic contribution'' that is driven by gluons with $l_{\perp}\sim Q_s$ corresponds to a significantly higher interval of rapidity evolution.

We stress, that thinking in terms of the evolution of the leading-order term is only mnemonic, since the ``evolution interval'' depends on momentum $l_{\perp}$, which itself does not characterize any of the kinematic variables in the leading-order expression.

\section{The results: all channels.}
The calculation for the contribution from all channels is presented in Appendix B. The final result is
\beq\label{all}
\frac{d\sigma^H}{d^2p_{h\perp} d\eta}&=&\frac{1}{(2\pi)^2}\sum_q \int_{x_F}^1 \frac{d\zeta}{\zeta^2}\: D^q_{H,\mu^2}(\zeta)\: \frac{x_F}{\zeta}\: f^q_{\mu^2}\left(\frac{x_F}{\zeta}\right) \int_{y\by}e^{i\frac{p_{h\perp}}{\zeta}(y-\by)}s[y,\by]+\Big(q\rightarrow \bar q,\ p_{h\perp}\rightarrow - p_{h\perp}\Big)\nonumber\\
&+&\frac{1}{(2\pi)^2} \int_{x_F}^1 \frac{d\zeta}{\zeta^2}\: D^g_{H,\mu^2}(\zeta)\: \frac{x_F}{\zeta}\: f^g_{\mu^2}\left(\frac{x_F}{\zeta}\right) \int_{y\by}e^{i\frac{p_{h\perp}}{\zeta}(y-\by)}s_A[y,\by]\nonumber\\
&&
+\sum_q \int_{x_F}^1\frac{d\zeta}{\zeta^2}\: D^q_{H,\mu^2}(\zeta)\:
\frac{d\bar{\sigma}^q}{d^2\pp d\eta}\left(\frac{p_{h\perp}}{\zeta},\frac{x_F}{\zeta}\right)+\Big(q\rightarrow \bar q,\ p_{h\perp}\rightarrow - p_{h\perp}\Big)\nonumber\\
&&+\int_{x_F}^1 \frac{d\zeta}{\zeta^2}\: D^g_{H,\mu^2}(\zeta)\: \frac{d\bar{\sigma}^g}{d^2\pp d\eta}\left(\frac{p_{h\perp}}{\zeta},\frac{x_F}{\zeta}\right)
\eeq
where summation over all quark flavors is assumed.
The adjoint dipole scattering amplitude $s_A(x,y)$ is
\beq
\label{adjdip}
s_A[x,y]=\frac{1}{N_c^2-1}{\rm tr} \Big[S_A(x)S^{\dagger}_A(y)\Big]\; .
\eeq
In the following we do not write out explicitly the antiquark contribution. As indicated in eq.(\ref{all}) it is obtained from the quark contribution by
 substituting the antiquark PDFs and FFs for the quark ones and flipping the sign of the transverse momentum $p_{h\perp}\rightarrow - p_{h\perp}$ in all the relavnt expressions (i.e. the leading order cross section and eqs. (\ref{eq:quark1}), (\ref{eq:quark2}) and (\ref{eq:quark3})).

The quark production cross section is
\beq
\frac{d\bar{\sigma}^q}{d^2\pp d\eta}=\frac{d\bar{\sigma}_1^{q\to q,{\rm r}}}{d^2\pp d\eta}+
\frac{d\bar{\sigma}_1^{q\to q,{\rm v}}}{d^2\pp d\eta}+\frac{d\bar{\sigma}_1^{g\to q,{\rm r}}}{d^2\pp d\eta}
\eeq
with
\beq
\frac{d\bar{\sigma}_1^{q\to q, \rm{r}}}{d^2\pp d\eta}&=&\frac{g^2}{(2\pi)^3} C_F\int_0^{1-x_p} d\xi \; \frac{x_p}{1-\xi}\; f^{q}_{\mu^2}\left(\frac{x_p}{1-\xi}\right) \;  \left[ \frac{1+(1-\xi)^2}{\xi} \right]\int _{y \by z} e^{i\pp(y-\by)} \nonumber\\
&&\times \bigg[ A^i_{\xi,\frac{x_p}{1-\xi}}(y-z)A^i_{\xi,\frac{x_p}{1-\xi}}(\by-z)-  C_{\mu^2}\left(\xi,\frac{x_p}{1-\xi},z\right) \bigg] \, \bigg\{  s[y,\by] +(1-\xi)^2 \,  s\big[(1-\xi)y,(1-\xi)\by\big] \bigg\} \nonumber\\
&&+\frac{g^2}{(2\pi)^3} \int_0^{1-x_p}  d\xi \;\frac{x_p}{1-\xi}\, f^{D,q}_{\mu^2}\left(\frac{x_p}{1-\xi}\right)  \;  \left[ \frac{1+(1-\xi)^2}{\xi} \right]\int _{y \by z} e^{i\pp(y-\by)} \nonumber\\
&&\times A^i_{\xi,\frac{x_p}{1-\xi}}(y-z)A^i_{\xi,\frac{x_p}{1-\xi}}(\by-z) \, \bigg\{ -\frac{N_c}{2}\bigg( s\big[ y-\xi(y-z), z \big] \, s[z,\by] + s\big[ z, \by-\xi(\by-z)\big] \, s[y,z] \bigg) \nonumber\\
&&\hspace{5cm}+\frac{1}{2N_c} \bigg( s \big[y-\xi(y-z),\by \big] + s\big[ y, \by-\xi(\by-z)\big] \bigg) \bigg\}\; ,
\label{eq:quark1}
\eeq

\beq
\frac{d\bar{\sigma}_1^{q\to q, \rm{v}}}{d^2\pp d\eta}&=&-\frac{g^2}{(2\pi)^3}C_F\; x_p\; f^{q}_{\mu^2}(x_p)\; \int_0^{1}  d\xi  \left[ \frac{1+(1-\xi)^2}{\xi} \right]
\int_{y\by z} e^{i\pp(y-\by)} s[y,\by]\nonumber\\
&&\times\bigg\{ A^i_{\xi,x_p}(y-z)A^i_{\xi,x_p}(y-z)+ A^i_{\xi,x_p}(\by-z)A^i_{\xi,x_p}(\by-z) -2C_{\mu^2}(\xi,x_p,z) \bigg\}\nonumber\\
&&+\frac{g^2}{(2\pi)^3}C_F\; x_p\; f^{D,q}_{\mu^2}(x_p)\; \int_0^{1}  d\xi  \left[ \frac{1+(1-\xi)^2}{\xi} \right]
\int_{y\by z} e^{i\pp(y-\by)}\nonumber\\
&&\times\Bigg\{
A^i_{\xi,x_p}(y-z)A^i_{\xi,x_p}(y-z) \bigg[ \frac{N_c}{2} \, s\Big[ y+\xi(y-z), z+\xi(y-z) \Big] \, s\Big[ z+\xi(y-z),\by\Big]\nonumber\\
&&-\frac{1}{2N_c}\, s\Big[ y+\xi(y-z), \by\Big] \Bigg]\nonumber\\
&&+  A^i_{\xi,x_p}(\by-z)A^i_{\xi,x_p}(\by-z) \bigg[ \frac{N_c}{2} \, s\Big[ z+\xi(\by-z), \by+\xi(\by-z) \Big] \, s\Big[ y,z+\xi(\by-z)\Big]\nonumber\\
&&-\frac{1}{2N_c}\, s\Big[ y, \by+\xi(\by-z)\Big] \Bigg] \Bigg\}\; ,
\label{eq:quark2}
\eeq

\beq
\frac{d\bar{\sigma}_1^{g\to q, \rm{r}}}{d^2\pp d\eta}&=&\frac{1}{2}\frac{g^2}{(2\pi)^3}\int_0^{1-x_p}  d\xi \, \frac{x_p}{1-\xi} f^{g}_{\mu^2}\left(\frac{x_p}{1-\xi}\right)\big[\xi^2+(1-\xi)^2\big] \int_{y\by z}e^{i\pp(y-\by)}\nonumber\\
&&\times\bigg\{ \bigg[A^i_{\xi,\frac{x_p}{1-\xi}}(y-z)A^i_{\xi,\frac{x_p}{1-\xi}}(\by-z) -C_{\mu^2}\left(\xi,\frac{x_p}{1-\xi},z\right)\bigg]\bigg[ s[y,\by]+(1-\xi)^2s_A\big[(1-\xi)y,(1-\xi)\by\big]\bigg]\nonumber\\
&&-\frac{N_c^2}{(N_c^2-1)}A^i_{\xi,\frac{x_p}{1-\xi}}(y-z)A^i_{\xi,\frac{x_p}{1-\xi}}(\by-z)\bigg[ s\big[y,(1-\xi)\by+\xi z\big]s\big[(1-\xi)\by+\xi z,z\big]-\frac{1}{N_c^2}s[y,z]\bigg]\nonumber\\
&&-\frac{N_c^2}{(N_c^2-1)}A^i_{\xi,\frac{x_p}{1-\xi}}(y-z)A^i_{\xi,\frac{x_p}{1-\xi}}(\by-z)\bigg[ s\big[z,(1-\xi)y+\xi z\big]s\big[(1-\xi)y+\xi z,\by\big]-\frac{1}{N_c^2}s[z,\by]\bigg]\bigg\} \; ,
\label{eq:quark3}
\eeq

The gluon production cross section is
\beq
\frac{d\bar{\sigma}^g}{d^2\pp d\eta}=\frac{d\bar{\sigma}_1^{q\to g, \rm{r}}}{d^2\pp d\eta}+\frac{d\bar{\sigma}_1^{g\to g, \rm{r}}}{d^2\pp d\eta}+\frac{d\bar{\sigma}_1^{g\to g, \rm{v}}}{d^2\pp d\eta}+\frac{d\bar{\sigma}_1^{g\to q, \rm{v}}}{d^2\pp d\eta}+\frac{d\bar{\sigma}_1^{g\to \bq, \rm{v}}}{d^2\pp d\eta}
\eeq
with

\beq
\frac{d\bar{\sigma}_1^{q\to g, \rm{r}}}{d^2\pp d\eta}&=&\frac{g^2}{(2\pi)^3} \int_{x_p}^{1}
d\xi \;\frac{x_p}{\xi} f^{q}_{\mu^2}\left(\frac{x_p}{\xi}\right)  \left[ \frac{1+(1-\xi)^2}{\xi} \right] \int _{z \bz y} e^{i\pp(z-\bz)} \nonumber\\
&&\times\bigg[ A^i_{\xi,\frac{x_p}{\xi}}(z-y)A^i_{\xi,\frac{x_p}{\xi}}(\bz-y)-C_{\mu^2}\left(\xi,\frac{x_p}{\xi},y\right)\bigg] \, \bigg\{ C_F \, s_A[z,\bz] +C_F \, \xi^2\; s\big[ \xi z, \xi\bz \big]\bigg\} \nonumber\\
&&+\frac{g^2}{(2\pi)^3} \int_{x_p}^{1}
d\xi \;\frac{x_p}{\xi} f^{q}_{\mu^2}\left(\frac{x_p}{\xi}\right)  \left[ \frac{1+(1-\xi)^2}{\xi} \right] \int _{z \bz y} e^{i\pp(z-\bz)}\nonumber\\
&&\times A^i_{\xi,\frac{x_p}{\xi}}(z-y)A^i_{\xi,\frac{x_p}{\xi}}(\bz-y)\bigg\{
-\frac{N_c}{2}\bigg( s\big[ z,(1-\xi)y+\xi \bz \big] \, s[y,z] + s[\bz,y] \; s\big[ (1-\xi)y+\xi z,\bz \big] \bigg)\nonumber\\
&&+\frac{1}{2N_c} \bigg( s \big[y,(1-\xi)y+\xi \bz \big] + s\big[ (1-\xi)y+\xi z,y \big] \bigg) \bigg\}\; ,
\eeq

\beq
\frac{d\bar{\sigma}_1^{g\to g, \rm{r}}}{d^2\pp d\eta}&=&\frac{g^2}{(2\pi)^3}2C_A\int_0^{1-x_p} d\xi\, \frac{x_p}{1-\xi}\; f^{g}_{\mu^2}\: \left(\frac{x_p}{1-\xi}\right) \, \bigg[ \frac{1-\xi}{\xi}+\frac{\xi}{1-\xi}+\xi(1-\xi)\bigg]\, \int_{y\by z} e^{i\pp(y-\by)}\nonumber\\
&&\times\bigg[A^i_{\xi,\frac{x_p}{1-\xi}}(y-z)A^i_{\xi,\frac{x_p}{1-\xi}}(\by-z)-C_{\mu^2}\left(\xi,\frac{x_p}{1-\xi},z\right)\bigg]\bigg\{ s_A[y,\by]+(1-\xi)^2\, s_A\Big[(1-\xi)y, (1-\xi)\by\Big]\bigg\}\nonumber\\
&&-\frac{g^2}{(2\pi)^3}\int_0^{1-x_p} d\xi\,\frac{x_p}{1-\xi}\:  f^{g}_{\mu^2}\left(\frac{x_p}{1-\xi}\right) \, \bigg[ \frac{1-\xi}{\xi}+\frac{\xi}{1-\xi}+\xi(1-\xi)\bigg]\nonumber\\
&&\times \int_{y\by z} e^{i\pp(y-\by)}A^i_{\xi,\frac{x_p}{1-\xi}}(y-z)A^i_{\xi,\frac{x_p}{1-\xi}}(\by-z)
\frac{N_c^3}{N_c^2-1}\bigg\{s[y,z] \, s\Big[(1-\xi)\by+\xi z,y\Big]\, s\Big[z,(1-\xi)\by+\xi z\Big]\nonumber\\
&&\hspace{1.5cm}+s[z,y] \, s\Big[y,(1-\xi)\by+\xi z \Big]\, s\Big[(1-\xi)\by+\xi z,z\Big]\nonumber\\
&&\hspace{1.5cm}+s\Big[\by,(1-\xi)y+\xi z\Big]\, s\Big[(1-\xi)y+\xi z,z\Big]\, s[z,\by]\nonumber\\
&&\hspace{1.5cm}+s\Big[(1-\xi)y+\xi z,\by\Big]\, s\Big[z, (1-\xi)y+\xi z\Big]\, s[\by,z]-\frac{1}{N_c^2}\overline{X}_{y\by z}\bigg]\bigg\}\; ,
\eeq
\beq
\frac{d\bar{\sigma}_1^{g\to g, \rm{v}}}{d^2\pp d\eta}&=&-\frac{g^2}{(2\pi)^3}C_A\; x_p\, f^{g}_{\mu^2}(x_p)\;\int_0^{1} d\xi \bigg[\frac{1-\xi}{\xi}+\frac{\xi}{1-\xi}+\xi(1-\xi)\bigg]\int_{y\by z}e^{i\pp(y-\by)}s_A[y,\by] \nonumber\\
&&\times\bigg\{A^i_{\xi,x_p}(y-z)A^i_{\xi,x_p}(y-z)+A^i_{\xi,x_p}(\by-z)A^i_{\xi,x_p}(\by-z)-2C_{\mu^2}(\xi,x_p,z)\bigg\}\nonumber\\
&&+\frac{g^2}{(2\pi)^3}\;x_p\, f^{g}_{\mu^2}(x_p)\,\int_0^{1} d\xi \bigg[\frac{1-\xi}{\xi}+\frac{\xi}{1-\xi}+\xi(1-\xi)\bigg]\int_{y\by z}e^{i\pp(y-\by)} \nonumber\\
&&\times\bigg\{A^i_{\xi,x_p}(y-z)A^i_{\xi,x_p}(y-z)\bigg[-\frac{N_c}{2(N_c^2-1)}\overline{X}'_{y\by z}\nonumber\\
&&+\frac{N_c^3}{2(N_c^2-1)}\bigg(s\Big[y+\xi(y-z),z+\xi(y-z)\Big] s\Big[z+\xi(y-z),\by\Big]s\Big[\by,y+\xi(y-z)\Big]\nonumber\\
&&\hspace{2cm}+s\Big[z+\xi(y-z),y+\xi(y-z)\Big]s\Big[\by,z+\xi(y-z)\Big] s\Big[y+\xi(y-z),\by\Big]\bigg)\bigg]\nonumber\\
&&+A^i_{\xi,x_p}(\by-z)A^i_{\xi,x_p}(\by-z)\bigg[-\frac{N_c}{2(N_c^2-1)}\overline{X}''_{y\by z}\nonumber\\
&&+\frac{N_c^3}{2(N_c^2-1)}\bigg(s\Big[\by+\xi(\by-z),z+\xi(\by-z)\Big]s\Big[z+\xi(\by-z),y\Big] s\Big[y,\by+\xi(\by-z)\Big]\nonumber\\
&&\hspace{2cm}+s\Big[z+\xi(\by-z),\by+\xi(\by-z)\Big] s\Big[y,z+\xi(\by-z)\Big]s\Big[\by+\xi(\by-z),y\Big]\bigg)\bigg]\bigg\}\; ,
\eeq

\beq
\frac{d\bar{\sigma}_1^{g\to q, \rm{v}}}{d^2\pp d\eta}&=&-\frac{1}{2}\frac{g^2}{(2\pi)^3}\,x_p\; f^{ g}_{\mu^2}(x_p)\; \int_0^{1} d\xi \Big[\xi^2+(1-\xi)^2\Big]\int_{y\by z}e^{i\pp(y-\by)}\nonumber\\
&&\times\bigg\{\bigg[A^i_{\xi,x_p}(y-z)A^i_{\xi,x_p}(y-z)+A^i_{\xi,x_p}(\by-z)A^i_{\xi,x_p}(\by-z)-2C_{\mu^2}(\xi,x_p,z)\bigg]s_A[y,\by]\nonumber\\
&&\hspace{0.5cm}-A^i_{\xi,x_p}(y-z)A^i_{\xi,x_p}(y-z)\bigg[\frac{N_c^2}{N_c^2-1}s\big[y+\xi(y-z),\by\big]s\big[\by,z+\xi(y-z)\big]\nonumber\\
&&\hspace{5cm}-\frac{1}{N_c^2-1}s\big[y+\xi(y-z),z+\xi(y-z)\big]\bigg]\nonumber\\
&&\hspace{0.5cm}-A^i_{\xi,x_p}(\by-z)A^i_{\xi;x_p}(\by-z)\bigg[\frac{N_c^2}{N_c^2-1}s\big[\by+\xi(\by-z),y\big]s\big[y,z+\xi(\by-z)\big]\nonumber\\
&&\hspace{5cm}-\frac{1}{N_c^2-1}s\big[\by+\xi(\by-z),z+\xi(\by-z)\big]\bigg]\bigg\}\; .
\eeq
Here the various combinations of six point functions are
\beq
\label{defX}
X[x,y,z,\bx,\by,\bz]=\frac{1}{N_c}\tr\Big[S^F(x)S^{F\,\dagger}(y)S^F(z)S^{F\,\dagger}(\bx)S^F(\by)S^{F\,\dagger}(\bz)\Big]\; ,
\eeq
\beq
\label{defXbar}
\overline{X}_{y\by z}&=&X\Big[z,y,(1-\xi)\by+\xi z,z,y,(1-\xi)\by+\xi z\Big] + X\Big[z, (1-\xi)\by+\xi z,y,z, (1-\xi)\by+\xi z,y\Big]\nonumber\\
&&+X\Big[(1-\xi)y+\xi z, z, \by, (1-\xi)y+\xi z,z, \by\Big]+X\Big[(1-\xi)y+\xi z, \by,\bz, (1-\xi)y+\xi z,\by,z\Big]\; ,
\eeq
\beq
\label{defXp}
\overline{X}'_{y\by z}&=&X\Big[ z-\xi(y-z), y-\xi(y-z),\by,z-\xi(y-z), y-\xi(y-z), \by\Big]\nonumber\\
&&+X\Big[ y-\xi(y-z),z-\xi(y-z),\by,y-\xi(y-z),z-\xi(y-z),\by\Big]\; ,\\
\label{defXpp}
\overline{X}''_{y\by z}&=&X\Big[ z-\xi(\by-z),\by-\xi(\by-z),y,z-\xi(\by-z),\by-\xi(\by-z),y\Big]\nonumber\\
&&+X\Big[ \by-\xi(\by-z),z-\xi(\by-z),y,\by-\xi(\by-z),z-\xi(\by-z),y\Big]\; .
\eeq
\subsection{Discarding power corrections: The gluon channel}
As for the quark channel, we can get rid of the power corrections eliminating the Ioffe time regulator in all terms which stay finite in the limit  $\xi\rightarrow 0$.
This can be done in $\frac{d\bar{\sigma}_1^{g\to q, \rm{r}}}{d^2\pp d\eta}$, $\frac{d\bar{\sigma}_1^{g\to q, \rm{v}}}{d^2\pp d\eta}$ and $\frac{d\bar{\sigma}_1^{q\to g, \rm{r}}}{d^2\pp d\eta}$.
Two gluon production terms are divergent, and we have to keep the $\xi$ dependence in the Ioffe time regulator in those.
However we can take the same route as in the previous section, adding and subtracting the expressions corresponding to the small $\xi$ limit in the integrand.
For small $\xi$ we have
\beq
\frac{d\bar{\sigma}_1^{g\to g, \rm{r}}}{d^2\pp d\eta}&\rightarrow& \frac{g^2}{(2\pi)^3}4
C_A\, x_p\, f^{g}_{\mu^2}(x_p)\, \int_0^{1} \frac{d\xi}{\xi}\, \int_{y\by z} e^{i\pp(y-\by)}
\bigg[A^i_{\xi}(y-z)A^i_{\xi}(\by-z)-C_{\mu^2}(\xi,x_p,z)\bigg] s_A[y,\by]\nonumber\\
&&-\frac{g^2}{(2\pi)^3}\frac{2N_c^3}{N_c^2-1}\, x_p\, f^{g}_{\mu^2}(x_p)\, \int_0^{1} \frac{d\xi}{\xi}
\int_{y\by z} e^{i\pp(y-\by)}A^i_{\xi}(y-z)A^i_{\xi}(\by-z)
\\
&&\times \bigg\{s[y,z] \, s[\by,y]\, s[z,\by]
+s[z,y]\, s[y,\by ]\, s[\by,z]
-\frac{1}{N_c^2}\overline{X}^A_{y\by z}\bigg\}\; ,\nonumber
\eeq

\beq
\frac{d\bar{\sigma}_1^{g\to g, \rm{v}}}{d^2\pp d\eta}&\rightarrow&-\frac{2g^2}{(2\pi)^3}C_A\, x_p \, f^{g}_{\mu^2}(x_p)\,\int_0^{1}  \frac{d\xi}{\xi}\int_{y\by z}e^{i\pp(y-\by)}s_A[y,\by]
\nonumber\\
&&\times
\bigg\{A^i_{\xi}(y-z)A^i_{\xi}(y-z)+A^i_{\xi}(\by-z)A^i_{\xi}(\by-z)-2C_{\mu^2}(\xi,x_p,z)\bigg\}
\nonumber\\
&&+
\frac{g^2}{(2\pi)^3}\frac{N_c^3}{(N_c^2-1)}\, x_p\,f^{g}_{\mu^2}(x_p)\, \int_0^{1} \frac{d\xi}{\xi}\int_{y\by z}e^{i\pp(y-\by) }\nonumber\\
&&\times\bigg[A^i_{\xi}(y-z)A^i_{\xi}(y-z)+A^i_{\xi}(\by-z)A^i_{\xi}(\by-z)\bigg]\nonumber\\
&&\times \bigg[ s[y,z] s[z,\by]s[\by,y]+s[z,y]s[\by,z] s[y,\by]-\frac{1}{N_c^2}\overline{X}^A_{y\by z}\bigg]
\eeq
where
\beq
\overline{X}^A_{y\by z}&=&X[ z, y,\by,z, y, \by]+X[ y,z,\by,y,z,\by]\; .
\eeq
Thus
\beq
\frac{d\bar{\sigma}_1^{g\to g, \rm{r}}}{d^2\pp d\eta}+\frac{d\bar{\sigma}_1^{g\to g, \rm{v}}}{d^2\pp d\eta}&\rightarrow&-\frac{g^2}{(2\pi)^3}\, x_p\, f^{g}_{\mu^2}(x_p)\, \int_0^{1} \frac{d\xi}{\xi}\, \int_{y\by z} e^{i\pp(y-\by)}
\bigg[A^i_{\xi}(y-z)-A^i_{\xi}(\by-z)\bigg]^2 \nonumber\\
&&\hspace{-2cm}\times\;\Bigg[2
C_A\, s_A[y,\by] -\frac{N_c^3}{N_c^2-1}\bigg\{s[y,z] \, s[\by,y]\, s[z,\by]
+s[z,y]\, s[y,\by ]\, s[\by,z]
-\frac{1}{N_c^2}\overline{X}^A_{y\by z}\bigg\}\Bigg]\; .
\eeq

Adding and subtracting this term from the gluon production amplitude, the final expression for hadron production can be written as
\beq
\frac{d\sigma^H}{d^2p_{h\perp} d\eta}&=&\frac{1}{(2\pi)^2}\int_{x_F}^1\frac{d\zeta}{\zeta^2}\, D^q_{H,\mu^2}(\zeta)\, \frac{x_F}{\zeta}\, f^q_{\mu^2}\left(\frac{x_F}{\zeta}\right) \int_{y\by}e^{i\frac{p_{h\perp}}{\zeta}(y-\by)}s[y,\by]
+\int_{x_F}^1\frac{d\zeta}{\zeta^2}\, D^q_{H,\mu^2}(\zeta)\, \frac{d\bar{\sigma}^q}{d^2\pp d\eta}\left(\frac{p_{h\perp}}{\zeta},\frac{x_F}{\zeta}\right)\nonumber\\
& &\hspace{-1cm}+\frac{1}{(2\pi)^2}\int_{x_F}^1\frac{d\zeta}{\zeta^2}\, D^g_{H,\mu^2}(\zeta)\, \frac{x_F}{\zeta}\, f^g_{\mu^2}\left(\frac{x_F}{\zeta}\right)
\int_{y\by}e^{i\frac{p_{h\perp}}{\zeta}(y-\by)}s_A[y,\by]
+\int_{x_F}^1\frac{d\zeta}{\zeta^2}\, D^g_{H,\mu^2}(\zeta)\, \frac{d\bar{\sigma}^g}{d^2\pp d\eta}\left(\frac{p_{h\perp}}{\zeta},\frac{x_F}{\zeta}\right)\nonumber\\
&&\hspace{-1cm}+\Big(q\rightarrow \bar q; \ p_{h\perp}\rightarrow -p_{h\perp}\Big)
\eeq
where the quark production cross section is

\beq
&&\frac{d\bar{\sigma}_1^{q}}{d^2\pp d\eta}(p_\perp, x_p)=\frac{g^2}{(2\pi)^3}
  \int_{y,\bar y,z} e^{ip_\perp(y-\bar y)}\Bigg\{
\int_0^{1-x_p} d\xi\frac{x_p}{1-\xi} f^q_{\mu^2}\left(\frac{x_p}{1-\xi}\right)\Bigg[
 \;\;\frac{1+(1-\xi)^2}{[\xi]_+}\\
&&\times\;  \Bigg( C_F\Big[A^i(y-z)A^i(\bar y-z)-\Big(A^i(z)A^i(z)\Big)_{\mu^2}\Big] \, \bigg\{  s[y,\by] +(1-\xi)^2 \,  s\big[(1-\xi)y,(1-\xi)\by\big] \bigg\} \nonumber\\
&&\hspace{0.8cm}-\frac{N_c}{2}A^i(y-z)A^i(\bar y-z) \, \bigg\{ \left[s\big[ y-\xi(y-z), z \big] \, s[z,\by] + s\big[ z, \by-\xi(\by-z)\big] \, s[y,z] \right] \nonumber\\
&&\hspace{5cm}-\frac{1}{N^2_c} \left[ s \big[y-\xi(y-z),\by \big] + s\big[ y, \by-\xi(\by-z)\big] \right]\bigg\}\Bigg)\nonumber\\
&&+\frac{1}{2}\big[\xi^2+(1-\xi)^2\big] \Bigg( \bigg[A^i(y-z)A^i(\by-z)-\Big(A^i(z)A^i(z)\Big)_{\mu^2}\bigg]\bigg[ s[y,\by]+(1-\xi)^2s_A\big[(1-\xi)y,(1-\xi)\by\big]\bigg]\nonumber\\
&&-\frac{N_c^2}{(N_c^2-1)}A^i(y-z)A^i(\by-z)\bigg[ s\big[y,(1-\xi)\by+\xi z\big]s\big[(1-\xi)\by+\xi z,z\big]-\frac{1}{N_c^2}s[y,z]\bigg]\nonumber\\
&&-\frac{N_c^2}{(N_c^2-1)}A^i(y-z)A^i(\by-z)\bigg[ s\big[z,(1-\xi)y+\xi z\big]s\big[(1-\xi)y+\xi z,\by\big]-\frac{1}{N_c^2}s[z,\by]\bigg]\Bigg)\Bigg]\nonumber\\
&&-\; x_p f^{q}_{\mu^2}(x_p)\int_0^{1} d\xi \Bigg[
\frac{1+(1-\xi)^2}{[\xi]_+}
\; \Bigg(  C_F\Big[A^i(y-z)A^i(y-z)+A^i(\bar y-z)A^i(\bar y-z)-2\Big(A^i(z)A^i(z)\Big)_{\mu^2}\Big]s[y,\by]
\nonumber\\
&&-\;
\frac{N_c}{2} A^i(y-z)A^i(y-z) \left[  s\Big[ y+\xi(y-z), z+\xi(y-z) \Big] \, s\Big[ z+\xi(y-z),\by\Big]-\frac{1}{N^2_c}\, s\Big[ y+\xi(y-z), \by\Big] \right]\nonumber\\
&&-\frac{N_c}{2} A^i(\bar y-z)A^i(\bar y-z) \left[ \, s\Big[ z+\xi(\by-z), \by+\xi(\by-z) \Big] \, s\Big[ y,z+\xi(\by-z)\Big]-\frac{1}{N^2_c}\, s\Big[ y, \by+\xi(\by-z)\Big] \right] \Bigg)\nonumber\\
&&+\frac{1}{2}\Big[\xi^2+(1-\xi)^2\Big]\Bigg(\bigg[A^i(y-z)A^i(y-z)+A^i(\by-z)A^i(\by-z)-2\Big(A^i(z)A^i(z)\Big)_{\mu^2}\bigg]s_A[y,\by]\nonumber\\
&&\hspace{0.5cm}-\frac{N_c^2}{N_c^2-1}A^i(y-z)A^i(y-z)\bigg[s\big[y+\xi(y-z),\by\big]s\big[\by,z+\xi(y-z)\big]-\frac{1}{N_c^2}s\big[y+\xi(y-z),z+\xi(y-z)\big]\bigg]\nonumber\\
&&\hspace{0.5cm}-\frac{N_c^2}{N_c^2-1}A^i(\by-z)A^i(\by-z)\bigg[s\big[\by+\xi(\by-z),y\big]s\big[y,z+\xi(\by-z)\big]-\frac{1}{N_c^2}s\big[\by+\xi(\by-z),z+\xi(\by-z)\big]\bigg]\Bigg)\Bigg]\Bigg\}\nonumber\\
&+&\frac{g^2}{(2\pi)^3}N_c x_p f^q_{\mu^2}(x_p)\int_0^1 \frac{d\xi}{\xi}\int_{y,\bar y,z}e^{ip_\perp(y-\bar y)}[A_{\xi}^i(y-z)-A_{\xi}^i(\bar y-z)][A_{\xi}^i(y-z)-A_{\xi}^i(\bar y-z)]\Big[s(y,z)s(z,\bar y)-s(y,\bar y)\Big]\; , \nonumber
\eeq
and the gluon production cross section is
\beq\label{gluonprodf}
\frac{d\bar{\sigma}_1^{g}}{d^2\pp d\eta}&=&\frac{g^2}{(2\pi)^3} \,  \int_{y\by z} e^{i\pp(y-\by)}\,\Bigg\{\int_0^{1-x_p} d\xi
\frac{x_p}{1-\xi}\, f^{g}_{\mu^2}\left(\frac{x_p}{1-\xi}\right) \Bigg[\, \left( \frac{1-\xi}{[\xi]_+}+\frac{\xi}{1-\xi}+\xi(1-\xi)\right) \\
&&\times\,\Bigg(2C_A\bigg[A^i(y-z)A^i(\by-z)-\Big(A^i(z)A^i(z)\Big)_{\mu^2}\bigg]\bigg\{ s_A[y,\by]+(1-\xi)^2\, s_A\Big[(1-\xi)y, (1-\xi)\by\Big]\bigg\}\nonumber\\
&&-\frac{N_c^3}{N_c^2-1}A^i(y-z)A^ i(\by-z)
\bigg\{s[y,z] \, s\Big[(1-\xi)\by+\xi z,y\Big]\, s\Big[z,(1-\xi)\by+\xi z\Big]\nonumber\\
&&\hspace{1.5cm}+s[z,y] \, s\Big[y,(1-\xi)\by+\xi z \Big]\, s\Big[(1-\xi)\by+\xi z,z\Big]\nonumber\\
&&\hspace{1.5cm}+s\Big[\by,(1-\xi)y+\xi z\Big]\, s\Big[(1-\xi)y+\xi z,z\Big]\, s[z,\by]\nonumber\\
&&\hspace{1.5cm}+s\Big[(1-\xi)y+\xi z,\by\Big]\, s\Big[z, (1-\xi)y+\xi z\Big]\, s[\by,z]-\frac{1}{N_c^2}\overline{X}_{y\by z}\bigg\}\Bigg)\nonumber\\
&&+\frac{1+\xi^2}{1-\xi}  \Bigg(C_F\bigg[ A^i(y-z)A^i(\by-z)-\Big(A^i(z)A^i(z)\Big)_{\mu^2}\bigg] \, \bigg\{  \, s_A[y,\by] + \, (1-\xi)^2\; s\big[ (1-\xi) y, (1-\xi)\by \big]\bigg\} \nonumber\\
&&-\frac{N_c}{2} A^i(y-z)A^i(\by-z)\bigg\{
s\big[ y,(1-\xi)\by +\xi z \big] \, s[z,y] + s[\by,z] \; s\big[ (1-\xi) y+\xi z,\by \big] \nonumber\\
&&-\frac{1}{N^2_c} \bigg( s \big[z,(1-\xi) \by +\xi z\big] + s\big[ (1-\xi) y+\xi z,z \big] \bigg) \bigg\}\Bigg)\Bigg]\nonumber\\
&&-x_p f^{g}_{\mu^2}(x_p)\, \int_0^{1} d\xi\left(\frac{1-\xi}{[\xi]_+}+\frac{\xi}{[1-\xi]_+}+\xi(1-\xi)\right) \Bigg[\nonumber\\
&&\times C_A \Big[A^i(y-z)A^i(y-z)+A^i(\by-z)A^i(\by-z)-2\Big(A^i(z)A^i(z)\Big)_{\mu^2}\Big]s_A[y,\by]\nonumber\\
&&+\frac{N_c^3}{2(N_c^2-1)}A^i(y-z)A^i(y-z)\bigg\{\frac{1}{N_c^2}\overline{X}'_{y\by z}\nonumber\\
&&-\bigg(s\Big[y+\xi(y-z),z+\xi(y-z)\Big] s\Big[z+\xi(y-z),\by\Big]s\Big[\by,y+\xi(y-z)\Big]\nonumber\\
&&\hspace{0.4cm}+s\Big[z+\xi(y-z),y+\xi(y-z)\Big]s\Big[\by,z+\xi(y-z)\Big] s\Big[y+\xi(y-z),\by\Big]\bigg)\bigg\}\nonumber\\
&&+\frac{N_c^3}{2(N_c^2-1)}A^i(\by-z)A^i(\by-z)\bigg\{\frac{1}{N_c^2}\overline{X}''_{y\by z}\nonumber\\
&&-\bigg(s\Big[\by+\xi(\by-z),z+\xi(\by-z)\Big]s\Big[z+\xi(\by-z),y\Big] s\Big[y,\by+\xi(\by-z)\Big]\nonumber\\
&&\hspace{0.4cm}+s\Big[z+\xi(\by-z),\by+\xi(\by-z)\Big] s\Big[y,z+\xi(\by-z)\Big]s\Big[\by+\xi(\by-z),y\Big]\bigg)\bigg\}\Bigg]\Bigg\}\nonumber\\
&&+\frac{g^2}{(2\pi)^3}\, x_p f^{g}_{\mu^2}(x_p)\, \int_0^1 \frac{d\xi}{\xi}\, \int_{y\by z} e^{i\pp(y-\by)}[A_{\xi}^i(y-z)-A_{\xi}^i(\bar y-z)]^2\nonumber\\
&&\times\Bigg[\frac{N_c^3}{N_c^2-1}\bigg\{s[y,z] \, s[\by,y]\, s[z,\by]
+s[z,y]\, s[y,\by ]\, s[\by,z]
-\frac{1}{N_c^2}\overline{X}^A_{y\by z}\bigg\}-2C_As_A[y,\by] \Bigg]\; .\nonumber
\eeq

This is the final result of this paper. All the discussion of the previous section applies to this result in the same measure. In particular the last term in eq.(\ref{gluonprodf}) should not be mistaken for an additional evolution of the adjoint dipole, since the ``effective evolution interval'' then depends on the momentum of the extra emitted gluon. This term does not appear in \cite{bowen,vitev}.

As stated before, in all the above formulae the PDFs and FFs are defined in the somewhat
unconventional scheme, defined explicitly by the subtraction in eqs. (\ref{drq2})
and (\ref{fragh2}). For numerical implementations it is convenient to use the
standard $\overline{MS}$ scheme instead. The conversion between the two schemes
is given by $f_{\mu^2}=f^{\overline{MS}}_{1.26\mu^2}$. This is discussed in
detail in Appendix C.

 \section*{Acknowledgments}
 We thank Bowen Xiao for useful remarks on the first version of this manuscript. AK and ML  express their gratitude to Universidade de Santiago de Compostela, TA, NA,  GB, and ML  to University of Connecticut, and TA and AK to Ben-Gurion University of the Negev, for warm hospitality during stays when parts of this work were done.
This research  was supported by the People Programme (Marie Curie Actions) of the European Union's Seventh Framework Programme FP7/2007-2013/ under REA
grant agreement \#318921; the DOE grant DE-FG02-13ER41989 (AK); the BSF grant \#2012124 and the  ISRAELI SCIENCE FOUNDATION grant \#87277111 (ML);    the European Research Council grant HotLHC ERC-2011-StG-279579, Ministerio de Ciencia e Innovaci\'on of Spain under project FPA2011-22776, Xunta de Galicia (Conseller\'{\i}a de Educaci\'on and Conseller\'\i a de Innovaci\'on e Industria - Programa Incite),  the Spanish Consolider-Ingenio 2010 Programme CPAN and  FEDER (TA, NA and GB).

\section{Appendix A: The Derivation. Quark to quark to Hadron channel.}
In this Appendix we describe in detail the derivation for hadron production cross section for the simplest channel, where the projectile quark produces an outgoing quark, which subsequently fragments into the hadron. In this setup we show how the Balitsky-Kovchegov evolution equation arises as the appropriate tool to evolve the leading-order amplitude. In  Appendix B we complete the derivation for all the other production channels.

\subsection{Eikonal production.}

The standard eikonal paradigm for propagation of the initial dressed quark with vanishing  transverse momentum through the target leads to the final state
\begin{eqnarray}\label{out}
\vert{\rm out},\alpha,s\rangle&=&
\int_x \bigg[A^qS^F_{\alpha\beta}(x)\vert ({\rm q})\; x_BP^+,\ x,\beta,s\rangle\\
&&\hspace{-1cm}+g\int_\Omega \frac{dLPS}{2\pi} \int_{yz} F_{({\rm qg})}(x_BP^+,\xi,y-x,z-x)_{s\bar s,i}\,t^a_{\alpha\beta}\,S^F_{\beta\gamma}(y)\,S^A_{ab}(z)\,\vert ({\rm q})\, p^+,y,\gamma,\bar s; ({\rm g})\, q^+,z,b,i\rangle\bigg]\nonumber\\
&=&\int_x \bigg\{S^F_{\alpha\beta}(x)\vert ({\rm q})\; x_BP^+,x,\beta,s\rangle_D\nonumber\\
&&\hspace{-1cm}+g\int_\Omega \frac{dLPS}{2\pi} \int_{yz} F_{({\rm qg})}(x_BP^+,\xi,y-x,z-x)_{s,\bar s,i}\left[t^a_{\alpha\beta}S^F_{\beta\gamma}(y)S^A_{ab}(z)-S^F_{\alpha\beta}(x)t^b_{\beta\gamma}\right]\vert ({\rm q})\, p^+,y,\gamma,\bar s; ({\rm g})\, q^+,z,b,i\rangle\bigg\}\; ,\nonumber
\end{eqnarray}
where $p^+= (1-\xi)x_BP^+$, $q^+=\xi x_BP^+$ and $S^F(x)$ is a unitary matrix in the fundamental representation - the eikonal scattering matrix of a projectile quark in the color field of the target, and $\Omega$ is the phase space for the splitting defined in eq. (\ref{eq:Omega}). As explained in Section III, $dLPS$ stands for the phase space in $+$ components for the splitting, corresponding to the $+$ component of the parent parton for the real terms, $dLPS=d\left[\frac{x_pP^+}{1-\xi}\right]$, and the $+$ momentum running in the loop for the virtual ones, $dLPS=d\left[\xi x_pP^+\right]$, with $\xi$ the $+$-momentum fraction taken by the emitted gluon. After this point, we will use the explicit expressions for $dLPS$.

It is convenient to rearrange this expression in terms of the dressed quark states. The single dressed quark state is defined in eq.(\ref{pair1}), and so the first term in rhs of eq.(\ref{out}) is already in the correct form. However the second term needs some work. The dressed quark-gluon state is different from the bare quark-gluon state at order $g$. In particular it has an admixture of a single bare quark state. This is easy to see, since it has to be orthogonal to the single dressed quark state, but without such a term it is not. To order $g$ the requirement of orthogonality is satisfied by
\beq
\vert ({\rm q})\, (1-\xi)x_BP^+,y,\alpha, s; ({\rm g})\, \xi x_BP^+,z,a,i\rangle_D&=&\vert ({\rm q})\, (1-\xi)x_BP^+,y, \gamma,\bar s; ({\rm g})\, \xi x_BP^+,z,b,i\rangle\nonumber\\
&&\hspace{-2cm}-g\int_x F_{({\rm qg})}(x_BP^+,\xi,y-x,z-x)_{s,\bar s,i}\,t^a_{\alpha\beta}\,\vert ({\rm q})\, x_BP^+, x,\ \beta, \bar s\rangle \; .
\eeq
Thus in terms of the dressed states we have
\begin{eqnarray}\label{out1}
\vert {\rm out},\alpha,s\rangle&=&\int_x\bigg\{S^F_{\alpha\beta}(x)\vert({\rm q})\;x_BP^+,x,\beta,s\rangle_D\\
&&\hspace{0cm}+\frac{g^2}{2\pi}\int_\Omega d\big[\xi x_pP^+\big]\int_{x'yz} F_{({\rm qg})}(x_BP^+,\xi,y-x,z-x)_{s,\bar s,i}\left[t^a_{\alpha\beta}S^F_{\beta\gamma}(y)S^A_{ab}(z)-S^F_{\alpha\beta}(x)t^b_{\beta\gamma}\right]
\nonumber\\
&&\hspace{3.5cm}\times\,
 F^*_{(\rm {qg})}(x_BP^+,\xi,y-x',z-x')_{\bar s,\bar {\bar s},i}\,t^b_{\gamma\delta}\,\vert ({\rm q})\, x_BP^+,x',\ \delta, \bar{\bar s}\rangle_D\nonumber\\
&&+\frac{g}{2\pi}\int_\Omega d\left[\frac{x_pP^+}{1-\xi}\right] \int_{yz} F_{({\rm qg})}(x_BP^+,\xi,y-x,z-x)_{s,\bar s,i}\,\left[t^a_{\alpha\beta}S^F_{\beta\gamma}(y)S^A_{ab}(z)-S^F_{\alpha\beta}(x)t^b_{\beta\gamma}\right]\nonumber\\
&&\hspace{3.5cm}\times\,\vert ({\rm q})\, (1-\xi)x_BP^+,y,\gamma,\bar s; ({\rm g})\; \xi x_BP^+,z,b,i\rangle_D\;\bigg\}\; .\nonumber
\end{eqnarray}
The integration over $x'$ identifies $x'$ with $x$ by realizing one of the $\delta$-functions. We can thus write
\begin{eqnarray}\label{out2}
&&\hspace{-1cm}\vert {\rm out},\alpha,s\rangle=\int_x\bigg\{S^F_{\alpha\beta}(x)\,\vert ({\rm q})\; x_BP^+,x,\beta,s\rangle_D\\
&&+\frac{g^2}{2\pi}\int_\Omega d\big[\xi x_pP^+\big]  \int_{yz} \left[t^a_{\alpha\beta}S^F_{\beta\gamma}(y)S^A_{ab}(z)-S^F_{\alpha\beta}(x)t^b_{\beta\gamma}\right] \bar{F}^2_{({\rm qg})}(x_BP^+,\xi,y-x,z-x)\, t^b_{\gamma\delta}\,\vert ({\rm q})\; x_BP^+,x, \delta,  s\rangle_D\nonumber\\
&&+\frac{g}{2\pi}\int_\Omega d\left[\frac{x_pP^+}{1-\xi}\right] \int_{yz} F_{({\rm qg})}(x_BP^+,\xi,y-x,z-x)_{s,\bar s,i}\left[t^a_{\alpha\beta}S^F_{\beta\gamma}(y)S^A_{ab}(z)-S^F_{\alpha\beta}(x)t^b_{\beta\gamma}\right]\nonumber\\
&&\hspace{3cm}\times \vert ({\rm q})\; (1-\xi)x_BP^+,y,\gamma,\bar s; ({\rm g})\;\xi x_BP^+, z, b,i\rangle_D\; \bigg\}\nonumber
\end{eqnarray}
where
\begin{equation}
\bar{F}^2_{({\rm qg})}(x_BP^+,\xi,y-x,z-x)=-
\frac{1}{ x_BP^+}\frac{1+(1-\xi)^2}{\xi}\delta^2\Big(x-[(1-\xi)y+\xi z]\Big)A_{\xi,x_B}^i(y-z)A_{\xi,x_B}^i(y-z)\; .
\end{equation}

Denoting the quark distribution function in the proton by $f^q_{\mu^2}(x_p)$, and the quark fragmentation function to hadron $H$ by $D^q_{H,\mu^2}(\zeta)$, we have for the hadronic level production cross section at leading order\footnote{As mentioned in the body of the text, our PDFs and FFs are defined in a given factorization scheme that does not coincide with the standard $\overline{MS}$ one. But their relation, discussed in detail in Appendix C, amounts to a mild rescaling of factorization scales.}
\begin{eqnarray}\label{cross}
\frac{d\sigma^{ H}}{d^2p_{h\perp} d\eta}&=& \int_{x_F}^1\frac{d\zeta}{\zeta^2}\: \frac{x_F}{\zeta}\; f^q_{\mu^2}\left(\frac{x_F}{\zeta}\right)\;D^q_{H,\mu^2}(\zeta)\;\frac{d\sigma^{q\to q}}{d^2p_\perp d\eta}\left(\frac{p_{h\perp}}{\zeta},{x_F\over \zeta}\right)\\
&=&
\frac{1}{(2\pi)^2}\int_{x_F}^1\frac{d\zeta}{\zeta^2}\, \frac{x_F}{\zeta}\; f^q_{\mu^2}\left(\frac{x_F}{\zeta}\right)\; D^q_{H,\mu^2}(\zeta)\int d^2xd^2y\; e^{i\frac{p_{h\perp}}{\zeta}(x-y)}\, s_{Y_T}(x,y)\nonumber \ ,
\end{eqnarray}
where $s_{Y_T}(x,y)$ is the eikonal dipole s-matrix evolved to rapidity $Y_T$.

The operator definitions of the distribution and fragmentation functions, which we will find useful in the following, is\footnote{The operators $q(k_\perp,x)$ and $q^\dagger(k_\perp,x)$ are bare quark  creation and annihilation operators
normalized in such a way that their commutator is proportional to $\delta^{(2)}(k_\perp)\delta(k^+)$.}
\begin{equation}\label{fq}
f^q_{\mu^2}(x)=P^+\int_{k_\perp^2<\mu^2} d^2k_\perp\langle P\vert q^\dagger(k_\perp,x)q(k_\perp,x)\vert P\rangle; \ \ \ \ \ D^q_{H,\mu^2}(\zeta)=\int_{k_\perp^2<\mu^2} d^2k_\perp\langle q(p)\vert H^\dagger(k_\perp,\zeta p)H(k_\perp,\zeta p)\vert q(p)\rangle
\end{equation}
Here $|P\rangle$ is the wave function of the proton while
$H$ is the operator annihilating the appropriate hadron\footnote{In practice one of course never defines the hadron annihilation operator $H$, but rather uses an equivalent definition of the fragmentation function as the square of the overlap between the quark and hadron states. We find it more convenient to use the notations of eq.(\ref{fq}).}, and the momentum $k_\perp$ is perpendicular to the momentum of the outgoing quark $p$.

Let us first set aside fragmentation and consider quark production at NLO.
The quark production cross section is given by the expectation value of the dressed quark number in the outgoing state, multiplied by the number of dressed quarks in the incoming wave function.
For the quark production we find
\begin{equation}\label{cross1}
\frac{d\sigma^{q}}{d^2p_\perp d\eta}=\frac{1}{(2\pi)^2}x_p\: f^D_{\mu^2}(x_p)\int d^2xd^2y e^{ip_\perp(x-y)} s_{Y_T}(x,y)+\frac{d\sigma_1^q}{d^2p_\perp d\eta}
\end{equation}
where $f^D$ is the number of {\it dressed quarks} in the projectile wave function, or equivalently the {\it leading-order} quark distribution function, and
\begin{equation}
\frac{d\sigma_1^q}{d^2p_\perp d\eta}=\frac{d\sigma_1^{ q\to q\rm{, r }}}{d^2p_\perp d\eta}+\frac{d\sigma_1^{ q\to q\rm{, v }}}{d^2p_\perp d\eta}\; .
\end{equation}
The real contribution to the cross section is
\beq
\frac{d\sigma_1^{ q\to q\rm{, r }}}{d^2\pp d\eta}&=&\frac{g^2}{(2\pi)^3} \int
\frac{dx_B}{x_B P^+} \; x_p P^+ \; f^{q}_{\mu^2}(x_B)  \int
d\left[ \frac{x_p P^+}{1-\xi}\right] \delta\left(x_B P^+-\frac{x_p P^+}{1-\xi}\right) \left[ \frac{1+(1-\xi)^2}{\xi} \right] \int _{y \by z} e^{i\pp(y-\by)} \nonumber\\
&&\times \Ai(y-z)\Ai(\by-z) \, \bigg\{ C_F \, s[y,\by] +C_F \, s\big[ (1-\xi)y+\xi z, (1-\xi)\by+\xi z \big] \nonumber\\
&&-\frac{N_c}{2}\bigg( s\big[ y-\xi(y-z), z \big] \, s[z,\by] + s\big[ z, \by-\xi(\by-z)\big] \, s[y,z] \bigg)\nonumber\\
&&+\frac{1}{2N_c} \bigg( s \big[y-\xi(y-z),\by \big] + s\big[ y, \by-\xi(\by-z)\big] \bigg) \bigg\}\; .
\eeq

Note that one can perform a change of variables in the second term by first changing $(1-\xi)y+\xi z\to y'$ and $(1-\xi)\by+\xi z\to \by'$ and then rescaling
$y'\mapsto(1-\xi)y$, $\by'\mapsto(1-\xi)\by$ and $z\mapsto(1-\xi)z$ to write the real contribution to the quark production in $q\mapsto qg$ as

\beq\label{dsigmaq}
\frac{d\sigma_1^{q\to q, \rm{r}}}{d^2\pp d\eta}&=&\frac{g^2}{(2\pi)^3} \int_0^1 dx_B \; f^{q}_{\mu^2}(x_B) \int_0^{1-x_p} d\xi \; \frac{x_p}{1-\xi}\; \delta\left( x_B-\frac{x_p}{1-\xi}\right) \;  \left[ \frac{1+(1-\xi)^2}{\xi} \right]\int _{y \by z} e^{i\pp(y-\by)} \nonumber\\
&&\times \Ai(y-z)\Ai(\by-z) \, \bigg\{ C_F \, s[y,\by] +(1-\xi)^2\, C_F\, s\big[(1-\xi)y,(1-\xi)\by\big] \nonumber\\
&&-\frac{N_c}{2}\bigg( s\big[ y-\xi(y-z), z \big] \, s[z,\by] + s\big[ z, \by-\xi(\by-z)\big] \, s[y,z] \bigg)\nonumber\\
&&+\frac{1}{2N_c} \bigg( s \big[y-\xi(y-z),\by \big] + s\big[ y, \by-\xi(\by-z)\big] \bigg) \bigg\} \; .
\eeq
The virtual part of the cross section is
\beq\label{dsigmaqv}
\frac{d\sigma_1^{q\to q, \rm{v}}}{d^2\pp d\eta}&=&\frac{g^2}{(2\pi)^3} \int
\frac{dx_B}{x_B P^+} \; x_p P^+ \; f^{q}_{\mu^2}(x_B) \; \delta\left(x_B P^+-x_p P^+\right) \int
d\left[ \xi x_p P^+\right] \left[ \frac{1+(1-\xi)^2}{\xi} \right] \int _{y \by z} e^{i\pp(y-\by)} \nonumber\\
&&\times\Bigg\{  \Ai(y-z)\Ai(y-z) \bigg[ -C_F\ s[y,\by] +\frac{N_c}{2} \, s\Big[ y+\xi(y-z), z+\xi(y-z) \Big] \, s\Big[ z+\xi(y-z),\by\Big]\nonumber\\
&&-\frac{1}{2N_c}\, s\Big[ y+\xi(y-z), \by\Big] \Bigg]\nonumber\\
&&+  \Ai(\by-z)\Ai(\by-z) \bigg[ -C_F\ s[y,\by] +\frac{N_c}{2} \, s\Big[ z+\xi(\by-z), \by+\xi(\by-z) \Big] \, s\Big[ y,z+\xi(\by-z)\Big]\nonumber\\
&&-\frac{1}{2N_c}\, s\Big[ y, \by+\xi(\by-z)\Big] \Bigg] \Bigg\}\,,
\eeq
which can be simplified to
\beq\label{dsigmaq2}
\frac{d\sigma_1^{q\to q, \rm{v}}}{d^2\pp d\eta}
&=&\frac{g^2}{(2\pi)^3} \int_0^1
dx_B \; x_p\; f^{q}_{\mu^2}(x_B) \; \delta\left(x_B-x_p\right) \int_0^1
d\xi \left[ \frac{1+(1-\xi)^2}{\xi} \right] \int _{y \by z} e^{i\pp(y-\by)} \nonumber\\
&&\times\Bigg\{  \Ai(y-z)\Ai(y-z) \bigg[ -C_F\ s[y,\by] +\frac{N_c}{2} \, s\Big[ y+\xi(y-z), z+\xi(y-z) \Big] \, s\Big[ z+\xi(y-z),\by\Big]\nonumber\\
&&-\frac{1}{2N_c}\, s\Big[ y+\xi(y-z), \by\Big] \Bigg]\nonumber\\
&&+  \Ai(\by-z)\Ai(\by-z) \bigg[ -C_F\ s[y,\by] +\frac{N_c}{2} \, s\Big[ z+\xi(\by-z), \by+\xi(\by-z) \Big] \, s\Big[ y,z+\xi(\by-z)\Big]\nonumber\\
&&-\frac{1}{2N_c}\, s\Big[ y, \by+\xi(\by-z)\Big] \Bigg] \Bigg\}\; .
\eeq

Here we have taken the initial quark to have zero transverse momentum (integrated over $x$ and $\bar x$) and have averaged over the spin and color in the initial state. In the order $\alpha_s$  term, there is no difference between the dressed and bare quark distribution and we use $f^{q}_{\mu^2}(x_B)$ for both.

\subsection{Collinear pieces}

The expressions in eqs.(\ref{dsigmaq},\ref{dsigmaqv}) contain collinear divergences. Those come from the infrared region of integration over the coordinate $z$. We assume that the dipole scattering matrix vanishes when the size of the dipole is larger than the inverse saturation momentum, in other words $s(x,z)_{x-z\rightarrow\infty}=0$. This behavior cuts off the large $z$ integration region except in three terms  in eqs.(\ref{dsigmaq},\ref{dsigmaqv}). The two divergent terms in the real part of the cross section and the divergent term in the virtual part are
\beq\label{coldiv}
I_1^r&=&\frac{g^2}{(2\pi)^3}C_F\int_0^1 dx_B \; f^{D,q}_{\mu^2}(x_B) \int_0^{1-x_p} d\xi \; \frac{x_p}{1-\xi}\; \delta\left( x_B-\frac{x_p}{1-\xi}\right) \;  \left[ \frac{1+(1-\xi)^2}{\xi} \right] \tilde{C}_{\mu^2}(\xi,x_B)\nonumber\\
&&\times\int_{y\by}e^{i\pp(y-\by)} s[y,\by]\; ,\\
I_2^r&=&\frac{g^2}{(2\pi)^3}C_F\int_0^1 dx_B \; f^{D,q}_{\mu^2}(x_B) \int_0^{1-x_p} d\xi \; \frac{x_p}{1-\xi}\; \delta\left( x_B-\frac{x_p}{1-\xi}\right) \;  \left[ \frac{1+(1-\xi)^2}{\xi} \right] (1-\xi)^2\tilde{C}_{\mu^2}(\xi,x_B)\nonumber\\
&&\times\int_{y\by}e^{i\pp(y-\by)} s\big[(1-\xi) y,(1-\xi) \by\big] \; , \nonumber\\
I^v&=&-2\frac{g^2}{(2\pi)^3}C_F\int_0^1 dx_B\; f^{D,q}_{\mu^2}(x_B)\; x_p \; \delta\left( x_B-x_p\right) \int_0^1 d\xi  \left[ \frac{1+(1-\xi)^2}{\xi} \right]
\tilde{C}_{\mu^2}(\xi,x_B)\nonumber\\
&&\times\int_{y\by}e^{i\pp(y-\by)} s[y,\by]\nonumber\; .
\eeq
where the integral over $z$ up to ``factorization scale'' $\mu$  is defined in eq.(\ref{fact1}).

By separating these divergent pieces, we can write the real contribution to the quark production as
\begin{equation}
\label{q2qgQPr}
\frac{d\sigma_1^{q\to q, \rm{r}}}{d^2\pp d\eta}=\frac{d\bar{\sigma}_1^{q\to q, \rm{r}}}{d^2\pp d\eta}+I_1^r+I_2^r%
\end{equation}
where the collinear finite part  reads
\beq\label{dsigmaqcol}
\frac{d\bar{\sigma}_1^{q\to q, \rm{r}}}{d^2\pp d\eta}&=&\frac{g^2}{(2\pi)^3} C_F\int_0^1 dx_B \; f^{D,q}_{\mu^2}(x_B) \int_0^{1-x_p} d\xi \; \frac{x_p}{1-\xi}\; \delta\left( x_B-\frac{x_p}{1-\xi}\right) \;  \left[ \frac{1+(1-\xi)^2}{\xi} \right]\int _{y \by z} e^{i\pp(y-\by)} \nonumber\\
&&\times \bigg[ \Ai(y-z)\Ai(\by-z)-  C_{\mu^2}(\xi,x_B,z) \bigg] \, \bigg\{  s[y,\by] +(1-\xi)^2 \,  s\big[(1-\xi)y,(1-\xi)\by\big] \bigg\} \nonumber\\
&&+\frac{g^2}{(2\pi)^3} \int_0^1 dx_B \; f^{D,q}_{\mu^2}(x_B) \int_0^{1-x_p} d\xi \; \frac{x_p}{1-\xi}\; \delta\left( x_B-\frac{x_p}{1-\xi}\right) \;  \left[ \frac{1+(1-\xi)^2}{\xi} \right]\int _{y \by z} e^{i\pp(y-\by)} \nonumber\\
&&\times \Ai(y-z)\Ai(\by-z) \, \bigg\{ -\frac{N_c}{2}\bigg( s\big[ y-\xi(y-z), z \big] \, s[z,\by] + s\big[ z, \by-\xi(\by-z)\big] \, s[y,z] \bigg) \nonumber\\
&&\hspace{5cm}+\frac{1}{2N_c} \bigg( s \big[y-\xi(y-z),\by \big] + s\big[ y, \by-\xi(\by-z)\big] \bigg) \bigg\}\; .
\eeq
A similar decomposition for the virtual part reads
\begin{equation}
\label{q2qgQPv}
\frac{d\sigma_1^{q\to q, \rm{v}}}{d^2\pp d\eta}=\frac{d\bar{\sigma}_1^{q\to q, \rm{v}}}{d^2\pp d\eta}+I^v
\end{equation}
where the collinear finite part of the cross section is
\beq\label{sigmavirtual1}
\frac{d\bar{\sigma}_1^{q\to q, \rm{v}}}{d^2\pp d\eta}&=&-\frac{g^2}{(2\pi)^3}C_F\int_0^1 dx_B\; f^{D,q}_{\mu^2}(x_B)\; x_p \; \delta\left( x_B-x_p\right) \int_0^1 d\xi  \left[ \frac{1+(1-\xi)^2}{\xi} \right]
\int_{y\by z} e^{i\pp(y-\by)} s[y,\by]\nonumber\\
&&\times\bigg\{ \Ai(y-z)\Ai(y-z)+ \Ai(\by-z)\Ai(\by-z)-2C_{\mu^2}(\xi,x_B,z) \bigg\}\nonumber\\
&&+\frac{g^2}{(2\pi)^3}C_F\int_0^1 dx_B\; f^{D,q}_{\mu^2}(x_B)\; x_p \; \delta\left( x_B-x_p\right) \int_0^1 d\xi  \left[ \frac{1+(1-\xi)^2}{\xi} \right]
\int_{y\by z} e^{i\pp(y-\by)}\nonumber\\
&&\times\Bigg\{
\Ai(y-z)\Ai(y-z) \bigg[ \frac{N_c}{2} \, s\Big[ y+\xi(y-z), z+\xi(y-z) \Big] \, s\Big[ z+\xi(y-z),\by\Big]\nonumber\\
&&-\frac{1}{2N_c}\, s\Big[ y+\xi(y-z), \by\Big] \Bigg]\nonumber\\
&&+  \Ai(\by-z)\Ai(\by-z) \bigg[ \frac{N_c}{2} \, s\Big[ z+\xi(\by-z), \by+\xi(\by-z) \Big] \, s\Big[ y,z+\xi(\by-z)\Big]\nonumber\\
&&-\frac{1}{2N_c}\, s\Big[ y, \by+\xi(\by-z)\Big] \Bigg] \Bigg\}\; .
\eeq

The collinear divergences of eq.(\ref{coldiv}) can be absorbed into NLO corrections to the quark PDF and the quark fragmentation functions.

Recall that the function $f^D$ that appears in the leading-order term is the number of ``dressed quarks'' in the proton, or equivalently, the leading-order quark distribution. Part of the $O(\alpha_s)$ terms complete it to the NLO PDF of bare quarks, the quantity that is familiar from DIS etc. It is straightforward to relate the two, since we simply have to calculate the number of bare quarks in the $O(\alpha_s)$ part of the wave function of the incoming proton.

In the collinear approximation (that is assuming that all incoming dressed quarks have zero transverse momentum) we have the dressed quark distribution
\beq
f^D_{\mu^2}(x_p)\equiv P^+\int_{k_{\perp}}\langle P\vert D^\dagger(x_p,k_\perp)D(x_p,k_\perp)\vert P \rangle \; .
\eeq
Here
$D(x_p,k_\perp)$ is the annihilation operator for the  dressed quark.
The simple probabilistic interpretation allows us to write the bare quark PDF defined in (\ref{fq}) in terms of the dressed quark PDF in the following way.
\beq
f^q_{\mu^2}(x_p)&&=\frac{P^+}{2\pi}\int d\left[\frac{x_pP^+}{1-\xi}\right]\int_{l_\perp} \langle P\vert D^\dagger\left(\frac{x_p}{1-\xi},l_\perp\right)D\left(\frac{x_p}{1-\xi},l_\perp\right)\vert P\rangle \int_{k_{\perp}}\frac{{}_D\langle \frac{x_p}{1-\xi},l_\perp\vert q^\dagger(x_p,k_\perp)q(x_p,k_\perp)\vert \frac{x_p}{1-\xi},l_\perp\rangle_D}{{}_D\langle \frac{x_p}{1-\xi},l_\perp\vert \frac{x_p}{1-\xi},l_\perp\rangle_D}\nonumber\\
&&=\frac{x_pP^+}{2\pi}\int \frac{d\xi}{(1-\xi)^2} f^D_{\mu^2}\left(\frac{x_p}{1-\xi}\right)\int_{k_{\perp}}\frac{{}_D\langle \frac{x_p}{1-\xi},0\vert q^\dagger(x_p,k_\perp)q(x_p,k_\perp)\vert \frac{x_p}{1-\xi},0\rangle_D}{{}_D\langle \frac{x_p}{1-\xi},0\vert \frac{x_p}{1-\xi},0\rangle_D} \; .
\end{eqnarray}
With the explicit form of the dressed quark wave function (\ref{pair}) we find
\begin{eqnarray}\label{drq}
f^q_{\mu^2}(x_p)&=&f^D_{\mu^2}(x_p)+\frac{g^2C_F}{2\pi}\int_0^{1-x_p} \frac{d\xi}{1-\xi} f^D_{\mu^2}\left(\frac{x_p}{1-\xi}\right) \frac{1+(1-\xi)^2}{\xi}
\tilde{C}_{\mu^2}\left(\xi,\frac{x_p}{1-\xi}\right)\nonumber\\
&-&\frac{g^2C_F}{2\pi}f^D_{\mu^2}(x_p)\int_0^1 d\xi\frac{1+(1-\xi)^2}{\xi}\tilde{C}_{\mu^2}\left(\xi,x_p\right)
\end{eqnarray}
where $A_{\xi}^i( y-z)$ is defined in eq.(\ref{axi}) and the second term is due to the virtual correction to the quark wave function eq.(\ref{A}).
Writing the last two terms together we get the usual + prescription for the splitting function, so the expression now contains no soft singularity and we can remove the regulator in the $\tilde{C}_{\mu^2}$, giving
\begin{equation}\label{drq2}
f^q_{\mu^2}(x_p)=f^D_{\mu^2}(x_p)+\frac{g^2C_F}{2\pi}\int_0^{1-x_p} \frac{d\xi}{1-\xi} f^D_{\mu^2}\left(\frac{x_p}{1-\xi}\right) \left[\frac{1+(1-\xi)^2}{\xi}\right]_+
\int_z \Big(A^i(z)A^i(z)\Big)_{\mu^2}.
\end{equation}
The last two terms in eq.(\ref{drq}) are equal to $I^r_1$ and half of $I^v$ in eq.(\ref{coldiv}).
Thus we have
\begin{equation}\label{crosscol}
\frac{d\sigma^q}{d^2p_\perp d\eta}=\frac{1}{(2\pi)^2}x_p\, f^q_{\mu^2}(x_p)\int d^2xd^2y e^{ip_\perp(x-y)} s_{Y_T}(x,y)+\frac{d\bar\sigma_1^{ q\to q\rm{, r }}}{d^2p_\perp d\eta}+\frac{d\bar \sigma_1^{ q\to q\rm{, v }}}{d^2p_\perp d\eta}+I_2^r+\frac{1}{2}I^v\; .
\end{equation}

\subsection{Fragmentation.}
The other collinearly divergent term in eq.(\ref{dsigmaq}) has a similar effect on the fragmentation function. Recall that we are interested in calculation of hadron production rather than quark production. On the other hand, eq. (\ref{dsigmaq}) gives the production cross section of {\it dressed} quarks. To convert it into the hadron production cross section we have to multiply it by the fragmentation function of a {\it dressed} quark.
The difference between the dressed and bare quark fragmentation is only important in the leading-order term. For hadron production from the final state quark we have
\beq\label{sigmah1}
\frac{d\sigma^H}{d^2p_{h\perp} d\eta}&=&\int_{x_F}^1\frac{d\zeta}{\zeta^2}\: D^{D,q}_{H,\mu^2}(\zeta)\: \frac{x_F}{\zeta}\: f^D_{\mu^2}\left(\frac{x_F}{\zeta}\right)\int d^2xd^2y\: e^{i\frac{p_{h\perp}}{\zeta}(x-y)} s_{Y_T}(x,y)\nonumber\\
&&+\int_{x_F}^1\frac{d\zeta}{\zeta^2}\: D^q_{H,\mu^2}(\zeta)\: \frac{d\sigma_1^q}{d^2p_\perp d\eta}\left(\frac{p_{h\perp}}{\zeta},\frac{x_F}{\zeta}\right)\; .
\eeq
In order to write this in terms of the standard bare quark fragmentation function we need to do a little calculation. The dressed and bare quark fragmentation functions are defined as
\begin{equation}
D^q_{H,\mu^2}(\zeta)=\int_{k_\perp^2<\mu^2} d^2k_\perp\langle q(p)\vert H^\dagger(k_\perp,\zeta p)H(k_\perp,\zeta p)\vert q(p)\rangle\; ,
\end{equation}
\begin{equation}
D^{D,q}_{H,\mu^2}(\zeta)=\int_{k_\perp^2<\mu^2} d^2k_\perp\ {}_D\langle q(p)\vert H^\dagger(k_\perp,\zeta p)H(k_\perp,\zeta p)\vert q(p)\rangle_D\; .
\end{equation}
To relate the two we write in the collinear approximation
\begin{eqnarray}
D^{D,q}_{H,\mu^2}(\zeta)&=&\int d\xi\int_{k_\perp^2<\mu^2} d^2k_\perp\ \langle q\vert H^\dagger(k_\perp,\frac{\zeta}{1-\xi} )H(k_\perp,\frac{\zeta}{1-\xi} )\vert q\rangle\nonumber\\
&&\times
\int_{l_\perp^2<\mu^2} d^2l_\perp\ {}_D\langle q\vert q^\dagger(l_\perp,(1-\xi)P^+ )q(l_\perp,(1-\xi)P^+ )\vert q\rangle_D\nonumber\\
&+&\int d\xi\int_{k_\perp^2<\mu^2} d^2k_\perp\ \langle g\vert H^\dagger(k_\perp,\frac{\zeta}{\xi} )H(k_\perp,\frac{\zeta}{\xi} )\vert g\rangle\times
\int_{l_\perp^2<\mu^2} d^2l_\perp\ {}_D\langle q\vert g^\dagger(l_\perp,\xi P^+ )g(l_\perp,\xi P^+ )\vert q\rangle_D
\end{eqnarray}
where the last term accounts for the fragmentation of the gluon component of the dressed quark state.

Given the explicit wave function of the dressed quark we find

\beq
\label{fragh}
D_{H,\mu^2}^{D,q}(\zeta)=D^q_{H,\mu^2}(\zeta)&+&\frac{g^2}{2\pi}C_F D^q_{H,\mu^2}(\zeta)\int_0^1 d\xi \frac{1+(1-\xi)^2}{\xi}\tilde{C}_{\mu^2}\left(\xi,\frac{x_F}{\zeta}\right)\nonumber\\
&-&\frac{g^2}{2\pi}C_F \int_{0}^{1-\zeta}\frac{d\xi}{1-\xi} D^q_{H,\mu^2}\left(\frac{\zeta}{1-\xi}\right)\frac{1+(1-\xi)^2}{\xi}\tilde{C}_{\mu^2}\left(\xi,\frac{x_F}{\zeta}\right)\nonumber\\
&-&\frac{g^2}{2\pi}C_F \int_{\zeta}^{1}\frac{d\xi}{\xi} D^g_{H,\mu^2}\left(\frac{\zeta}{\xi}\right)\frac{1+(1-\xi)^2}{\xi}\tilde{C}_{\mu^2}\left(\xi,\frac{x_F}{\zeta}\right)\; .
\eeq
Again, as we did when going from eq. (\ref{drq}) to eq. (\ref{drq2}), we can write
\beq
\label{fragh2}
D_{H,\mu^2}^{D,q}(\zeta)=D^q_{H,\mu^2}(\zeta)&-&\frac{g^2}{2\pi}C_F \int_{0}^{1-\zeta}\frac{d\xi}{1-\xi} D^q_{H,\mu^2}\left(\frac{\zeta}{1-\xi}\right)\left[\frac{1+(1-\xi)^2}{\xi}\right]_+\int_z \Big(A^i(z)A^i(z)\Big)_{\mu^2}\nonumber\\
&-&\frac{g^2}{2\pi}C_F \int_{\zeta}^{1}\frac{d\xi}{\xi} D^g_{H,\mu^2}\left(\frac{\zeta}{\xi}\right)\frac{1+(1-\xi)^2}{\xi}\int_z \Big(A^i(z)A^i(z)\Big)_{\mu^2}\; .
\eeq
 Here the last term corresponds to the fragmentation of the gluon component of the dressed quark. This term pairs up with the collinear divergent piece in the production cross section from the final state gluon, and we will postpone its discussion for later.

Note that the Weiszacker-Williams field in eq.(\ref{fragh}) is defined with the constraint corresponding to the longitudinal momentum of the incoming quark $x_p=\frac{x_F}{\zeta}$. This is because the total momentum of the dressed quark which fragments does not depend on the fraction of momentum carried by the ``soft'' gluon.

Using eq.(\ref{fragh}) in eq.(\ref{sigmah1}) we find that all remaining collinear divergences combine into the quark fragmentation function with the final result:
\beq
\frac{d\sigma^H}{d^2p_{h\perp} d\eta}&=&\frac{1}{(2\pi)^2}\int_{x_F}^1\frac{d\zeta}{\zeta^2}\: D^q_{H,\mu^2}(\zeta)
\int_0^1 dx_B\: \delta\left( x_B-\frac{x_F}{\zeta}\right)\frac{x_F}{\zeta}\: f^q_{\mu^2}(x_B)\int_{y\by}e^{i\frac{p_{h\perp}}{\zeta}(y-\by)}s[y,\by]\nonumber\\
&&+\int_{x_F}^1\frac{d\zeta}{\zeta^2}\: D^q_{H,\mu^2}(\zeta)\: \frac{d\bar{\sigma}^q}{d^2\pp d\eta}\left(\frac{p_{h\perp}}{\zeta},\frac{x_F}{\zeta}\right)\; .
\eeq

Collecting all the collinear divergent pieces into the bare quark PDF and fragmentation function leads to the expressions eqs.(\ref{dsigmaqcol1},\ref{sigmavirtual2}).

\subsection{Evolution.}
Now we have to understand how evolution figures into this. All the expressions so far contain dipole scattering amplitude. The way we set up the calculation, this amplitude depends on rapidity. It is evolved up to rapidity $Y_T$ starting with an initial condition provided at $Y^0_T$. So far we have not specified what is the  evolution equation that governs this evolution. The consistency of the calculation presented above determines this equation.

There are several equivalent ways of deriving the evolution equation.
A simple physical way to do this, is to require that the increase in total energy of the collision can be either encoded in the increase of $P^+_P$ or in the additional boost of the target with the same effect.
Let us consider the production at fixed $x_p$ but at higher energy (note that this means that we are also increasing the rapidity of the observed hadron in the center of mass frame), such that
\begin{equation}
P^+\rightarrow P^+e^\Delta\approx P^+(1+\Delta)\; .
\end{equation}
This must be equivalent to evolving the $s$-matrix without changing $P^+$. This requirement leads to the equation
\begin{equation}\label{e2}
\frac{\partial}{\partial \ln P^+}\frac{d\sigma}{d^2p_\perp d\eta}=\frac{d s_{Y_T}}{d Y_T}\frac{\delta}{\delta s_{Y_T}}\frac{d\sigma}{d^2p_\perp d\eta}\; .
\end{equation}

Note, that while writing eq.(\ref{e2}) we have assumed that the parameter $\tau$ does not depend on the rapidity of the target. This is something worth clarifying, especially since our interpretation of the parameter $\tau$ was the thickness of the target. One would naively expect that if the target is boosted, the time $\tau$ should experience Lorentz contraction. Indeed, since the cross section depends only on the combination $P^+/\tau$, it is obvious that
\begin{equation}
\frac{d\sigma}{d^2p_\perp d\eta}(P^+e^\Delta,\tau)=\frac{d\sigma}{d^2p_\perp d\eta}(P^+,\tau e^{-\Delta})\; .
\end{equation}
Increasing $P^+$ increases the number of quark-gluon pairs that live long enough to scatter coherently off a fixed target (time dilation of the projectile). Alternatively, decreasing the thickness of the target (Lorentz contraction of the target) increases the time resolution, and therefore now more quark-gluon pairs are resolved in the wave function of a fixed projectile. The two descriptions are indeed equivalent.

However what we mean by evolution of the target is the effective change in the scattering matrix on the target, which, without any functional change in the expression for $\sigma$, completely accounts for the change of the cross section with energy. This means that we want to keep ``parameter'' $\tau$ fixed through the evolution. This is yet the third description of the evolution consistent with the idea, that when one boosts the target wave function, effectively its thickness remains unchanged due to emission of softer gluons. However this emission of softer gluons modified the dipole scattering amplitude on the target. Thus we can either think of a projectile dipole accompanied by fast fluctuations in its wave function, in which case the cross section grows because the faster moving target resolves more fluctuations, or alternatively think about a ``bare'' dipole, whose scattering amplitude changes due to emission of extra soft gluons in the target wave function.

The corollary of this discussion, is that the constant (rapidity independent) parameter $\tau$ has to be interpreted as the thickness of the target at initial rapidity $Y^0_T$, since it does not depend on the total energy. This naturally gives $\tau^{-1}=P^{-0}_T$, and therefore $P^+/\tau=s_0/2$ as used in the text.

With this interpretation of the cutoff parameter in hand, we can derive the evolution from a different point of view.
The result for our calculation of the cross section should not depend on our choice of $s_0$. After all our choice of $s_0$ was arbitrary except for the requirement that it should be large enough so that eikonal approximation is valid. If we start with a higher $s_0$, the scattering amplitude would have to be evolved over a smaller range of rapidities. The final result should not care which $s_0$ we choose if we evolve the dipole cross section appropriately.

The dependence on $s_0$ enters explicitly through the cutoff on the phase space and through the dependence of the scattering amplitude on the amount of evolution $Y_T$. The evolution therefore has to be such that the following equation is satisfied
\begin{equation}\label{e1}
s_0\frac{d}{d s_0}\left[\frac{d\sigma}{d^2p_\perp d\eta}\right]=\left[s_0\frac{\partial}{\partial s_0}-\frac{d s_{Y_T}}{d Y_T}\frac{\delta}{\delta s_{Y_T}}\right]\frac{d\sigma}{d^2p_\perp d\eta} =0\,,
\end{equation}
which is equivalent to eq. (\ref{e2}).

We can now derive the evolution equation satisfied by the dipole amplitude. The only dependence on $P^+$ in the production cross section is in integration limit in the next-to-leading-order term. We can change integration limits, by integrating over $\xi$ first, and over the transverse momenta $l_\perp$ and $m_\perp$ later. The production cross section  can then be written in the form
\begin{equation}\label{exch}
\int d^2l_\perp d^2m_\perp \int_{\xi>\xi_{min}}\frac{d \xi}{\xi}\Sigma_i \Phi_i(\xi,l_\perp,m_\perp),
\end{equation}
where the $i$-th term  in the expression for the production cross section is represented by some regular function $\Phi_i(\xi,l_\perp,m_\perp)$.
Here
\begin{equation}
\xi_{min,r}={\rm max}\left\{\frac{l_\perp^2}{x_ps_0},\frac{m_\perp^2}{x_ps_0}\right\}
\end{equation}
for the real part of the cross section, and
\begin{equation}
\xi_{min,v}(1-\xi_{min,v})={\rm max}\left\{\frac{l_\perp^2}{x_ps_0},\frac{m_\perp^2}{x_ps_0}\right\}
\end{equation}
for the virtual part. The momenta $l_\perp$ and $m_\perp$ are the momenta in the Fourier transforms of the two factors of $A_i$ present in every term.
Differentiating with respect to $\ln s_0$ sets $\xi$ at the lower limit of the integration and gets rid of the explicit factor $1/\xi$. We then have
\begin{equation}\label{xmin}
s_0\frac{\partial}{\partial s_0}\frac{d\sigma}{d^2p_\perp d\eta}=\int d^2l_\perp d^2m_\perp\Sigma_i \Phi_i(\xi_{min,i},l_\perp,m_\perp)\; .
\end{equation}
Note that the minimal value $\xi_{min}\ll 1$, since the relevant momentum in the Weiszacker-Williams field is either $p_\perp$ or $Q_s$, and both must be much smaller than the initial energy $s_0$ in order for our initial eikonal approximation to hold. Since $F_i(\xi,l,m)$ is a regular function of $\xi$, we can set $\xi_{min}=0$ in eq.(\ref{xmin}) up to corrections of order $p^2_\perp/s_0$ or $Q_s^2/s_0$.  This simplifies the expressions in eq.(\ref{xmin}) considerably. The real and virtual contributions then combine into
\begin{equation}
s_0\frac{\partial}{\partial s_0}\left[\frac{d\sigma}{d^2p_\perp d\eta}\right]=-\frac{\alpha_sN_c}{\pi}x_p f^q_{\mu^2}(x_p)\int_{y,\bar y,z}\frac{1}{(2\pi)^3}e^{ip_\perp(y-\bar y)}\frac{(y-\bar y)^2}{(y-z)^2(\bar y-z)^2}\Big [s(y,\bar y)-s(y,z)s(z,\bar y)\Big].
\end{equation}
This, by virtue of eq. (\ref{cross1}) and eq.(\ref{e1}) means that the dipole amplitude evolves according to the BK equation.

Note that in this derivation we have assumed that the number of dressed quarks in the proton  $f^D_{\mu^2}(x_p)$ does not have significant dependence on $s_0$ and therefore its derivative with respect to $s_0$ can be neglected. This point is discussed in more detail in Section III.

\section{Appendix B: The Derivation. Other channels.}

In this appendix we provide the derivation for the cross section for all other channels. First we derive expressions for the partonic level cross sections, explicitly separating collinearly divergent terms. Then we show that these perturbative collinear divergent terms combine into final results in terms of partonic PDF's and fragmentation functions.

\subsection{Quark to gluon channel.}
Production cross section of the final state gluon from initial state quark is given by
\beq
\frac{d\sigma_1^{q\to g, \rm{r}}}{d^2\pp d\eta}&=&\frac{g^2}{(2\pi)^3} \int
\frac{dx_B}{x_B P^+} \; x_p P^+ \; f^{q}_{\mu^2}(x_B)  \int
d\left[ \frac{x_p P^+}{\xi}\right]  \delta\left(x_B P^+-\frac{x_p P^+}{\xi}\right)\left[ \frac{1+(1-\xi)^2}{\xi} \right] \int _{z \bz y} e^{i\pp(z-\bz)} \nonumber\\
&&\times \Ai(z-y)\Ai(\bz-y) \, \bigg\{ C_F \, s_A[z,\bz] +C_F \, s\big[ (1-\xi)y+\xi z, (1-\xi)y+\xi \bz \big] \nonumber\\
&&-\frac{N_c}{2}\bigg( s\big[ z,(1-\xi)y+\xi \bz \big] \, s[y,z] + s_A[\bz,y] \; s\big[ (1-\xi)y+\xi z,\bz \big] \bigg)\nonumber\\
&&+\frac{1}{2N_c} \bigg( s \big[y,(1-\xi)y+\xi \bz \big] + s\big[ (1-\xi)y+\xi z,y \big] \bigg) \bigg\}
\eeq
where the adjoint dipole scattering amplitude is defined in eq.(\ref{adjdip})
Naturally there is no virtual contribution in this channel.
Performing in the second term change of variables $(1-\xi)y+\xi z\mapsto z'$ and $(1-\xi)y+\xi \bz\mapsto \bz'$ and then rescaling $z'\mapsto\xi z$, $\bz'\mapsto\xi\bz$ and $y\mapsto\xi y$ this can be written as

\beq
\frac{d\sigma_1^{q\to g, \rm{r}}}{d^2\pp d\eta}&=&\frac{g^2}{(2\pi)^3} \int_0^1
dx_B \; f^{q}_{\mu^2}(x_B)  \int_{x_p}^1
d\xi\;  \frac{x_p}{\xi} \; \delta\left(x_B-\frac{x_p}{\xi}\right) \left[ \frac{1+(1-\xi)^2}{\xi} \right] \int _{z \bz y} e^{i\pp(z-\bz)} \nonumber\\
&&\times \Ai(z-y)\Ai(\bz-y) \, \bigg\{ C_F \, s_A[z,\bz] +C_F \, \xi^2\; s\big[ \xi z, \xi\bz \big] \nonumber\\
&&-\frac{N_c}{2}\bigg( s\big[ z,(1-\xi)y+\xi \bz \big] \, s[y,z] + s[\bz,y] \; s\big[ (1-\xi)y+\xi z,\bz \big] \bigg)\nonumber\\
&&+\frac{1}{2N_c} \bigg( s \big[y,(1-\xi)y+\xi \bz \big] + s\big[ (1-\xi)y+\xi z,y \big] \bigg) \bigg\}\; .
\eeq

Separating the collinearly divergent pieces this can be written as

\beq
\label{q2gqGP}
\frac{d\sigma_1^{q\to g, \rm{r}}}{d^2\pp d\eta}&=&\frac{d\bar{\sigma}_1^{q\to g, \rm{r}}}{d^2\pp d\eta}
+\frac{g^2}{(2\pi)^3}\, C_F \int_0^1 dx_B\;  f^{q}_{\mu^2}(x_B)\int_{x_p}^1 d\xi \; \frac{x_p}{\xi} \; \delta\left( x_B-\frac{x_p}{\xi} \right) \left[\frac{1+(1-\xi)^2}{\xi} \right] \nonumber\\
&&\hspace{3.3cm}\times \tilde{C}_{\mu^2}(\xi,x_B)\int_{z\bz}e^{i\pp(z-\bz)} s_A[z,\bz]\nonumber\\
&&+\frac{g^2}{(2\pi)^3}\, C_F \int_0^1 dx_B\;  f^{q}_{\pp}(x_B)\int_{x_p}^1 d\xi \; \frac{x_p}{\xi} \; \delta\left( x_B-\frac{x_p}{\xi} \right) \left[\frac{1+(1-\xi)^2}{\xi} \right] \; \xi^2\nonumber\\
&&\hspace{1.5cm}\times \tilde{C}_{\mu^2}(\xi,x_B) \int_{z\bz}e^{i\pp(z-\bz)} s[\xi z,\xi\bz],
\eeq
with the collinearly finite part
\beq
\frac{d\bar{\sigma}_1^{q\to g, \rm{r}}}{d^2\pp d\eta}&=&\frac{g^2}{(2\pi)^3} \int_0^1 dx_B \; f^{q}_{\mu^2}(x_B)  \int_{x_p}^1 d\xi\;  \frac{x_p}{\xi} \; \delta\left(x_B-\frac{x_p}{\xi}\right) \left[ \frac{1+(1-\xi)^2}{\xi} \right] \int _{z \bz y} e^{i\pp(z-\bz)} \nonumber\\
&&\times\bigg[ \Ai(z-y)\Ai(\bz-y)-C_{\mu^2}(\xi,x_B,y)\bigg] \, \bigg\{ C_F \, s_A[z,\bz] +C_F \, \xi^2\; s\big[ \xi z, \xi\bz \big]\bigg\} \nonumber\\
&&+\frac{g^2}{(2\pi)^3} \int_0^1 dx_B \; f^{q}_{\mu^2}(x_B)  \int_{x_p}^1 d\xi\;  \frac{x_p}{\xi} \; \delta\left(x_B-\frac{x_p}{\xi}\right) \left[ \frac{1+(1-\xi)^2}{\xi} \right] \int _{z \bz y} e^{i\pp(z-\bz)}\nonumber\\
&&\times\Ai(z-y)\Ai(\bz-y)\bigg\{
-\frac{N_c}{2}\bigg( s\big[ z,(1-\xi)y+\xi \bz \big] \, s[y,z] + s[\bz,y] \; s\big[ (1-\xi)y+\xi z,\bz \big] \bigg)\nonumber\\
&&+\frac{1}{2N_c} \bigg( s \big[y,(1-\xi)y+\xi \bz \big] + s\big[ (1-\xi)y+\xi z,y \big] \bigg) \bigg\}\; .
\eeq

The first and second collinearly divergent terms in eq.(\ref{q2gqGP}) combine into a part of the NLO corrections to the gluon PDF and quark fragmentation functions respectively, as we will see later.

\subsection{Gluon to gluon channel.}

We next discuss hadron production from the projectile gluon. Just like in the quark case, we first consider the production on partonic level and take care of fragmentation later.
At leading order the projectile gluon scatters elastically producing a gluon in the final state.
The cross section for this process is
\beq
\frac{d\sigma^g}{d^2\pp d\eta}=\frac{1}{(2\pi)^2}\, x_p\, f^{D,g}_{\mu^2}(x_p)\int_{y,\by}e^{i\pp(y-\by)}s_A[y,\by],
\eeq
where $f^{D,g}_{\mu^2}(x_p)$ is the dressed gluon distribution function.

At next-to-leading order the gluon in the projectile wave function splits into a gluon-gluon configuration and also into a quark-antiquark configuration. Thus, the "dressed" gluon state has both a two gluon state component and a quark-antiquark state component.
\beq
&&|({\rm g})\; x_BP^+,x,a,k\rangle_D=\int d^2x \, e^{ik_{\perp} x}\bigg\{ A^g\vert ({\rm g})\; x_BP^+,x,a,k\rangle\\
&&+ g\int_{\Omega} \frac{dLPS}{2\pi}\int_{yz}\, F_{({\rm gg})}(x_BP^+,\xi,y-x,z-x)_{ijk}\, if^{abc}\, \vert ({\rm g})\, p^+=(1-\xi)x_BP^+,y,b,i;  ({\rm g})\, q^+=\xi x_BP^+,z,c, j\rangle\nonumber\\
&&+g\int_{\Omega} \frac{dLPS}{2\pi}\int_{yz} F_{({\rm q\bq})}(x_BP^+,\xi,y-x,z-x)_{ss',k}\, t^a_{\alpha\beta}\, \vert ({\rm q})\, p^+=(1-\xi)x_BP^+,y, \alpha,s; ({\rm \bq})\,q^+=\xi x_BP^+,z,\beta,s'\rangle\bigg\},\nonumber
\eeq
where the measure $dLPS$ stands for the longitudinal phase and it is defined as in the case of quark production. The normalization constant $A^g$ is
\beq
A^g=1&-&g^2\frac{C_A}{2S}x_BP^+\int_{\Omega}d\xi\, d^2x \, d^2\bx \, d^2y \, d^2z F_{(\rm gg)}(x_BP^+,\xi,x,y,z)F^*_{(\rm gg)}(x_BP^+,\xi,\bx,y,z) \nonumber\\
&-&g^2\frac{C_F}{2S}x_BP^+\int_{\Omega}d\xi\, d^2x \, d^2\bx \, d^2y \, d^2z F_{(\rm q\bq)}(x_BP^+,\xi,x,y,z)F^*_{(\rm q\bq)}(x_BP^+,\xi,\bx,y,z) \; .
\eeq

The two gluon state and quark-antiquark state are orthogonal to each other, and the virtual corrections (normalization factors ) to the "bare" gluon state due to these orthogonal states are additive. Thus, we can consider $gg$ component and $q\bq$ components of the "dressed" gluon state separately.

The two gluon component of the "dressed" gluon state with color $a$, transverse index $k$ and vanishing transverse momentum component after passing through the target is given by

\beq
&&\hspace{-0.8cm}|{\rm out}, a, k\rangle = \int_x\bigg\{ S^{ab}_A(x) |({\rm g})\; x_BP^+,x,b,k\rangle_D \nonumber\\
&&\hspace{-0.5cm}+ \frac{g^2}{2\pi}\int_\Omega d\big[\xi x_pP^+\big] \int_{yz}\bigg[ if^{abc}\, S^{cd}_A(y)\,S^{be}_A(z)-if^{bed}\, S^{ab}_A(x)\bigg] \bar{F}^2_{({\rm gg})}(x_BP^+,\xi,y-x,z-x)\, if^{lde}\, |({\rm g})\;x_BP^+,x,l,k\rangle_D\nonumber\\
&&\hspace{-0.5cm}+\frac{g}{2\pi}\int_\Omega d\left[\frac{x_pP^+}{1-\xi}\right] \int_{yz} F_{({\rm gg})}(x_BP^+,\xi,y-x,z-x)_{ijk}\bigg[if^{abc}\,S^{cd}_A(y)\,S^{be}_A(z)-if^{bed}\,S^{ab}_A(x)\bigg]\nonumber\\
&&\hspace{2.5cm} \times\; |({\rm g})\; (1-\xi)x_BP^+,y,d,j; ({\rm g})\; \xi x_BP^+,z,e,i\rangle_D\bigg\}\; .
\eeq

The function $F_{({\rm gg})}(x_BP^+,\xi,y-x,z-x)_{ijk}$ can be read off, for example from  \cite{production}
\beq
\label{Fgg}
F_{({\rm gg})}(x_BP^+,\xi,y-x,z-x)_{ijk}&=&\frac{-i}{\sqrt{2\xi(1-\xi)x_BP^+}}\bigg\{ (1-\xi)\delta_{im}\delta_{jk}+\xi\delta_{jm}\delta_{ik}-\xi(1-\xi)\delta_{km}\delta_{ij}\bigg\}\nonumber\\
&&\hspace{2cm}\times\delta^{(2)}\Big[(1-\xi)y+\xi z-x\Big]\; \Am(y-z)
\eeq
and similarly
\beq
&&\hspace{-0.5cm}\bar{F}^2_{({\rm gg})}(x_BP^+,\xi,y-x,z-x)=\frac{1}{2x_BP^+}\bigg[ \frac{1-\xi}{\xi}+\frac{\xi}{1-\xi}+\xi(1-\xi)\bigg]\delta^{(2)}\Big[(1-\xi)y+\xi z-x\Big]\nonumber\\
&&\hspace{4cm}\times\;\Ai(y-z)\Ai(y-z)\; .
\eeq
\subsubsection{The real contribution}
The real contribution to gluon production cross section is
\beq
\label{g2ggR0}
\frac{d\sigma_1^{g\to g, \rm{r}}}{d^2\pp d\eta}&=&\frac{4g^2}{(2\pi)^3} \int
dx_B \; x_p P^+ \; f^{g}_{\mu^2}(x_B)  \int d\left[\frac{x_pP^+}{1-\xi}\right]  \delta\left(x_B P^+-\frac{x_p P^+}{1-\xi}\right)\int_{x\bx,y\by,z}e^{i\pp(y-\by)}\nonumber\\
&&\times\frac{1}{2(N_c^2-1)}\tr\bigg[F_{({\rm gg})}(x_BP^+,\xi,y-x,z-x)_{ijk}F^*_{({\rm gg})}(x_BP^+,\xi, \by-\bx,z-\bx)_{ijk}\bigg]\nonumber\\
&&\times \bigg[if^{abc}S^{cd}_A(y)S^{be}_A(z)-if^{bed}S^{ab}_A(x)\bigg]\bigg[-if^{a\bb\bc}S^{\bc d}_A(\by)S^{\bb e}_A(z)+if^{\bb ed}S^{a\bb}_A(\bx)\bigg]\; .
\eeq
Noting that the $SU(N_c)$ generator in adjoint representation $T^a_{bc}=-if^{abc}$, and using color algebra to write the result in terms of fundamental traces, we obtain
\beq
\label{g2ggR2}
\frac{d\sigma_1^{g\to g, \rm{r}}}{d^2\pp d\eta}&=&\frac{g^2}{(2\pi)^3}2C_A\int_0^1 dx_B\, f^{g}_{\mu^2}(x_B)\, \int_0^{1-x_p} d\xi \,\frac{x_p}{1-\xi}\, \delta\left(x_B-\frac{x_p}{1-\xi}\right) \, \bigg[ \frac{1-\xi}{\xi}+\frac{\xi}{1-\xi}+\xi(1-\xi)\bigg]\nonumber\\
&&\times\int_{y\by z} e^{i\pp(y-\by)}\Ai(y-z)\Ai(\by-z)\bigg\{ s_A[y,\by]+s_A\Big[(1-\xi)y+\xi z, (1-\xi)\by+\xi z\Big]\nonumber\\
&&-\frac{N_c^2}{2(N_c^2-1)}\bigg[s[y,z] \, s\Big[(1-\xi)\by+\xi z,y\Big]\, s\Big[z,(1-\xi)\by+\xi z\Big]\nonumber\\
&&\hspace{2cm}+s[z,y] \, s\Big[y,(1-\xi)\by+\xi z \Big]\, s\Big[(1-\xi)\by+\xi z,z\Big]\nonumber\\
&&\hspace{2cm}+s\Big[\by,(1-\xi)y+\xi z\Big]\, s\Big[(1-\xi)y+\xi z,z\Big]\, s[z,\by]\nonumber\\
&&\hspace{2cm}+s\Big[(1-\xi)y+\xi z,\by\Big]\, s\Big[z, (1-\xi)y+\xi z\Big]\, s[\by,z]-\frac{1}{N_c^2}\overline{X}_{y\by z}\bigg]\bigg\},
\eeq
 where the definitions of $X[x,y,z,\bx,\by,\by]$ and $\overline{X}_{y\by z}$ are given in eqs.(\ref{defX}) and (\ref{defXbar}) respectively.

By performing the same change of variables in the second adjoint dipole and rescaling the variables exactly in the same way as in the case of the quark production in $q\mapsto qg$, we finally obtain
\beq
\label{g2ggR3}
\frac{d\sigma_1^{g\to g, \rm{r}}}{d^2\pp d\eta}&=&\frac{g^2}{(2\pi)^3}2C_A\int_0^1 dx_B\, f^{g}_{\mu^2}(x_B)\, \int_0^{1-x_p} d\xi \,\frac{x_p}{1-\xi}\, \delta\left(x_B-\frac{x_p}{1-\xi}\right) \, \bigg[ \frac{1-\xi}{\xi}+\frac{\xi}{1-\xi}+\xi(1-\xi)\bigg]\nonumber\\
&&\times\int_{y\by z} e^{i\pp(y-\by)}\Ai(y-z)\Ai(\by-z)\bigg\{ s_A[y,\by]+(1-\xi)^2\, s_A\Big[(1-\xi)y, (1-\xi)\by\Big]\nonumber\\
&&-\frac{N_c^2}{2(N_c^2-1)}\bigg[s[y,z] \, s\Big[(1-\xi)\by+\xi z,y\Big]\, s\Big[z,(1-\xi)\by+\xi z\Big]\nonumber\\
&&\hspace{2cm}+s[z,y] \, s\Big[y,(1-\xi)\by+\xi z \Big]\, s\Big[(1-\xi)\by+\xi z,z\Big]\nonumber\\
&&\hspace{2cm}+s\Big[\by,(1-\xi)y+\xi z\Big]\, s\Big[(1-\xi)y+\xi z,z\Big]\, s[z,\by]\nonumber\\
&&\hspace{2cm}+s\Big[(1-\xi)y+\xi z,\by\Big]\, s\Big[z, (1-\xi)y+\xi z\Big]\, s[\by,z]-\frac{1}{N_c^2}\overline{X}_{y\by z}\bigg]\bigg\}\; .
\eeq

Separating the collinearly divergent terms we obtain
\beq
\label{g2ggRF}
\frac{d\sigma_1^{g\to g, \rm{r}}}{d^2\pp d\eta}=\frac{d\bar{\sigma}_1^{g\to g, \rm{r}}}{d^2\pp d\eta}
&+&\frac{g^2}{(2\pi)^3}2C_A\int_0^1 dx_B\, f^{g}_{\mu^2}(x_B)\, \int_0^{1-x_p} d\xi \,\frac{x_p}{1-\xi}\, \delta\left(x_B-\frac{x_p}{1-\xi}\right) \, \bigg[ \frac{1-\xi}{\xi}+\frac{\xi}{1-\xi}+\xi(1-\xi)\bigg]\nonumber\\
&&\times \tilde{C}_{\mu^2}(\xi,x_B)\int_{y\by} e^{i\pp(y-\by)} s_A[y,\by]\nonumber\\
&&+\frac{g^2}{(2\pi)^3}2C_A\int_0^1 dx_B\, f^g_{\mu^2}(x_B)\, \int_0^{1-x_p} d\xi \,\frac{x_p}{1-\xi}\, \delta\left(x_B-\frac{x_p}{1-\xi}\right) \, \bigg[ \frac{1-\xi}{\xi}+\frac{\xi}{1-\xi}+\xi(1-\xi)\bigg]\nonumber\\
&&\times(1-\xi)^2\tilde{C}_{\mu^2}(\xi,x_B)\int_{y\by} e^{i\pp(y-\by)} \; s_A\Big[(1-\xi)y,(1-\xi)\by\Big],
\eeq
where  the collinearly finite part of the cross section is
\beq
\frac{d\bar{\sigma}_1^{g\to g, \rm{r}}}{d^2\pp d\eta}&=&\frac{g^2}{(2\pi)^3}2C_A\int_0^1 dx_B\, f^{g}_{\mu^2}(x_B)\, \int_0^{1-x_p} d\xi \,\frac{x_p}{1-\xi}\, \delta\left(x_B-\frac{x_p}{1-\xi}\right) \, \bigg[ \frac{1-\xi}{\xi}+\frac{\xi}{1-\xi}+\xi(1-\xi)\bigg]\, \int_{y\by z} e^{i\pp(y-\by)}\nonumber\\
&&\times\bigg[\Ai(y-z)\Ai(\by-z)-C_{\mu^2}(\xi,x_B,z)\bigg]\bigg\{ s_A[y,\by]+(1-\xi)^2\, s_A\Big[(1-\xi)y, (1-\xi)\by\Big]\bigg\}\nonumber\\
&&-\frac{g^2}{(2\pi)^3}\int_0^1 dx_B\, f^{g}_{\mu^2}(x_B)\, \int_0^{1-x_p} d\xi \,\frac{x_p}{1-\xi}\, \delta\left(x_B-\frac{x_p}{1-\xi}\right) \, \bigg[ \frac{1-\xi}{\xi}+\frac{\xi}{1-\xi}+\xi(1-\xi)\bigg]\nonumber\\
&&\times\int_{y\by z} e^{i\pp(y-\by)}\Ai(y-z)\Ai(\by-z)
\frac{N_c^3}{N_c^2-1}\bigg\{s[y,z] \, s\Big[(1-\xi)\by+\xi z,y\Big]\, s\Big[z,(1-\xi)\by+\xi z\Big]\nonumber\\
&&\hspace{1.5cm}+s[z,y] \, s\Big[y,(1-\xi)\by+\xi z \Big]\, s\Big[(1-\xi)\by+\xi z,z\Big]\nonumber\\
&&\hspace{1.5cm}+s\Big[\by,(1-\xi)y+\xi z\Big]\, s\Big[(1-\xi)y+\xi z,z\Big]\, s[z,\by]\nonumber\\
&&\hspace{1.5cm}+s\Big[(1-\xi)y+\xi z,\by\Big]\, s\Big[z, (1-\xi)y+\xi z\Big]\, s[\by,z]-\frac{1}{N_c^2}\overline{X}_{y\by z}\bigg]\bigg\}\; .
\eeq
\subsubsection{The virtual contribution.}
The virtual contribution to gluon production cross section in $g\to gg$, can be written as
\beq
\label{g2ggV}
\frac{d\sigma_1^{g\to g, \rm{v}}}{d^2\pp d\eta}&=&\frac{2g^2}{(2\pi)^3}\int dx_B\, f^{g}_{\mu^2}(x_B)\, x_pP^+\, \delta\left( x_BP^+-x_pP^+\right)\,\int d\big[\xi x_pP^+\big] \int_{x\bx yz }\, e^{i\pp(x-\bx)} \bar{F}^2_{({\rm gg})}(x_BP^+,\xi,y-x,z-x)\nonumber\\
&&\times \frac{1}{N_c^2-1}S^{a\bb}_A(\bx)\bigg[ if^{abc}S^{cd}_A(y)S^{be}_A(z)-if^{bed}S^{ab}_A(x)\bigg]if^{\bb de}+c.c.\; \; .
\eeq
By using the explicit expression for $\bar{F}^2_{({\rm gg})}(x_BP^+,\xi,y-x,z-x)$ eq.\eqref{g2ggV}, changing variables $y\to y+\xi(y-z)$ and $z\to z+\xi(y-z)$ and writing the octet dipole in terms of fundamental traces, we obtain

\beq
\frac{d\sigma_1^{g\to g, \rm{v}}}{d^2\pp d\eta}&=&\frac{g^2}{(2\pi)^3}\int_0^1 dx_B\, f^{g}_{\mu^2}(x_B)\, x_p\, \delta(x_B-x_p)\int_0^1 d\xi \bigg[\frac{1-\xi}{\xi}+\frac{\xi}{1-\xi}+\xi(1-\xi)\bigg]\int_{y\by z}e^{i\pp(y-\by) }\nonumber\\
&&\times\bigg\{\Ai(y-z)\Ai(y-z)\bigg[-N_c\,s_A[y,\by] -\frac{N_c}{2(N_c^2-1)}\overline{X}'_{y\by z}\nonumber\\
&&+\frac{N_c^3}{2(N_c^2-1)}\bigg(s\Big[y+\xi(y-z),z+\xi(y-z)\Big] s\Big[z+\xi(y-z),\by\Big]s\Big[\by,y+\xi(y-z)\Big]\nonumber\\
&&\hspace{2cm}+s\Big[z+\xi(y-z),y+\xi(y-z)\Big]s\Big[\by,z+\xi(y-z)\Big] s\Big[y+\xi(y-z),\by\Big]\bigg)\bigg]\nonumber\\
&&+\Ai(\by-z)\Ai(\by-z)\bigg[-N_c\,s_A[y,\by] -\frac{N_c}{2(N_c^2-1)}\overline{X}''_{y\by z}\nonumber\\
&&+\frac{N_c^3}{2(N_c^2-1)}\bigg(s\Big[\by+\xi(\by-z),z+\xi(\by-z)\Big]s\Big[z+\xi(\by-z),y\Big] s\Big[y,\by+\xi(\by-z)\Big]\nonumber\\
&&\hspace{2cm}+s\Big[z+\xi(\by-z),\by+\xi(\by-z)\Big] s\Big[y,z+\xi(\by-z)\Big]s\Big[\by+\xi(\by-z),y\Big]\bigg)\bigg]\bigg\},
\eeq
where the definitions of 6-point function, $\overline{X}'_{y\by z}$ and $\overline{X}''_{y\by z}$ are given in eqs. (\ref{defX}), (\ref{defXp}) and (\ref{defXpp}) respectively.

Separating the collinearly divergent piece, we obtain
\beq
\label{g2ggVF}
\frac{d\sigma_1^{g\to g, \rm{v}}}{d^2\pp d\eta}&=&\frac{d\bar{\sigma}_1^{g\to g, \rm{v}}}{d^2\pp d\eta}-\frac{g^2}{(2\pi)^3}2C_A\int_0^1 dx_B \, f^{g}_{\mu^2}(x_B) \, x_p \, \delta(x_B-x_p)\int_0^1 d\xi \bigg[ \frac{1-\xi}{\xi}+\frac{\xi}{1-\xi}+\xi(1-\xi)\bigg]\nonumber\\
&&\hspace{3.4cm}\times \tilde{C}_{\mu^2}(\xi,x_B)\int_{y\by}e^{i\pp(y-\by)}s_A[y,\by],
\eeq
with the collinear finite piece
\beq
\frac{d\bar{\sigma}_1^{g\to g, \rm{v}}}{d^2\pp d\eta}&=&-\frac{g^2}{(2\pi)^3}C_A\int_0^1 dx_B\, f^{g}_{\mu^2}(x_B)\, x_p\, \delta(x_B-x_p)\int_0^1 d\xi \bigg[\frac{1-\xi}{\xi}+\frac{\xi}{1-\xi}+\xi(1-\xi)\bigg]\int_{y\by z}e^{i\pp(y-\by)}s_A[y,\by] \nonumber\\
&&\times\bigg\{\Ai(y-z)\Ai(y-z)+\Ai(\by-z)\Ai(\by-z)-2C_{\mu^2}(\xi,x_B,z)\bigg\}\nonumber\\
&&+\frac{g^2}{(2\pi)^3}\int_0^1 dx_B\, f^{g}_{\mu^2}(x_B)\, x_p\, \delta(x_B-x_p)\int_0^1 d\xi \bigg[\frac{1-\xi}{\xi}+\frac{\xi}{1-\xi}+\xi(1-\xi)\bigg]\int_{y\by z}e^{i\pp(y-\by)} \nonumber\\
&&\times\bigg\{\Ai(y-z)\Ai(y-z)\bigg[-\frac{N_c}{2(N_c^2-1)}\overline{X}'_{y\by z}\nonumber\\
&&+\frac{N_c^3}{2(N_c^2-1)}\bigg(s\Big[y+\xi(y-z),z+\xi(y-z)\Big] s\Big[z+\xi(y-z),\by\Big]s\Big[\by,y+\xi(y-z)\Big]\nonumber\\
&&\hspace{2cm}+s\Big[z+\xi(y-z),y+\xi(y-z)\Big]s\Big[\by,z+\xi(y-z)\Big] s\Big[y+\xi(y-z),\by\Big]\bigg)\bigg]\nonumber\\
&&+\Ai(\by-z)\Ai(\by-z)\bigg[-\frac{N_c}{2(N_c^2-1)}\overline{X}''_{y\by z}\nonumber\\
&&+\frac{N_c^3}{2(N_c^2-1)}\bigg(s\Big[\by+\xi(\by-z),z+\xi(\by-z)\Big]s\Big[z+\xi(\by-z),y\Big] s\Big[y,\by+\xi(\by-z)\Big]\nonumber\\
&&\hspace{2cm}+s\Big[z+\xi(\by-z),\by+\xi(\by-z)\Big] s\Big[y,z+\xi(\by-z)\Big]s\Big[\by+\xi(\by-z),y\Big]\bigg)\bigg]\bigg\}\; .
\eeq

\subsection{Gluon to quark channel.}

The quark-antiquark component of the "dressed" gluon state with color $a$, transverse index $k$ and vanishing transverse momentum component reads after passing through the target
\beq
&&\hspace{-0.8cm}|{\rm out}, a, k\rangle = \int_x\bigg\{ S^{ab}_A(x) |({\rm g})\; x_BP^+,x,b,k\rangle_D \nonumber\\
&&\hspace{-0.6cm}
+ \frac{g^2}{2\pi}\int_\Omega d\big[\xi x_pP^+\big] \int_{yz}\bigg\{\Big[S^{\dagger}_F(z)t^aS_F(y)\Big]_{\delta \gamma}-S^{ab}_A(x)t^a_{\delta\gamma}\bigg\} \bar{F}^2_{({\rm q\bq})}(x_BP^+,\xi,y-x,z-x)\, t^{\bb}_{\gamma\delta}\, |({\rm g})\; x_BP^+,x,\bb,k\rangle_D\nonumber\\
&&\hspace{-0.6cm}+\frac{g}{2\pi}\int_\Omega d\left[\frac{x_pP^+}{1-\xi}\right] \int_{yz} F_{({\rm q\bq})}(x_BP^+,\xi,y-x,z-x)_{ss',j}\bigg\{\Big[S^{\dagger}_F(z)t^aS_F(y)\Big]_{\delta \gamma}-S^{ab}_A(x)t^a_{\delta\gamma}\bigg\} \nonumber\\
&&\hspace{4cm} \times\; |({\rm q})\; (1-\xi)x_BP^+,y,\gamma,s; ({\rm \bq})\; \xi x_BP^+,z,\delta,s'\rangle_D\bigg\}\; .
\eeq
The function $F_{({\rm q\bq})}(x_BP^+,\xi,y-x,z-x)_{ss',j}$ can be read off from \cite{production} and its explicit form reads
\beq
\label{Fqbq}
\hspace{-0.5cm}F_{({\rm q\bq})}(x_BP^+,\xi,y-x,z-x)_{ss',j}&=&\frac{-i}{\sqrt{2x_BP^+}}\bigg[-(2\xi-1)\delta_{ij}\delta_{-s',s}+i\epsilon^{ij}\sigma^3_{-s',s}\bigg]\nonumber\\
&&\times\; \delta^{(2)}\big[(1-\xi)y+\xi z-x\big]\Ai(y-z)\; ,
\eeq
and similarly
\beq
\bar{F}^2_{({\rm q\bq})}(x_BP^+,\xi,y-x,z-x)=\frac{1}{x_BP^+}\big[\xi^2+(1-\xi)^2\big]\delta^{(2)}\big[(1-\xi)y+\xi z-x\big]\Ai(y-z)\Ai(y-z)\,.
\eeq

The $q\bq$ component of the "dressed" gluon state contributes  to quark, antiquark and gluon production. At the level of the cross section the real term corresponds  to the quark (and antiquark) production cross section, while the virtual term corresponds to the quark loop contribution to gluon production.
\subsubsection{The real contribution.}

The real contribution to the  quark production cross section reads
\beq
\label{g2qbq}
\frac{d\sigma_1^{g\to q, \rm{r}}}{d^2\pp d\eta}&=&\frac{g^2}{(2\pi)^3}\int dx_B\,  f^{g}_{\mu^2}(x_B)\,x_pP^+\int d\left[\frac{x_pP^+}{1-\xi}\right]\, \delta\left[ x_BP^+-\frac{x_pP^+}{1-\xi}\right]\int_{x\bx y\by z}e^{i\pp(y-\by)}\nonumber\\
&&\times\frac{1}{2(N_c^2-1)}\tr\bigg[F_{({\rm q\bq})}(x_BP^+,\xi,y-x,z-x)_j F^*_{({\rm q\bq})}(x_BP^+,\xi,\by-\bx,z-\bx)_j\bigg]\nonumber\\
&&\times \bigg\{ \Big[ S^{\dagger}_F(z)t^aS_F(y)\Big]_{\delta\gamma}-S^{ab}_A(x)t^b_{\delta\gamma}\bigg\}\bigg\{  \Big[ S^{\dagger}_F(\by)t^aS_F(z)\Big]_{\gamma\delta}-S^{a\bb}_A(\bx)t^{\bb}_{\gamma\delta}\bigg\}\; .
\eeq
We now use the explicit form of the function $F_{({\rm q\bq})}(x_BP^+,\xi,y-x,z-x)_{ss',j}$ given in eq.\eqref{Fqbq}, use color algebra to express adjoint matrices in terms of fundamental ones,
and  perform the same change of variables as in the previous cases to obtain
\beq
\label{g2qbq3}
\frac{d\sigma_1^{g\to q, \rm{r}}}{d^2\pp d\eta}&=&\frac{1}{2}\frac{g^2}{(2\pi)^3}\int_0^1 dx_B\,  f^{g}_{\mu^2}(x_B)\int_0^{1-x_p} d\xi\, \frac{x_p}{1-\xi}\delta\left(x_B-\frac{x_p}{1-\xi}\right)\big[\xi^2+(1-\xi)^2\big] \int_{y\by z}e^{i\pp(y-\by)}\nonumber\\
&&\times\Ai(y-z)\Ai(\by-z)\bigg\{ s[y,\by]+(1-\xi)^2\, s_A\big[(1-\xi)y,(1-\xi)\by\big]\nonumber\\
&&-\frac{N_c^2}{(N_c^2-1)}\bigg[ s\big[y,(1-\xi)\by+\xi z\big]s\big[(1-\xi)\by+\xi z,z\big]-\frac{1}{N_c^2}s[y,z]\bigg]\nonumber\\
&&-\frac{N_c^2}{(N_c^2-1)}\bigg[ s\big[z,(1-\xi)y+\xi z\big]s\big[(1-\xi)y+\xi z,\by\big]-\frac{1}{N_c^2}s[z,\by]\bigg]\bigg\}\; .
\eeq
Separating the collinearly divergent pieces this becomes
\beq
\label{g2qbqfR}
\frac{d\sigma_1^{g\to q, \rm{r}}}{d^2\pp d\eta}=\frac{d\bar{\sigma}_1^{g\to q\bq, \rm{r}}}{d^2\pp d\eta}&+&\frac{1}{2}\frac{g^2}{(2\pi)^3}\int_0^1 dx_B\,  f^{g}_{\mu^2}(x_B)\int_0^{1-x_p} d\xi\, \frac{x_p}{1-\xi}\delta\left(x_B-\frac{x_p}{1-\xi}\right)\big[\xi^2+(1-\xi)^2\big] \nonumber\\
&&\hspace{3cm}\times \tilde{C}_{\mu^2}(\xi,x_B)\int_{y\by}e^{i\pp(y-\by)}s[y,\by]\nonumber\\
&+&\frac{1}{2}\frac{g^2}{(2\pi)^3}\int_0^1 dx_B\,  f^{g}_{\mu^2}(x_B)\int_0^{1-x_p} d\xi\, \frac{x_p}{1-\xi}\delta\left(x_B-\frac{x_p}{1-\xi}\right)\big[\xi^2+(1-\xi)^2\big] (1-\xi)^2\nonumber\\
&&\hspace{3cm}\times \tilde{C}_{\mu^2}(\xi,x_B)\int_{y\by}e^{i\pp(y-\by)}s_A\big[(1-\xi)y,(1-\xi)\by\big],
\eeq
with the collinear finite part
\beq
\frac{d\bar{\sigma}_1^{g\to q, \rm{r}}}{d^2\pp d\eta}&=&\frac{1}{2}\frac{g^2}{(2\pi)^3}\int_0^1 dx_B\,  f^{g}_{\mu^2}(x_B)\int_0^{1-x_p} d\xi\, \frac{x_p}{1-\xi}\delta\left(x_B-\frac{x_p}{1-\xi}\right)\big[\xi^2+(1-\xi)^2\big] \int_{y\by z}e^{i\pp(y-\by)}\nonumber\\
&&\times\bigg\{ \bigg[\Ai(y-z)\Ai(\by-z)-C_{\mu^2}(\xi,x_B,z)\bigg]\bigg[ s[y,\by]+(1-\xi)^2s_A\big[(1-\xi)y,(1-\xi)\by\big]\bigg]\nonumber\\
&&-\frac{N_c^2}{(N_c^2-1)}\Ai(y-z)\Ai(\by-z)\bigg[ s\big[y,(1-\xi)\by+\xi z\big]s\big[(1-\xi)\by+\xi z,z\big]-\frac{1}{N_c^2}s[y,z]\bigg]\nonumber\\
&&-\frac{N_c^2}{(N_c^2-1)}\Ai(y-z)\Ai(\by-z)\bigg[ s\big[z,(1-\xi)y+\xi z\big]s\big[(1-\xi)y+\xi z,\by\big]-\frac{1}{N_c^2}s[z,\by]\bigg]\bigg\}\; .
\eeq
\subsubsection{The virtual contribution.}

The virtual term in the cross section corresponds to the quark loop contribution to gluon production and it reads
\beq
\frac{d\sigma_1^{g\to q, \rm{v}}}{d^2\pp d\eta}&=&\frac{g^2}{(2\pi)^3}\int dx_B\, f^{g}_{\mu^2}(x_B)\, x_pP^+\, \delta\left( x_BP^+-x_pP^+\right)\,\int d\big[\xi x_pP^+\big] \int_{x\bx yz }\, e^{i\pp(x-\bx)} \bar{F}^2_{({\rm q\bq})}(x_BP^+,\xi,y-x,z-x)\nonumber\\
&&\times\frac{1}{N_c^2-1}\bigg\{S^{a\bb}_A(\bx)\Big[S^{\dagger}_F(z)t^aS_F(y)\Big]_{\delta \gamma}-S^{ab}_A(x)t^b_{\delta\gamma}\bigg\}t^{\bb}_{\delta \gamma}+c.c.\; \; .
\eeq
Using the explicit expression for $ \bar{F}^2_{({\rm q\bq})}(\xi,x_B,y-x,z-x)$, shifting the variables $y\to y+\xi(y-z)$ and $z\to z+\xi(y-z)$, writing the crossed terms in fundamental representation and separating the collinear divergencies, the virtual contribution becomes
\beq
\label{g2qqGPv}
\frac{d\sigma_1^{g\to q, \rm{v}}}{d^2\pp d\eta}&=&\frac{d\bar{\sigma}_1^{g, \rm{virtual}}}{d^2\pp d\eta}-\frac{g^2}{(2\pi)^3}\int_0^1 dx_B\, f^{g}_{\mu^2}(x_B)\, x_p\delta(x_B-x_p)\int_0^1 d\xi \Big[\xi^2+(1-\xi)^2\Big]\nonumber\\
&&\hspace{3cm}\times \tilde{C}_{\mu^2}(\xi,x_B)\int_{y\by}e^{i\pp(y-\by)}s_A[y,\by],
\eeq
where the finite part is
\beq
\frac{d\bar{\sigma}_1^{g\to q, \rm{v}}}{d^2\pp d\eta}&=&-\frac{1}{2}\frac{g^2}{(2\pi)^3}\int_0^1 dx_B\, f^{ g}_{\mu^2}(x_B)\, x_p\, \delta\left( x_B-x_p\right)\int_0^1 d\xi \Big[\xi^2+(1-\xi)^2\Big]\int_{y\by z}e^{i\pp(y-\by)}\nonumber\\
&&\times\bigg\{\bigg[\Ai(y-z)\Ai(y-z)+\Ai(\by-z)\Ai(\by-z)-2C_{\mu^2}(\xi,x_B,z)\bigg]s_A[y,\by]\nonumber\\
&&\hspace{0.5cm}-\Ai(y-z)\Ai(y-z)\bigg[\frac{N_c^2}{N_c^2-1}s\big[y+\xi(y-z),\by\big]s\big[\by,z+\xi(y-z)\big]\nonumber\\
&&\hspace{5cm}-\frac{1}{N_c^2-1}s\big[y+\xi(y-z),z+\xi(y-z)\big]\bigg]\nonumber\\
&&\hspace{0.5cm}-\Ai(\by-z)\Ai(\by-z)\bigg[\frac{N_c^2}{N_c^2-1}s\big[\by+\xi(\by-z),y\big]s\big[y,z+\xi(\by-z)\big]\nonumber\\
&&\hspace{5cm}-\frac{1}{N_c^2-1}s\big[\by+\xi(\by-z),z+\xi(\by-z)\big]\bigg]\bigg\}\; .
\eeq

\subsection{PDF's and Fragmentation functions.}
Here we present the relation between the PDF's and fragmentation functions of "`dressed"' and "`bare"' quarks and gluons. The derivation follows the same route as the one that leads to eq.(\ref{drq}) and eq.(\ref{fragh}).

The NLO expression for the quark PDF which solves the DGLAP equations is:
\beq
\label{qpdf}
f^q_{\mu^2}(x_p)=f^{D,q}_{\mu^2}(x_p)&+&\frac{g^2}{2\pi}C_F\int_0^{1-x_p}\frac{d \xi}{1-\xi}f^{D,q}_{\mu^2}\left(\frac{x_p}{1-\xi}\right)\bigg[\frac{1+(1-\xi)^2}{\xi}\bigg]\tilde{C}_{\mu^2}\left(\xi,\frac{x_p}{1-\xi}\right)\nonumber\\
&-&\frac{g^2}{2\pi}C_F f^{D,q}_{\mu^2}(x_p)\int_0^1d \xi \bigg[\frac{1+(1-\xi)^2}{\xi}\bigg] \tilde{C}_{\mu^2}(\xi,x_p)\nonumber\\
&+& \frac{g^2}{2\pi}\frac{1}{2}\int_0^{1-x_p}\frac{d\xi}{1-\xi} f^{D,g}_{\mu^2}\left(\frac{x_p}{1-\xi}\right)\Big[\xi^2+(1-\xi)^2\Big]\tilde{C}_{\mu^2}\left(\xi,\frac{x_p}{1-\xi}\right)\; .
\eeq
As we have shown in the previous section, this relation arises when using the operator definition of the dressed and bare parton distribution functions. This is identical to the standard perturbative relation when $f^{D,q}$ is understood as the leading-order PDF. The only subtlety here is that the collinearly divergent integral $C_{\mu^2}$ in eq.(\ref{qpdf}) depends on the momentum fraction $\xi$ through the Ioffe time cutoff. This dependence is very weak, and shows up only as a power correction of the type $p^2/s_0$ to the logarithmically divergent integral. Therefore, as we did when going from eq. (\ref{drq}) to eq. (\ref{drq2}), we can write
\beq
\label{qpdf2}
f^q_{\mu^2}(x_p)=f^{D,q}_{\mu^2}(x_p)&+&\frac{g^2}{2\pi}C_F\int_0^{1-x_p}\frac{d \xi}{1-\xi}f^{D,q}_{\mu^2}\left(\frac{x_p}{1-\xi}\right)\bigg[\frac{1+(1-\xi)^2}{\xi}\bigg]_+\int_z \Big(A^i(z)A^i(z)\Big)_{\mu^2}\nonumber\\
&+& \frac{g^2}{2\pi}\frac{1}{2}\int_0^{1-x_p}\frac{d\xi}{1-\xi} f^{D,g}_{\mu^2}\left(\frac{x_p}{1-\xi}\right)\Big[\xi^2+(1-\xi)^2\Big]\int_z \Big(A^i(z)A^i(z)\Big)_{\mu^2}\; .
\eeq

The analogous expression for the gluon PDF is given by
\beq
\label{gpdf}
f^g_{\mu^2}(x_p)=f^{D,g}_{\mu^2}(x_p)&+&\frac{g^2}{2\pi}2C_A\int_0^{1-x_p}\frac{d\xi}{1-\xi}f^{D,g}_{\mu^2}\left(\frac{x_p}{1-\xi}\right)
\bigg[\frac{1-\xi}{\xi}+\frac{\xi}{1-\xi}+\xi(1-\xi)\bigg]\tilde{C}_{\mu^2}\left(\xi,\frac{x_p}{1-\xi}\right)\nonumber\\
&-&\frac{g^2}{2\pi}C_A\,f^{D,g}_{\mu^2}(x_p)\int_0^{1}d\xi
\bigg[\frac{1-\xi}{\xi}+\frac{\xi}{1-\xi}+\xi(1-\xi)\bigg]\tilde{C}_{\mu^2}(\xi,x_p)\nonumber\\
&-&\frac{g^2}{2\pi}N_f\, \frac{1}{2}\, f^{D,g}_{\mu^2}(x_p)\int_0^{1}d\xi \Big[\xi^2+(1-\xi)^2\Big]\tilde{C}_{\mu^2}(\xi,x_p)\nonumber\\
&+&\frac{g^2}{2\pi}C_F\int_{x_p}^1\frac{d\xi}{\xi}\bigg[f^{D,q}_{\mu^2}\left(\frac{x_p}{\xi}\right)
+f^{D,\bar{q}}_{\mu^2}\left(\frac{x_p}{\xi}\right)\bigg]\frac{1+(1-\xi)^2}{\xi}\tilde{C}_{\mu^2}(\xi,x_p)
\eeq
that, again, can be written as
\beq
\label{gpdf2}
f^g_{\mu^2}(x_p)=f^{D,g}_{\mu^2}(x_p)&+&\frac{g^2}{2\pi}2C_A\int_0^{1-x_p}\frac{d\xi}{1-\xi}f^{D,g}_{\mu^2}\left(\frac{x_p}{1-\xi}\right)
\bigg[\left(\frac{1-\xi}{\xi}+\frac{1}{2}\xi(1-\xi)\right)_++\frac{\xi}{1-\xi}+\frac{1}{2}\xi(1-\xi)\bigg]\int_z \Big(A^i(z)A^i(z)\Big)_{\mu^2}\nonumber\\
&-&\frac{g^2}{2\pi}N_f\, \frac{1}{2}\, f^{D,g}_{\mu^2}(x_p)\int_0^{1}d\xi \Big[\xi^2+(1-\xi)^2\Big]\int_z \Big(A^i(z)A^i(z)\Big)_{\mu^2}\nonumber\\
&+&\frac{g^2}{2\pi}C_F\int_{x_p}^1\frac{d\xi}{\xi}\bigg[f^{D,q}_{\mu^2}\left(\frac{x_p}{\xi}\right)
+f^{D,\bar{q}}_{\mu^2}\left(\frac{x_p}{\xi}\right)\bigg]\frac{1+(1-\xi)^2}{\xi}\int_z \Big(A^i(z)A^i(z)\Big)_{\mu^2}\; .
\eeq

The analysis analogous to that preceding eq.(\ref{fragh}) leads to the following NLO relation between the dressed and bare quark and gluon fragmentation functions:
\beq
\label{qFF}
D_{H,\mu^2}^{D,q}(\zeta)=D^q_{H,\mu^2}(\zeta)&+&\frac{g^2}{2\pi}C_F D^q_{H,\mu^2}(\zeta)\int_0^1 d\xi \frac{1+(1-\xi)^2}{\xi}\tilde{C}_{\mu^2}\left(\xi,\frac{x_p}{\zeta}\right)\nonumber\\
&-&\frac{g^2}{2\pi}C_F \int_{0}^{1-\zeta}\frac{d\xi}{1-\xi} D^q_{H,\mu^2}\left(\frac{\zeta}{1-\xi}\right)\frac{1+(1-\xi)^2}{\xi}\tilde{C}_{\mu^2}\left(\xi,\frac{x_p}{\zeta}\right)\nonumber\\
&-&\frac{g^2}{2\pi}C_F \int_{\zeta}^{1}\frac{d\xi}{\xi} D^g_{H,\mu^2}\left(\frac{\zeta}{\xi}\right)\frac{1+(1-\xi)^2}{\xi}\tilde{C}_{\mu^2}\left(\xi,\frac{x_p}{\zeta}\right)\; ,
\eeq
\beq
\label{qFF2}
D_{H,\mu^2}^{D,q}(\zeta)=D^q_{H,\mu^2}(\zeta)&-&\frac{g^2}{2\pi}C_F \int_{0}^{1-\zeta}\frac{d\xi}{1-\xi} D^q_{H,\mu^2}\left(\frac{\zeta}{1-\xi}\right)\left[\frac{1+(1-\xi)^2}{\xi}\right]_+\int_z \Big(A^i(z)A^i(z)\Big)_{\mu^2}\nonumber\\
&-&\frac{g^2}{2\pi}C_F \int_{\zeta}^{1}\frac{d\xi}{\xi} D^g_{H,\mu^2}\left(\frac{\zeta}{\xi}\right)\frac{1+(1-\xi)^2}{\xi}\int_z \Big(A^i(z)A^i(z)\Big)_{\mu^2}\; ,
\eeq
\beq
\label{gFF}
D_{H,\mu^2}^{D,g}(\zeta)=D_{H,\mu^2}^{g}(\zeta)&+&\frac{g^2}{2\pi}C_A\,D_{H,\mu^2}^g(\zeta)\int_0^1 d\xi \bigg[ \frac{1-\xi}{\xi}+\frac{\xi}{1-\xi}+\xi(1-\xi)\bigg]\tilde{C}_{\mu^2}\left(\xi,\frac{x_p}{\zeta}\right)\nonumber\\
&-&\frac{g^2}{2\pi}2C_A\int_0^{1-\zeta} \frac{d\xi}{1-\xi}D_{H,\mu^2}^g\left(\frac{\zeta}{1-\xi}\right) \bigg[ \frac{1-\xi}{\xi}+\frac{\xi}{1-\xi}+\xi(1-\xi)\bigg]\tilde{C}_{\mu^2}\left(\xi,\frac{x_p}{\zeta}\right)\nonumber\\
&+&\frac{g^2}{2\pi}N_f\, \frac{1}{2}\, D_{H,\mu^2}^g(\zeta)\int_0^1 d\xi \Big[\xi^2+(1-\xi)^2\Big]\tilde{C}_{\mu^2}\left(\xi,\frac{x_p}{\zeta}\right)\\
&-&\frac{g^2}{2\pi}\frac{1}{2}\int_0^{1-\zeta}\frac{d\xi}{1-\xi}\bigg[ D^q_{H,\mu^2}\left(\frac{\zeta}{1-\xi}\right)+D^{\bq}_{H,\mu^2}\left(\frac{\zeta}{1-\xi}\right)\bigg]\Big[\xi^2+(1-\xi)^2\Big]\tilde{C}_{\mu^2}\left(\xi,\frac{x_p}{\zeta}\right)\; ,\nonumber
\eeq
\beq
\label{gFF2}
D_{H,\mu^2}^{D,g}(\zeta)=D_{H,\mu^2}^{g}(\zeta)&-&\frac{g^2}{2\pi}2C_A\int_0^{1-\zeta} \frac{d\xi}{1-\xi}D_{H,\mu^2}^g\left(\frac{\zeta}{1-\xi}\right)\bigg[\left(\frac{1-\xi}{\xi}+\frac{1}{2}\xi(1-\xi)\right)_++\frac{\xi}{1-\xi}+\frac{1}{2}\xi(1-\xi)\bigg]\int_z \Big(A^i(z)A^i(z)\Big)_{\mu^2}\nonumber\\
&+&\frac{g^2}{2\pi}N_f\, \frac{1}{2}\, D_{H,\mu^2}^g(\zeta)\int_0^1 d\xi \Big[\xi^2+(1-\xi)^2\Big]\int_z \Big(A^i(z)A^i(z)\Big)_{\mu^2}\\
&-&\frac{g^2}{2\pi}\frac{1}{2}\int_0^{1-\zeta}\frac{d\xi}{1-\xi}\bigg[ D^q_{H,\mu^2}\left(\frac{\zeta}{1-\xi}\right)+D^{\bq}_{H,\mu^2}\left(\frac{\zeta}{1-\xi}\right)\bigg]\Big[\xi^2+(1-\xi)^2\Big]\int_z \Big(A^i(z)A^i(z)\Big)_{\mu^2}\; .\nonumber
\eeq

\subsection{Including PDFs}
\subsubsection{Quark production}
The leading-order quark production cross section reads
\beq\label{qlead}
\frac{d\sigma^q}{d^2\pp d\eta}=\frac{1}{(2\pi)^2}\int dx_B\, x_p\delta(x_B-x_p)\, f^{D,q}_{\mu^2}(x_B)\int_{y,\by}e^{i\pp(y-\by)}s[y,\by]\; .
\eeq
Combining this with the second term in eq.\eqref{q2qgQPr}, half of the second term in eq.\eqref{q2qgQPv} and the second term in eq.\eqref{g2qbqfR} turns the dressed quark PDF $f^{D,q}_{\mu^2}(x_p)$ into the standard quark  PDF $f^{q}_{\mu^2}(x_p)$ in eq.(\ref{qlead}). Thus, the quark production cross section at NLO is
\beq
\frac{d\sigma^q}{d^2\pp d\eta}&=&\frac{1}{(2\pi)^2}\int_0^1 dx_B\, x_p\delta(x_B-x_p)\, f^{q}_{\mu^2}(x_B)\int_{y,\by}e^{i\pp(y-\by)}s[y,\by]
+\frac{d\bar{\sigma}^q}{d^2\pp d\eta}\nonumber\\
&&+\frac{g^2}{(2\pi)^3}C_F\int_0^1 dx_B \; f^{q}_{\mu^2}(x_B) \int_0^{1-x_p} d\xi \; \frac{x_p}{1-\xi}\; \delta\left( x_B-\frac{x_p}{1-\xi}\right) \;  \left[ \frac{1+(1-\xi)^2}{\xi} \right] (1-\xi)^2\nonumber\\
&&\hspace{3cm}\times
 \tilde{C}_{\mu^2}(\xi,x_B)\int_{y\by}e^{i\pp(y-\by)} s\big[ (1-\xi)y, (1-\xi)\by\big] \nonumber\\
&&-\frac{g^2}{(2\pi)^3}C_F\int_0^1 dx_B\; f^{q}_{\mu^2}(x_B)\; x_p \; \delta\left( x_B-x_p\right) \int_0^1 d\xi  \left[ \frac{1+(1-\xi)^2}{\xi} \right]\nonumber\\
&&\hspace{3.5cm}\times \tilde{C}_{\mu^2}(\xi,x_B)\int_{y\by}e^{i\pp(y-\by)}s[y,\by]\nonumber\\
&&+\frac{1}{2}\frac{g^2}{(2\pi)^3}\int_0^1 dx_B\,  f^{g}_{\mu^2}(x_B)\int_0^{1-x_p} d\xi\, \frac{x_p}{1-\xi}\delta\left(x_B-\frac{x_p}{1-\xi}\right)\big[\xi^2+(1-\xi)^2\big] (1-\xi)^2\nonumber\\
&&\hspace{3cm}\times \tilde{C}_{\mu^2}(\xi,x_B)\int_{y\by}e^{i\pp(y-\by)}s_A\big[(1-\xi)y,(1-\xi)\by\big],
\eeq
where
\beq
\frac{d\bar{\sigma}^q}{d^2\pp d\eta}=\frac{d\bar{\sigma}_1^{q\to q,{\rm r}}}{d^2\pp d\eta}+\frac{d\bar{\sigma}_1^{q\to q,{\rm v}}}{d^2\pp d\eta}+\frac{d\bar{\sigma}_1^{g\to q,{\rm r}}}{d^2\pp d\eta}
\eeq
defined above.

\subsubsection{Gluon production}

Similarly, the leading-order gluon production cross section is

\beq
\frac{d\sigma^g}{d^2\pp d\eta}=\frac{1}{(2\pi)^2}\int_0^1 dx_B\, x_p\delta(x_B-x_p)\, f^{D,g}_{\mu^2}(x_B)\int_{y,\by}e^{i\pp(y-\by)}s_A[y,\by]\; .
\eeq

 Combining this with the second term in eq. \eqref{q2gqGP}, second term in eq.\eqref{g2ggRF}, half of the second term in eq.\eqref{g2ggVF} and half of the second term in eq. \eqref{g2qqGPv} turns the dressed gluon PDF  $f^{D,g}_{\mu^2}(x_B)$ into the standard gluon PDF $f^{g}_{\mu^2}(x_B)$. Thus,  at next-to-leading-order we have
\beq
\frac{d\sigma^g}{d^2\pp d\eta}&=&\frac{1}{(2\pi)^2}\int_0^1 dx_B\, x_p \delta(x_B-x_p)\, f^{g}_{\mu^2}(x_B)\int_{y,\by}e^{i\pp(y-\by)}s_A[y,\by]
+\frac{d\bar{\sigma}^g}{d^2\pp d\eta}\nonumber\\
&&+\frac{g^2}{(2\pi)^3}\, C_F \int_0^1 dx_B\;  f^{q}_{\mu^2}(x_B)\int_{x_p}^1 d\xi \; \frac{x_p}{\xi} \; \delta\left( x_B-\frac{x_p}{\xi} \right) \left[\frac{1+(1-\xi)^2}{\xi} \right] \; \xi^2\tilde{C}_{\mu^2}(\xi,x_B) \nonumber\\
&&\hspace{1.5cm}\times \int_{y\by}e^{i\pp(y-\by)} s[\xi y,\xi\by]\nonumber\\
&&+\frac{g^2}{(2\pi)^3}2C_A\int_0^1 dx_B\, f^g_{\mu^2}(x_B)\, \int_0^{1-x_p} d\xi \,\frac{x_p}{1-\xi}\, \delta\left(x_B-\frac{x_p}{1-\xi}\right) \, \bigg[ \frac{1-\xi}{\xi}+\frac{\xi}{1-\xi}+\xi(1-\xi)\bigg](1-\xi)^2\nonumber\\
&&\hspace{1.5cm}\times \tilde{C}_{\mu^2}(\xi,x_B)\int_{y\by} e^{i\pp(y-\by)} \; s_A\Big[(1-\xi)y,(1-\xi)\by\Big]\nonumber\\
&&-\frac{g^2}{(2\pi)^3}C_A\int_0^1 dx_B \, f^{g}_{\mu^2}(x_B) \, x_p \, \delta(x_B-x_p)\int_0^1 d\xi \bigg[ \frac{1-\xi}{\xi}+\frac{\xi}{1-\xi}+\xi(1-\xi)\bigg]\tilde{C}_{\mu^2}(\xi,x_B)\nonumber\\
&&\hspace{1.5cm}\times\int_{y\by}e^{i\pp(y-\by)}s_A[y,\by] \nonumber\\
&&-\frac{1}{2}\frac{g^2}{(2\pi)^3}\int_0^1 dx_B\, f^{g}_{\mu^2}(x_B)\, x_p\delta(x_B-x_p)\int_0^1 d\xi \Big[\xi^2+(1-\xi)^2\Big]\tilde{C}_{\mu^2}(\xi,x_B)\nonumber\\
&&\hspace{1.5cm}\times\int_{y\by}e^{i\pp(y-\by)}s_A[y,\by],
\eeq
where
\beq
\frac{d\bar{\sigma}^g}{d^2\pp d\eta}=\frac{d\bar{\sigma}_1^{q\to g, \rm{r}}}{d^2\pp d\eta}+\frac{d\bar{\sigma}_1^{g\to g, \rm{r}}}{d^2\pp d\eta}+\frac{d\bar{\sigma}_1^{g\to g, \rm{v}}}{d^2\pp d\eta}+\frac{d\bar{\sigma}_1^{g\to q, \rm{v}}}{d^2\pp d\eta}+\frac{d\bar{\sigma}_1^{g\to \bq, \rm{v}}}{d^2\pp d\eta}\; .
\eeq

\subsection{Fragmentation}
The hadron production cross section from the final state quark is
\beq
\frac{d\sigma^H_q}{d^2p_{h\perp} d\eta}&=&\frac{1}{(2\pi)^2}\int_{x_F}^1\frac{d\zeta}{\zeta^2}\: D^{D,q}_{H,\mu^2}(\zeta)
\int_0^1 dx_B\: \frac{x_F}{\zeta}\delta\left(x_B-\frac{x_F}{\zeta}\right)\: f^{D,q}_{\mu^2}(x_B)
\int_{y\by}e^{i\frac{p_{h\perp}}{\zeta}(y-\by)}s[y,\by] \nonumber\\
&+&\int_{x_F}^1\frac{d\zeta}{\zeta^2}D^q_{H,\mu^2}(\zeta)\frac{d\sigma^q_1}{d^2\pp d\eta}\left(\frac{p_{h\perp}}{\zeta},\frac{x_F}{\zeta}\right)\; .
\eeq
Using eq.\eqref{qFF}, this becomes
\beq
\label{FqF}
\frac{d\sigma^H_q}{d^2p_{h\perp} d\eta}&=&\frac{1}{(2\pi)^2}\int_{x_F}^1\frac{d\zeta}{\zeta^2}\: D^q_{H,\mu^2}(\zeta)
\int_0^1 dx_B\: \delta\left( x_B-\frac{x_F}{\zeta}\right)\frac{x_F}{\zeta}\: f^q_{\mu^2}(x_B)\int_{y\by}e^{i\frac{p_{h\perp}}{\zeta}(y-\by)}s[y,\by]\nonumber\\
&&+\int_{x_F}^1\frac{d\zeta}{\zeta^2}\: D^q_{H,\mu^2}(\zeta)\: \frac{d\bar{\sigma}^q}{d^2\pp d\eta}\left(\frac{p_{h\perp}}{\zeta},\frac{x_F}{\zeta}\right)\nonumber\\
&&-\frac{g^2}{(2\pi)^3}C_F\int_{x_F}^1\frac{d\zeta}{\zeta^2}\int_0^1 dx_B\: \int_{\zeta}^{1}\; \frac{d\xi}{\xi}\; D^g_{H,\mu^2}\left(\frac{\zeta}{\xi}\right)\: \frac{x_F}{\zeta}\: \delta\left(x_B-\frac{\zeta}{\xi}\right)\: \bigg[\frac{1+(1-\xi)^2}{\xi}\bigg]\: f^q_{\mu^2}(x_B)\nonumber\\
&&\times \tilde{C}_{\mu^2}(\xi,x_B)\int_{y\by}e^{i\frac{p_{h\perp}}{\zeta}(y-\by)}s[y,\by]\nonumber\\
&&+\frac{g^2}{(2\pi)^3}\frac{1}{2}\int_{x_F}^1\frac{d\zeta}{\zeta^2} \int_0^1 dx_B\: D^q_H(\zeta)\: f^g_{\mu^2}(x_B) \int_0^{1-\frac{x_F}{\zeta}} d\xi \frac{x_F}{\zeta(1-\xi)}\: \delta\left(x_B-\frac{x_F}{\zeta(1-\xi)}\right)\: \Big[\xi^2+(1-\xi)^2\Big](1-\xi)^2\nonumber\\
&&\times \tilde{C}_{\mu^2}(\xi,x_B)\int_{y\by}e^{i\frac{p_{h\perp}}{\zeta}(y-\by)}s_A[(1-\xi)y,(1-\xi)\by]\; .
\eeq
The hadron production cross section from the final state gluon is
\beq
\frac{d\sigma^H_g}{d^2p_{h\perp} d\eta}&=&\frac{1}{(2\pi)^2}\int_{x_F}^1\frac{d\zeta}{\zeta^2}\: D^{D,g}_{H,\mu^2}(\zeta) \int_0^1 dx_B\: \frac{x_F}{\zeta}\: \delta\left(x_B-\frac{x_F}{\zeta}\right)\: f^{D,g}_{\mu^2}(x_B) \int_{y\by}e^{i\frac{p_{h\perp}}{\zeta}(y-\by)}s_A[y,\by] \nonumber\\
&+&\int_{x_F}^1\frac{d\zeta}{\zeta^2}\: D^g_{H,\mu^2}(\zeta)\: \frac{d\sigma^g_1}{d^2\pp d\eta}\left(\frac{p_{h\perp}}{\zeta},\frac{x_F}{\zeta}\right)\; .
\eeq
Using eq.\eqref{gFF}, we have
\beq
\label{FgF}
\frac{d\sigma^H_g}{d^2p_{h\perp} d\eta}&=&\frac{1}{(2\pi)^2}\int_{x_F}^1\frac{d\zeta}{\zeta^2}\: D^{g}_{H,\mu^2}(\zeta) \int_0^1 dx_B\: \frac{x_F}{\zeta}\: \delta\left(x_B-\frac{x_F}{\zeta}\right)\: f^{g}_{\mu^2}(x_B) \int_{y\by}e^{i\frac{p_{h\perp}}{\zeta}(y-\by)}s_A[y,\by] \nonumber\\
&&+\int_{x_F}^1\frac{d\zeta}{\zeta^2}\: D^g_{H,\mu^2}(\zeta)\: \frac{d\bar{\sigma}^g}{d^2\pp d\eta}\left(\frac{p_{h\perp}}{\zeta},\frac{x_F}{\zeta}\right)\nonumber\\
&&-\frac{g^2}{(2\pi)^3}\frac{1}{2}\int_{x_F}^1 \frac{d\zeta}{\zeta^2}\int_0^{1-\zeta}\frac{d\xi}{1-\xi}\: D^q_{H,\mu^2}\left(\frac{\zeta}{1-\xi}\right) \int_0^1 dx_B \Big[\xi^2+(1-\xi)^2\Big] \frac{x_F}{\zeta}\: \delta\left(x_B-\frac{x_F}{\zeta}\right)\: f^g_{\mu^2}(x_B)\nonumber\\
&&\times \tilde{C}_{\mu^2}(\xi,x_B) \int_{y\by} e^{i\frac{p_{h\perp}}{\zeta}(y-\by)}s_A[y,\by]\nonumber\\
&&+\frac{g^2}{(2\pi)^3}C_F \int_{x_F}^1\frac{d\zeta}{\zeta^2}\: D^g_{H,\mu^2}(\zeta) \int_0^1 dx_B\: f^q_{\mu^2}(x_B) \int_{\frac{x_F}{\zeta}}^1 d\xi\: \frac{x_F}{\zeta\xi}\: \delta\left(x_B-\frac{x_F}{\xi\zeta}\right) \bigg[\frac{1+(1-\xi)^2}{\xi}\bigg] \xi^2\nonumber\\
&&\times \tilde{C}_{\mu^2}(\xi,x_B) \int_{y\by} e^{i\frac{p_{h\perp}}{\zeta}(y-\by)}s[\xi y,\xi\by]\; .
\eeq
The remaining collinear divergent terms cancel between eq.\eqref{FqF} and eq.\eqref{FgF}. The final formula reads
\beq
\frac{d\sigma^H}{d^2p_{h\perp} d\eta}&=&\frac{1}{(2\pi)^2} \int_{x_F}^1 \frac{d\zeta}{\zeta^2}\: D^q_{H,\mu^2}(\zeta)\: \frac{x_F}{\zeta}\: f^q_{\mu^2}\left(\frac{x_F}{\zeta}\right) \int_{y\by} e^{i\frac{p_{h\perp}}{\zeta}(y-\by)} s[y,\by]
+\int_{x_F}^1\frac{d\zeta}{\zeta^2}\: D^q_{H,\mu^2}(\zeta)\: \frac{d\bar{\sigma}^q}{d^2\pp d\eta}\left(\frac{p_{h\perp}}{\zeta},\frac{x_F}{\zeta}\right)\nonumber\\
&&\hspace{-1cm}+\frac{1}{(2\pi)^2} \int_{x_F}^1 \frac{d\zeta}{\zeta^2}\: D^g_{H,\mu^2}(\zeta)\: \frac{x_F}{\zeta}\: f^g_{\mu^2}\left(\frac{x_F}{\zeta}\right)\: \int_{y\by} e^{i\frac{p_{h\perp}}{\zeta}(y-\by)} s_A[y,\by]
+\int_{x_F}^1\frac{d\zeta}{\zeta^2}\: D^g_{H,\mu^2}(\zeta)\: \frac{d\bar{\sigma}^g}{d^2\pp d\eta}\left(\frac{p_{h\perp}}{\zeta},\frac{x_F}{\zeta}\right)\; .
\eeq

In the final account we have to add the contribution from antiquark production, and also in principle the initial state antiquark, although it is small for  forward hadron production. This final expression is given in the body of the paper.


\section{Appendix C: Relating our collinear factorisation scheme to  $\overline{MS}$.}

\subsection{Isolating the collinear contributions in D dimensions}

We  focus on the quark to quark channel only, but the other channels work in an analogous way.
In this channel, before any type of collinear or low-$x$ resummation, there are four types of terms at NLO: real or virtual and containing just one dipole scattering amplitude or more than one. None of the (real or virtual) terms with more than one dipole scattering amplitude have any collinear divergence, so we  ignore them in this discussion. All the real NLO terms with just one dipole contain a collinear divergence. On the other hand, the virtual NLO terms with just one dipole can be split into a piece which is only collinearly divergent, and a BK-like piece which is  divergent in the ultraviolet (UV) and low-$x$ sensitive but collinearly finite. Discarding the latter for the moment, one can write together the LO contribution and the collinearly divergent NLO contributions to the $q\rightarrow q$ channel for the cross section in dimensional regularisation as
\begin{eqnarray}\label{sigma_coll_0_bis}
& &\left.\frac{d\sigma^{p\to H}}{d^2 p_{h\perp} d\eta}\right|_{q\rightarrow q, \, \textrm{LO + coll NLO}}
 =\lim_{D\rightarrow 4} \frac{1}{(2\pi)^{D-2}}  \int d^{D-2}y \int d^{D-2}\by  \; s\big[y,\by\big]   \int_{0}^1 \frac{d\zeta}{\zeta^{D-2}}\; D^{0,q}_{H}(\zeta)   \int_0^1 dx_B \; f^{0,q}(x_B) \nonumber\\
& & \hspace{1.2cm} \times\, \bigg\{ \frac{x_F}{\zeta}\; \delta\left( x_B\!-\!\frac{x_F}{\zeta}\right)\; e^{i(y-\by)p_{h\perp}/\zeta}
+\frac{g^2}{(2\pi)}\,  {\cal I}^{\textrm{coll}}_{y \by}
 \int_0^{1} d\xi \; \frac{x_F}{\zeta (1\!-\!\xi)}\; \delta\left( x_B\!-\!\frac{x_F}{\zeta (1\!-\!\xi)}\right) \;  P_{qq}(1\!-\!\xi;D) \nonumber\\
&& \hspace{2.5cm} \times \left[e^{i(y-\by)p_{h\perp}/\zeta}+\frac{1}{(1\!-\! \xi)^2}\, e^{i(y-\by)p_{h\perp} x_B/x_F} \right]  \bigg\}\; ,
\end{eqnarray}
where we have defined
\begin{equation}
{\cal I}^{\textrm{coll}}_{y \by}\equiv \left(\mu_{DR}^2\right)^{2-\frac{D}{2}} \int d^{D-2}z \;\;  A^i(y\!-\!z)\; A^i(\by\!-\!z) \label{I_coll_def}
\end{equation}
 and the $D$-dimensional splitting function
 \begin{equation}
P_{qq}(x;D)\equiv C_F\,  \left[ \frac{1+x^2}{1-x}+\frac{(D\!-\!4)}{2}(1-x) \right]_+ \, , \label{Pqq_D}
\end{equation}
and $\mu_{DR}$ is the scale used to keep the coupling dimensionless in dimensional regularization. The $+$ prescription comes, as usual, from the combination of the real and virtual terms together.

The explicit calculation of the integral ${\cal I}^{\textrm{coll}}_{y \by}$ can be done for example by introducing the momentum space representation of the
Weizs\"acker-Williams fields
and gives (with $D=4-2\epsilon$) \cite{Collins:2011zzd}
\begin{eqnarray}
{\cal I}^{\textrm{coll}}_{y \by}&=&\frac{1}{4\pi}\;  \Gamma\left(\frac{D}{2}\!-\!2\right) \Big[\pi\, \mu_{DR}^2\,  (y\!-\!\by)^2 \Big]^{2-\frac{D}{2}}
\nonumber\\
&=&-\frac{1}{4\pi} \bigg[\frac{S_{\epsilon}}{\epsilon}+\log \left(\mu_{DR}^2 \frac{(y\!-\!\by)^2}{4}\right) -2\Psi(1) +O(\epsilon)
\bigg]  \, , \label{I_coll_result}
\end{eqnarray}
where $S_{\epsilon}$ is the usual factor taking care of the universal constants in the $\overline{MS}$
renormalization scheme:
\begin{equation}
S_{\epsilon}\equiv  \frac{(4\pi)^{\epsilon}}{\Gamma(1\!-\!\epsilon)}=\epsilon \left[\frac{1}{\epsilon}+\Psi(1)+\ln 4\pi +\cal{O}(\epsilon)\right]\, .  \label{S_epsilon}
\end{equation}
Remember that the $\epsilon\rightarrow 0$ pole in ${\cal I}^{\textrm{coll}}_{y \by}$ is a collinear pole, not an UV one.

\subsection{Generic collinear factorization scheme in D dimensions}

The next step is to reabsorb the collinear divergences by a redefinition of the PDFs and FFs.
However, there is some freedom in this process, so that the renormalized distributions are scheme-dependent. In a generic collinear factorization scheme "$S$", the relations between the bare and dressed PDFs and FFs read
\begin{eqnarray}
  f^{q}_S(x_B;\mu_{DR}^2,\cdots)&\equiv & f^{D}(x_B) + \frac{g^2}{(2\pi)}\; \tilde{C}_S(\cdots)\: \int_0^{1-x_B} \frac{ d\xi}{(1\!-\!\xi)} \;  P_{qq}(1\!-\!\xi;D)\;  f^{D}\left(\frac{x_B}{(1\!-\!\xi)}\right) \nonumber\\
  & &\qquad + q \leftarrow g \textrm{ channel} \quad + O(g^4)\label{def_r_pdf_S}
\end{eqnarray}
and
\begin{eqnarray}
  D^{q}_{H,S}(\zeta;\mu_{DR}^2,\cdots)&\equiv &  D^{D,q}_{H}(\zeta) + \frac{g^2}{(2\pi)}\; \tilde{C}_S(\cdots)\: \int_{0}^{1-\zeta} \frac{ d\xi}{(1\!-\!\xi)^{5-D}} \;  P_{qq}(1\!-\!\xi;D)\; D^{D,q}_{H}\left(\frac{\zeta}{(1\!-\!\xi)}\right) \nonumber\\
  & &\qquad + g \leftarrow q \textrm{ channel} \quad + O(g^4)\, .\label{def_r_ff_S}
\end{eqnarray}
In these two relations, $\tilde{C}_S(\cdots)$ plays the role of a counter-term for collinear divergence. Hence, it must include a collinear single pole for $\epsilon\rightarrow 0$. Apart from that, $\tilde{C}_S(\cdots)$ is quite arbitrary, and the choice of a precise $\tilde{C}_S(\cdots)$ defines a collinear factorization scheme "S". $\tilde{C}_S(\cdots)$ can depend on some parameters, hence the "$\cdots$".Since the relations \eqref{def_r_pdf_S} and \eqref{def_r_ff_S} are written in $D$ dimensions, the renormalized distributions $f^{q}_S$ and $D^{q}_{H,S}$ a priori depend on the scale $\mu_{DR}$ (and on $D$). They also depend on the parameters present in the definition of $\tilde{C}_S$, if any.

One then gets for the cross section
\begin{eqnarray}\label{sigma_coll_r_pdf_r_ff_4D}
& &\left.\frac{d\sigma^{p\to H}}{d^2 p_{h\perp} d\eta}\right|_{q\rightarrow q, \, \textrm{LO + coll NLO}}
 = \frac{1}{(2\pi)^2}  \int d^2y \int d^2\by  \; s\big[y,\by\big]   \int_{0}^1 \frac{d\zeta}{\zeta^2}\; D^{q}_{H,S}(\zeta;\mu_{DR}^2,\cdots)    \nonumber\\
& & \hspace{1.2cm} \times\, \int_0^1 dx_B \; f^{q}_S(x_B;\mu_{DR}^2,\cdots) \Bigg\{ \frac{x_F}{\zeta}\; \delta\left( x_B\!-\!\frac{x_F}{\zeta}\right)\; e^{i(y-\by)p_{h\perp}/\zeta} \nonumber\\
& & \hspace{1.6cm} +\frac{g^2}{(2\pi)}\;\bigg[ \lim_{D\rightarrow 4}\Big[{\cal I}^{\textrm{coll}}_{y \by}- \tilde{C}_S(\cdots)\Big] \bigg]
 \int_0^{1} d\xi \; \frac{x_F}{\zeta (1\!-\!\xi)}\; \delta\left( x_B\!-\!\frac{x_F}{\zeta (1\!-\!\xi)}\right) \;  P_{qq}(1\!-\!\xi)\nonumber\\
& & \hspace{1.6cm}
\times \left( e^{i(y-\by)p_{h\perp}/\zeta}+\frac{1}{(1\!-\! \xi)^2}\, e^{i(y-\by)p_{h\perp} x_B/x_F} \right)
   \Bigg\}\; ,
\end{eqnarray}
where $P_{qq}(x)\equiv P_{qq}(x,4)$ is the usual $4$-dimensional splitting function.

The $q\rightarrow q$ NLO terms not included explicitly in eq. \eqref{sigma_coll_r_pdf_r_ff_4D} are collinear finite, as already discussed.
They contain UV divergencies, but these divergencies cancel between the virtual terms with one and more than one dipole scattering amplitude.

Now, all what is left to do is to discuss possible choices of collinear factorization schemes, or equivalently choices for the counter-term $\tilde{C}_S(\cdots)$.

\subsection{Generalization of our scheme with cut-off in position space}

In order to discuss our scheme in this setup, one has to find a $D$-dimensional generalization of our counter-term $\tilde{C}$. By comparison with eq. \eqref{I_coll_def}, it is clear that one should take
\begin{equation}
\tilde{C}_X(\mu_{CO}^2)\equiv \left(\mu_{DR}^2\right)^{2-\frac{D}{2}} \int d^{D-2}r \;\;  A^i(r)\; A^i(r)\;  \theta (\mu_{CO}^2\, r^2-1)  \, . \label{C_X_def}
\end{equation}
We call this scheme $X$, because it involves a cut-off $\mu_{CO}$ in position space. This cut-off has nothing to do with dimensional regularization, so that the scales $\mu_{CO}$ and $\mu_{DR}$ have to be independent. In this case, $\mu_{CO}$ plays the role of an external parameter in the definition of $\tilde{C}_X$. Using this scheme "$X$" both for the PDFs and FFs, one expects a priori renormalized PDFs and FFs which depends on both $\mu_{CO}$ and $\mu_{DR}$.

In $D$ dimensions, the Weizs\"acker-Williams field reads
\begin{eqnarray}\label{WW}
A^i(r)&\equiv &-i\int
\frac{d^{D-2}l_\perp}{(2\pi)^{D-2}}\; \frac{l_\perp^i}{l_\perp^2}\; e^{-il_\perp r}
=-\frac{\pi^{2-\frac{D}{2}}}{2\pi}\; \Gamma\left(\frac{D}{2}\!-\!1\right)\; \frac{r^i}{(r^2)^{\frac{D}{2}-1}}
  \, .
\end{eqnarray}
Plugging this expression into the definition \eqref{C_X_def}, it is straightforward to get
\begin{equation}
\tilde{C}_X(\mu_{CO}^2)= -\frac{1}{4\pi}\; \frac{\Gamma\left(\frac{D}{2}\!-\!1\right)}{\left(2\!-\!\frac{D}{2}\right)}\; \left(\frac{\pi\, \mu_{DR}^2}{\mu_{CO}^2}\right)^{2-\frac{D}{2}}= -\frac{1}{4\pi}\,
\bigg[ \frac{S_{\epsilon}}{\epsilon}+\log \left( \frac{\mu_{DR}^2}{4 \mu_{CO}^2}\right) -2\Psi(1) +O(\epsilon)
\bigg]
  \, . \label{C_X_result}
\end{equation}
The collinear pole at $\epsilon\rightarrow 0$ indeed cancel between $\tilde{C}_X(\mu_{CO}^2)$ and ${\cal I}^{\textrm{coll}}_{y \by}$ so that
\begin{equation}
\lim_{D\rightarrow 4}\Big[{\cal I}^{\textrm{coll}}_{y \by}- \tilde{C}_X(\cdots)\Big]=  -\frac{1}{4\pi}\; \log \left(\mu_{CO}^2\, (y\!-\!\by)^2 \right)\, .
\end{equation}
It is important to note that the scale $\mu_{DR}$ drops in this difference. It comes from the fact that it was not really necessary to go to dimensional regularization to define the scheme "$X$". An important consequence is that the renormalized PDFs and FFs in the scheme "$X$" should not depend on $\mu_{DR}^2$, after all, but only on the parameter $\mu_{CO}^2$. Hence, one has
\begin{eqnarray}\label{sigma_coll_X}
& &\left.\frac{d\sigma^{p\to H}}{d^2 p_{h\perp} d\eta}\right|_{q\rightarrow q, \, \textrm{LO + coll NLO}}
 = \frac{1}{(2\pi)^2}  \int d^2y \int d^2\by  \; s\big[y,\by\big]   \int_{0}^1 \frac{d\zeta}{\zeta^2}\; D^{q}_{H,X}(\zeta;\mu_{CO}^2)    \nonumber\\
& & \hspace{1.2cm} \times\, \int_0^1 dx_B \; f^{q}_X(x_B;\mu_{CO}^2) \Bigg\{ \frac{x_F}{\zeta}\; \delta\left( x_B\!-\!\frac{x_F}{\zeta}\right)\; e^{i(y-\by)p_{h\perp}/\zeta} \nonumber\\
& & \hspace{1.6cm} -\frac{g^2}{2(2\pi)^2}\;  \log \left(\mu_{CO}^2\, (y\!-\!\by)^2 \right)
 \int_0^{1} d\xi \; \frac{x_F}{\zeta (1\!-\!\xi)}\; \delta\left( x_B\!-\!\frac{x_F}{\zeta (1\!-\!\xi)}\right) \;  P_{qq}(1\!-\!\xi)\nonumber\\
& & \hspace{1.6cm}
\times \left( e^{i(y-\by)p_{h\perp}/\zeta}+\frac{1}{(1\!-\! \xi)^2}\, e^{i(y-\by)p_{h\perp} x_B/x_F} \right)
   \Bigg\}\; .
\end{eqnarray}
That expression (together with the other channels) is independent of the value of $\mu_{CO}$ (up to NNLO contributions) if the PDFs and FFs to satisfy the DGLAP equations with respect to $\mu_{CO}^2$. In that case, one is even free to evolve the PDFs and FFs independently, to two different factorization scales $\mu_{F}$ and $\mu_{\textrm{frag}}$, and get
\begin{eqnarray}\label{sigma_coll_X_bis}
& &\left.\frac{d\sigma^{p\to H}}{d^2 p_{h\perp} d\eta}\right|_{q\rightarrow q, \, \textrm{LO + coll NLO}}
 = \frac{1}{(2\pi)^2}  \int d^2y \int d^2\by  \; s\big[y,\by\big]   \int_{0}^1 \frac{d\zeta}{\zeta^2}\; D^{q}_{H,X}(\zeta;\mu_{\textrm{frag}}^2)    \nonumber\\
& & \hspace{1.2cm} \times\, \int_0^1 dx_B \; f^{q}_X(x_B;\mu_{F}^2) \Bigg\{ \frac{x_F}{\zeta}\; \delta\left( x_B\!-\!\frac{x_F}{\zeta}\right)\; e^{i(y-\by)p_{h\perp}/\zeta} \nonumber\\
& & \hspace{1.6cm} -\frac{g^2}{2(2\pi)^2}\;  \log \left(\mu_{F}^2\, (y\!-\!\by)^2 \right)\; e^{i(y-\by)p_{h\perp}/\zeta}
 \int_0^{1} d\xi \; \frac{x_F}{\zeta (1\!-\!\xi)}\; \delta\left( x_B\!-\!\frac{x_F}{\zeta (1\!-\!\xi)}\right) \;  P_{qq}(1\!-\!\xi)\nonumber\\
& & \hspace{1.6cm} -\frac{g^2}{2(2\pi)^2}\;  \log \left(\mu_{\textrm{frag}}^2\, (y\!-\!\by)^2 \right)\; e^{i(y-\by)p_{h\perp} x_B/x_F}
 \int_0^{1} d\xi \; \frac{x_F}{\zeta (1\!-\!\xi)^3}\; \delta\left( x_B\!-\!\frac{x_F}{\zeta (1\!-\!\xi)}\right) \;  P_{qq}(1\!-\!\xi)
   \Bigg\}\; .
\end{eqnarray}
That expression is indeed independent, up to NNLO terms, on the values of $\mu_{F}$ and of $\mu_{\textrm{frag}}$ separately, thanks to the initial state and final state DGLAP equations.

\subsection{$\overline{MS}$ scheme}

The most commonly used factorization scheme for the PDFs and FFs is the $\overline{MS}$ scheme. In that case, the counterterm $\tilde{C}_{\overline{MS}}$ is taken to cancel only the collinear pole of ${\cal I}^{\textrm{coll}}_{y \by}$ and the universal constants:
\begin{equation}
\tilde{C}_{\overline{MS}}\equiv  -\frac{1}{4\pi}\; \frac{S_{\epsilon}}{\epsilon} \, . \label{C_MSbar_def}
\end{equation}
Hence, one has
\begin{equation}
\lim_{D\rightarrow 4}\Big[{\cal I}^{\textrm{coll}}_{y \by}- \tilde{C}_{\overline{MS}}\Big]= -\frac{1}{4\pi} \bigg[\log \left(\mu_{DR}^2 \frac{(y\!-\!\by)^2}{4}\right) -2\Psi(1) \bigg] \, .\label{subtr_MSbar}
\end{equation}
It is necessary to go to dimensional regularization to define this scheme, and indeed the scale $\mu_{DR}$ stays in the difference \eqref{subtr_MSbar}.
In the $\overline{MS}$ scheme, the equation \eqref{sigma_coll_r_pdf_r_ff_4D} then reads
\begin{eqnarray}\label{sigma_coll_MSbar}
& &\left.\frac{d\sigma^{p\to H}}{d^2 p_{h\perp} d\eta}\right|_{q\rightarrow q, \, \textrm{LO + coll NLO}}
 = \frac{1}{(2\pi)^2}  \int d^2y \int d^2\by  \; s\big[y,\by\big]   \int_{0}^1 \frac{d\zeta}{\zeta^2}\; D^{q}_{H,\overline{MS}}(\zeta;\mu_{DR}^2)    \nonumber\\
& & \hspace{1.2cm} \times\, \int_0^1 dx_B \; f^{q}_{\overline{MS}}(x_B;\mu_{DR}^2) \Bigg\{ \frac{x_F}{\zeta}\; \delta\left( x_B\!-\!\frac{x_F}{\zeta}\right)\; e^{i(y-\by)p_{h\perp}/\zeta} \nonumber\\
& & \hspace{1.6cm} -\frac{g^2}{2(2\pi)^2}\;  \bigg[\log \left(\mu_{DR}^2 \frac{(y\!-\!\by)^2}{4}\right) -2\Psi(1) \bigg]
 \int_0^{1} d\xi \; \frac{x_F}{\zeta (1\!-\!\xi)}\; \delta\left( x_B\!-\!\frac{x_F}{\zeta (1\!-\!\xi)}\right) \;  P_{qq}(1\!-\!\xi)\nonumber\\
& & \hspace{1.6cm}
\times \left( e^{i(y-\by)p_{h\perp}/\zeta}+\frac{1}{(1\!-\! \xi)^2}\, e^{i(y-\by)p_{h\perp} x_B/x_F} \right)
   \Bigg\}\; .
\end{eqnarray}
It is now $\mu_{DR}^2$ which plays the role of evolution variable for the DGLAP evolution of the PDFs and FFs. Using the DGLAP equations to evolve the PDFs and FFs to the scales $\mu_{F}$ and $\mu_{\textrm{frag}}$ respectively, one gets
\begin{eqnarray}\label{sigma_coll_MSbar_bis}
& &\left.\frac{d\sigma^{p\to H}}{d^2 p_{h\perp} d\eta}\right|_{q\rightarrow q, \, \textrm{LO + coll NLO}}
 = \frac{1}{(2\pi)^2}  \int d^2y \int d^2\by  \; s\big[y,\by\big]   \int_{0}^1 \frac{d\zeta}{\zeta^2}\; D^{q}_{H,\overline{MS}}(\zeta;\mu_{\textrm{frag}}^2)    \nonumber\\
& & \hspace{0cm} \times\, \int_0^1 dx_B \; f^{q}_{\overline{MS}}(x_B;\mu_{F}^2) \Bigg\{ \frac{x_F}{\zeta}\; \delta\left( x_B\!-\!\frac{x_F}{\zeta}\right)\; e^{i(y-\by)p_{h\perp}/\zeta} \nonumber\\
& & \hspace{0.5cm} -\frac{g^2}{2(2\pi)^2}\;  \bigg[\log \left(\mu_{F}^2 \frac{(y\!-\!\by)^2}{4}\right) -2\Psi(1) \bigg]\; e^{i(y-\by)p_{h\perp}/\zeta}
 \int_0^{1} d\xi \; \frac{x_F}{\zeta (1\!-\!\xi)}\; \delta\left( x_B\!-\!\frac{x_F}{\zeta (1\!-\!\xi)}\right) \;  P_{qq}(1\!-\!\xi)\\
& & \hspace{0.5cm} -\frac{g^2}{2(2\pi)^2}\;  \bigg[\log \left(\mu_{\textrm{frag}}^2 \frac{(y\!-\!\by)^2}{4}\right) -2\Psi(1) \bigg]\; e^{i(y-\by)p_{h\perp} x_B/x_F}
 \int_0^{1} d\xi \; \frac{x_F}{\zeta (1\!-\!\xi)^3}\; \delta\left( x_B\!-\!\frac{x_F}{\zeta (1\!-\!\xi)}\right) \;  P_{qq}(1\!-\!\xi)
   \Bigg\}\; .\nonumber
\end{eqnarray}

\subsection{Matching of the factorization schemes}

The renormalized PDFs and FFs are scheme dependent, as well as the NLO "coefficient functions", but observables like the single inclusive cross section should be scheme independent (up to higher orders in $g^2$ than the one of interest). Hence, one can understand the translation from one scheme to another by comparing eqs. \eqref{sigma_coll_X_bis} and \eqref{sigma_coll_MSbar_bis}\footnote{The NLO terms omitted from these equations are indeed collinear safe, and thus coincide in the two collinear factorization schemes, up to NNLO effects induced by the use of different PDFs and FFs.}. It is clear that, up to NNLO correction, it is consistent to assume that the PDFs and FFs in the two schemes are functionally related to each other by a mere rescaling of their evolution variable, as
\begin{eqnarray}
f^{q}_X(x_B;\mu_{F}^2) &=&  f^{q}_{\overline{MS}}(x_B;R^2\, \mu_{F}^2)   \label{conversion_pdf},\\
D^{q}_{H,X}(\zeta;\mu_{\textrm{frag}}^2) &=& D^{q}_{H,\overline{MS}}(\zeta;R^2\, \mu_{\textrm{frag}}^2)   \label{conversion_ff}\, .
\end{eqnarray}
One can use these relations to rewrite eq. \eqref{sigma_coll_X_bis} with  $\overline{MS}$-scheme PDFs and FFs, and then run the DGLAP equations to remove the rescaling factor $R$ from the PDFs and FFs. As a result, one gets
\begin{eqnarray}\label{sigma_coll_X_to_MSbar}
& &\left.\frac{d\sigma^{p\to H}}{d^2 p_{h\perp} d\eta}\right|_{q\rightarrow q, \, \textrm{LO + coll NLO}}
 = \frac{1}{(2\pi)^2}  \int d^2y \int d^2\by  \; s\big[y,\by\big]   \int_{0}^1 \frac{d\zeta}{\zeta^2}\; D^{q}_{H,\overline{MS}}(\zeta;\mu_{\textrm{frag}}^2)    \nonumber\\
& & \hspace{1.2cm} \times\, \int_0^1 dx_B \; f^{q}_{\overline{MS}}(x_B;\mu_{F}^2) \Bigg\{ \frac{x_F}{\zeta}\; \delta\left( x_B\!-\!\frac{x_F}{\zeta}\right)\; e^{i(y-\by)p_{h\perp}/\zeta} \nonumber\\
& & \hspace{1.6cm} -\frac{g^2}{2(2\pi)^2}\;  \log \left(\frac{\mu_{F}^2\, (y\!-\!\by)^2}{R^2} \right)\; e^{i(y-\by)p_{h\perp}/\zeta}
 \int_0^{1} d\xi \; \frac{x_F}{\zeta (1\!-\!\xi)}\; \delta\left( x_B\!-\!\frac{x_F}{\zeta (1\!-\!\xi)}\right) \;  P_{qq}(1\!-\!\xi)\\
& & \hspace{1.6cm} -\frac{g^2}{2(2\pi)^2}\;  \log \left(\frac{\mu_{\textrm{frag}}^2\, (y\!-\!\by)^2}{R^2} \right)\; e^{i(y-\by)p_{h\perp} x_B/x_F}
 \int_0^{1} d\xi \; \frac{x_F}{\zeta (1\!-\!\xi)^3}\; \delta\left( x_B\!-\!\frac{x_F}{\zeta (1\!-\!\xi)}\right) \;  P_{qq}(1\!-\!\xi)
   \Bigg\}\; . \nonumber
\end{eqnarray}
Due to the scheme-independence of the cross section, the expression \eqref{sigma_coll_X_to_MSbar} has to be identical to the one orginally obtained in the $\overline{MS}$ scheme, eq. \eqref{sigma_coll_MSbar_bis}. This allows us to determine the correct rescaling factor $R$ to be
\begin{equation}
R= 2\, e^{\Psi(1)}\simeq 1.1229 \; . \label{R_result}
\end{equation}
With this value, the functional relations \eqref{conversion_pdf} and \eqref{conversion_ff} provide the correct correspondence between the renormalized distributions in the $\overline{MS}$ scheme and in the "$X$" scheme.

The relations \eqref{conversion_pdf} and \eqref{conversion_ff} generalize to the gluon distributions, all with the same rescaling factor $R$ given in eq. \eqref{R_result}.


\end{document}